\documentclass[aps,amsmath,amssymb,mathrsfs,showpacs,showkeys]{revtex4-2}
\allowdisplaybreaks
\usepackage[dvips]{graphicx}
\usepackage{times}
\usepackage{braket}
\usepackage{xcolor}
\usepackage{orcidlink}
\usepackage{hyperref}
\usepackage{csquotes}
\usepackage{multirow}
\usepackage{rotating}
\hypersetup{
	colorlinks=true,
	urlcolor=magenta,
	linkcolor=red,
	citecolor=blue
}
\DeclareMathOperator{\sech}{sech}

\begin{document}
\title{Analysis of radial and quasiradial oscillations in quark stars for
different equations of state}
	
\author{Nivamani Rajbongshi}
\email[Email: ]{nrajbongshi909@gmail.com}
\affiliation{Department of Physics, Dibrugarh University, Dibrugarh 786004, 
Assam, India}
	
\author{Umananda Dev Goswami\orcidlink{0000-0003-0012-7549}}
\email[Email: ]{umananda@dibru.ac.in}
\affiliation{Department of Physics, Dibrugarh University, Dibrugarh 786004, 
Assam, India}

\begin{abstract}
We investigate quark stars (QSs) described by three quark matter (QM) 
equations of state (EoSs) and analyse their radial and quasiradial oscillation 
modes within the slowly rotating relativistic approximation. The considered 
EoSs describe massless quarks, massive quarks, and massive quarks with a 
temperature-dependent gluon mass. Our analysis of radial modes shows that the 
mode separation depends on the stellar mass and radius for a given 
compactness, as well as on the stellar temperature. The considered QS models 
satisfy the slowly rotating relativistic condition of Hartle, Thorne and 
Chitre formalism, allowing rotation to be treated as a small perturbation 
of the non-rotating configuration. We find that, although more massive and 
larger stars can exhibit greater surface eccentricity and quadrupole 
deformation, their stronger self-gravity results in a lower tendency to 
deform than less massive and smaller stars. For all models, the variation in 
mass distribution increases towards the stellar surface, while the variations 
in pressure, energy density, and number density remain small near the center 
and become more pronounced towards the surface. More compact stars exhibit 
stronger twisting of the local inertial frame but weaker frame dragging for 
a given angular velocity. Rotation shifts the oscillation modes from their 
non-rotating values and leads to different vibrational decay times at 
different angular velocities. The fundamental modes considered here exhibit 
frequencies below $2$ kHz and show non-continuous behaviour across the 
investigated rotational configurations.
\end{abstract}
	
\keywords{Quark stars; Radial oscillations; Quasiradial oscillations.}  
	
\maketitle    

\section{Introduction}
The core of neutron stars (NSs) can reach extremely high densities, where 
nuclear matter (NM) may undergo a phase transition into quark matter (QM), 
leading to the possible formation of quark stars (QSs). A wide range of 
equations of state (EoSs) has been proposed to describe such QM configurations 
\cite{chodos_1974, farhi_1984, witten_1984, haensel_1986, bora_2022, 
alcock_1986, aziz_2019, deb_2010, numba_1961, numbda_1961_ii, maurya_2022}. 
Despite extensive theoretical work, current observational evidence is still 
insufficient to confirm or rule out the existence of QSs. However, strong 
experimental indications of quark-gluon plasma formation in high-energy 
heavy-ion collision experiments \cite{adams_2005, adcox_2005, back_2005, 
arsene_2005, adam_2003, ackermann_2001, gyulassy_2005, romatschke_2007, 
aamodt_2010, chatrchyan_2011, alice_2016, alice_2008, shuryak_2005, 
heinz_2013} provide motivation to explore the possibility of QM existing in 
extreme astrophysical environments, including the interiors of compact stars. 
If QSs do exist, a key question is what theoretical framework can accurately 
describe their internal dynamics and what the relevant degrees of freedom are, 
including the properties such as rest masses and interactions of the 
constituent QM. Addressing these issues is crucial not only for understanding 
such hypothetical objects but also for extending the evolutionary sequence of 
compact stars beyond NSs, potentially linking NS physics with black hole (BH) 
formation. In this work, we focus on QSs and investigate their internal 
dynamics based on the properties of constituent QM and their 
thermodynamic behavior within the framework of general relativity.

Numerous studies have investigated the dynamical and thermodynamical 
properties of non-rotating NSs within the framework of general relativity 
\cite{zheng_2023,zhao_2022,zhang_2024,bora_2021,mannarelli_2014,
mannarelli_2018,pani_2018}. In this context, the derivation of the 
Tolman-Oppenheimer-Volkoff (TOV) equation \cite{tolman_1939,oppenheimer_1939} 
marked a major milestone in describing the internal structure of static 
relativistic stars. Subsequently, Chandrasekhar developed the formalism for 
studying the radial pulsations of compact stars \cite{chandrasekhar_1964}. 
Although radial oscillations are difficult to observe directly, the inclusion 
of rotation can lead to the emission of gravitational waves (GWs) through 
the changes in the internal structure and oscillation properties of the stars. 
In this direction, the formalism developed by Hartle in 1967 
\cite{hartle_1967}, followed by Hartle and Thorne in 1968 \cite{hartle_1968} 
provide a framework for describing slowly rotating relativistic stars, while 
in 1972, Hartle, Thorne and Chitre (HTC) \cite{hartle_1972} derived 
the equations governing quasiradial oscillation (QRO) modes. These quasiradial 
modes contain complex oscillation frequencies, where the real component
gives the actual oscillation frequency and the imaginary component 
describes the temporal evolution and damping of the oscillation modes due to 
various mechanisms. In our work, we assumed that the origin of QSs is from 
the NSs with mass greater than $2 M_{\odot}$. This assumption is based on the
fact that the observational results of the Fermi gamma-ray telescope 
\cite{smith_2023} have shown that there exist some pulsars, such as PSR 
J2215+5135, PSR J1311-3430, PSR J1810+1744 and PSR J0952-0607, which have 
masses above $2 M_{\odot}$ and pulsating times in the range between $2.61$ ms 
to $1.41$ ms \cite{sullivan_2024,romani_2012, romani_2021,romani_2026,
smith_2023,salmi_2024,sanchez_2022}.

In 1984, E.~Witten predicted that NSs could contain a small nucleus 
of QM when they formed \cite{witten_1984}. The study of NS's core by E.~Annala 
et~al.~\cite{annala_2020} in a method based on the behavior of the speed of 
sound inside the star, in a model-independent way by combining astrophysical 
observations and theoretical ab initio calculations, reveals that the matter 
inside maximally massive stable NSs $\sim 2 M_{\odot}$ displays 
characteristics of the deconfined quark phase. Additionally, they predicted a 
$6.5$ km quark core for NSs with radius $12$ km and  mass $2 M_{\odot}$, in a 
subconformal ($c_{s} < 1/3$, where $c_{s}$ is the speed of sound) EoS. 
Assuming the presence of QM inside, NSs may evolve into the QSs or hybrid 
stars (HSs). J.~Bora and U.~D.~Goswami studied strange stars (SSs) with
different EoSs to understand radial oscillations and gravitational wave 
echoes (GWEs) of these stars \cite{bora_2021}. They found that the SSs with 
compactness between $1/3$ and $4/9$ can emanate GWEs at different frequencies 
based on their mass and radius relation \cite{bora_2021,mannarelli_2018}. 
M.~Mannarelli and F.~Tonelli studied the GWEs emission from SSs within the MIT 
bag model. In their model, they obtained characteristic GWE frequencies of 
the order of $10$ kHz \cite{mannarelli_2018}. Furthermore, T.~Zhao et 
al.~\cite{zhao_2022} studied the crossover model under Gibbs construction and 
Cowling approximation for HSs. Z.~Y.~Zheng et~al.~\cite{zheng_2023} 
investigated the nonradial oscillations of pure and hybrid NSs. They found 
that the $g$-mode of nonradial oscillations may be detectable at a distance of 
approximately $10-15$ kpc, based on the energy range of the emitted 
GWs. Their results indicate that near the stellar core, 
hybrid NSs exhibit weaker damping of GW signals and larger background 
perturbations compared to the pure NSs. The above mentioned studies primarily 
focus on non-rotating stellar configurations based on the TOV equations. While 
discussing the hydrostatic equilibrium solutions of static stars, R.~C.~Tolman 
pointed out the physical limitations of such static solutions in 1939 
\cite{tolman_1939}. He noted that from the perspective of an external 
observer, a static configuration may exhibit a quasi-static character, as the 
changes occurring at the stellar center take place at an extremely slow rate 
\cite{tolman_1939}. In 1967, J.~B.~Hartle developed a formalism for slowly 
rotating relativistic stars, in which uniform rotation is treated as a small 
perturbation of a known spherically symmetric, non-rotating stellar 
configuration \cite{hartle_1967}. Subsequently, in 1968, H.~D.~Wahlquist 
derived an exact closed-form solution to the Einstein field equations for a 
stationary, axially symmetric, Petrov type-D spacetime \cite{wahlquist_1968}. 
However, the Wahlquist solution has significant physical limitations compared 
with Hartle's formalism for rotating stars. In particular, the mathematical 
rigidity of the Wahlquist solution imposes a highly restrictive relationship 
between pressure and energy density, which results in a prolate rather than an 
oblate stellar configuration. Moreover, the Wahlquist solution cannot be 
matched to an exterior Kerr metric, and no other vacuum exterior solution has 
been shown to provide a satisfactory matching at its boundary 
\cite{wahlquist_1968}. In contrast, Hartle's formalism is considerably more 
flexible and can accommodate one-parameter EoS \cite{hartle_1967,hartle_1968,
hartle_1972}. Within this framework, rotation naturally deforms the star into 
an oblate configuration, while the exterior spacetime is described by a 
perturbation of the Schwarzschild metric characterized by the stellar mass, 
angular momentum, and mass quadrupole moment. Therefore, incorporating 
rotation is essential for constructing a physically realistic description of 
rotating relativistic stars. Rotation not only deforms the stellar fluid 
configuration but also influences radial oscillations, GW emission, and other 
dynamical properties of compact stars. Thus, incorporating rotational 
effects is essential for obtaining a more accurate description of the stellar 
interior and the gravitational field surrounding rotating compact objects. 
H.~Heiselberg and M.~H.~Jensen investigated phase transitions in slowly 
rotating NSs within the framework of general relativity 
\cite{heiselberg_1998}. Using Hartle's formalism, they demonstrated that 
the effects of a phase transition can manifest through changes in the stellar 
moment of inertia and, consequently, in the angular momentum. More recently, 
Zheng et al.~investigated fundamental QROs of HSs 
and pure NSs within the slow rotation approximation \cite{zheng_2026}. They 
found that as a rotating NS spins down, its central density increases and 
eventually a hadron-quark phase transition takes place if the star is massive 
enough. However, distinguishing HSs from pure NSs becomes challenging because 
fundamental QRO frequency decreases as the rotating configuration approaches 
its non-rotating limit. Their analysis further revealed a distinct kink 
followed by a steep rise in the derivative of fundamental QRO frequency with 
respect to rotation frequency at the onset of the mixed phase. This provides a 
observational signature that differentiates HSs from their pure NS 
counterparts.\\
\indent As the baryon number density inside NSs can reach several times the 
nuclear saturation density near the core, one requires special EoSs to 
understand such ultra-dense compact stars. Various EoSs \cite{wei_2020, 
fraga_2001, togashi_2013, togashi_2014, togashi_2016, lu_2019, 
koliogiannis_2021} have been proposed to describe the interior composition 
and properties of QM stars. Here, we analyze the interior of the QSs with 
massive and massless quark flavors. We also checked the temperature dependence 
of QS's oscillation behaviours based on models of massive quarks. We study the 
radial oscillations of those stars in a rotating framework and calculate their 
QROs. We also study the correction terms and deformation 
parameters due to the rotation of the stars based on Hartle and Thorne's 
formalism of slowly rotating relativistic stars 
\cite{hartle_1967,hartle_1968}.\\ 
\indent The rest of the paper is organised as follows. We describe the 
non-rotating and rotating stellar frameworks in Section~\ref{secII}, followed 
by a discussion of the EoSs in Section~\ref{secIII}. A brief overview of radial 
and QROs is presented in Section~\ref{secIV}. The 
numerical results are discussed and concluding remarks are presented in 
Section~\ref{secV} and Section \ref{secVI}, respectively. Throughout this 
work, the geometrized unit system ($G = c = 1$) is used for solving the 
stellar dynamic equations, such as the TOV equations, while the EoSs and 
related quantities are expressed in the natural unit system ($\hbar = c = 1$). Physical observables are converted to CGS or SI units 
wherever appropriate. However, 
certain physical quantities are retained in geometrized or natural units to 
maintain clarity and suitable numerical order; such quantities are explicitly 
indicated along with their corresponding units.

\section{Theoretical framework for rotating stars} \label{secII}
The rotating configuration of a star can be acquired from the non-rotating
configuration as the rotating part of the metric is the perturbation to the 
non-rotating metric, and hence it is essential to understand the non-rotating
configuration of a star. The line element of the non-rotating system is 
given by the spherically symmetric Schwarzschild metric, which is given as
\begin{equation}
	\label{schw_metric}
	ds^{2} = -\,e^{\nu}dt^{2} + e^{\lambda} dr^{2} + r^{2}\! \left(d\theta^{\,2} + \sin^{2}d\phi^{2} \right),
\end{equation}
where 
\begin{equation}
	e^{\nu} = e^{-\lambda} = \left(1 - \frac{2M(r)}{r}\right)
\end{equation}
with $M(r)$ as the mass of the star enclosed by the radius $r$. 
In the stellar interior, the matter part can be considered as perfect fluid, 
and can be expressed using local pressure $p$ and energy density $\epsilon$ in
the form of the perfect fluid energy-momentum tensor as
\begin{equation}
T^{\varrho \varphi} = -\, p\,g^{\varrho \varphi} + \left(p + \epsilon\right)u^{\varrho}u^{\varphi}.
\label{emt}
\end{equation}
For the metric \eqref{schw_metric} with the energy-momentum tensor 
\eqref{emt}, the TOV equation can be obtained as
\begin{equation}\label{TOV_equation}
\frac{dp}{dr} = -\,\frac{1}{r^2}(\epsilon + p)\left[M(r) + 4 \pi p\, r^{3}\right]e^{\lambda} = -\,\frac{1}{r^2}(\epsilon + p)\left[M(r) + 4 \pi p\, r^{3}\right]\!\left(1 - \frac{2M(r)}{r}\right)^{-1}.
\end{equation}
By solving this TOV equation, one can get the interior hydrostatic equilibrium 
and mass-radius relations of a star. However, for this solution, it is 
required to have the following two coupled differential equations along with
the TOV equation:
\begin{align}
&\frac{dM(r)}{dr} = 4 \pi \epsilon\, r^{2}, \label{TOV_mass} \\[5pt]
&\frac{d\nu}{dr} = -\,\frac{2}{\epsilon + p}\,\frac{dp}{dr}. \label{TOV_metric}
\end{align}
For a given EoS, these equations, i.e.,~Eqs.~\eqref{TOV_equation},
\eqref{TOV_mass} and \eqref{TOV_metric}, which form a initial value problem, 
can be solved numerically with suitable boundary conditions. These suitable
boundary condtions are, (i) at the center ($r = 0$), $M(r) = 0$, $p = p_{c}$ 
and (ii) at boundary ($r = R$), $M(r)=M(R)\equiv M$, $p = 0$. Specifically, 
under these boundary 
conditions, Eqs.~\eqref{TOV_equation} and \eqref{TOV_mass} along with the 
given EoS can be solved to get the value of mass $M$ and radius $R$ of a star. 
Then the values of $M$ and $R$ can be used to calculate boundary value of the 
metric term $\nu$, which is given as
\begin{equation}
	\nu(R) = \ln \left(1 - \frac{2M}{R}\right).
\end{equation}
This boundary condition on $\nu$ helps in the determination of its central 
value. Then one can solve Eqs.~\eqref{TOV_equation}, \eqref{TOV_mass} and 
\eqref{TOV_metric} numerically to determine the mass, pressure and metric term 
$\nu$ distributions at different radial lengths from the center to the radius 
of a star.

In our work, we use a slowly rotating relativistic star framework given by 
J.~B.~Hartle and K.~S.~Thorne \cite{hartle_1967,hartle_1968}. Here, as 
mentioned already, the rotation part of the metric is considered as the 
perturbation to the non-rotating metric, and uses the perturbations up to 
second-order only in the analysis. The line element of the perturbed geometry 
of a stationary, slowly rotating, axially symmetric system, under the 
inertial frame dragging condition, is given as
\begin{equation}
	\label{rot_metric}
	ds^{2} = -\,e^{\nu}(1+2h) dt^{2} + e^{\lambda}\left[1 + \frac{2m}{r-2M(r)}\right] dr^{2} + r^{2} (1 + 2k) \left[ d\theta^{2} + \sin^{2}\theta \left(d\phi - \omega dt\right)^{2}\right],
\end{equation}
where perturbation terms $h$, $m$ and $k$ are functions of $r$ and $\theta$, 
and are expressed in the forms up to the quadrupole ($l=2$) term as
\begin{align}
		&h(r,\theta) = h_0(r) + h_2(r)P_2(\theta) + \hdots, \\[5pt]
		&m(r,\theta) = m_0(r) + m_2(r)P_2(\theta) + \hdots, \\[5pt]
		&k(r,\theta) = k_0(r) + k_2(r)P_2(\theta) + \hdots,
\end{align}
with $h_0$, $m_0$ and $k_0$ are monopole ($l=0$), and $h_2$, $m_2$ and $k_2$ 
are quadrupole ($l=2$) correction or perturbation terms. $P_{2}(\theta) 
\equiv P_{2}(\cos\theta) = (3\cos^2\theta -1)/2$ is the Legendre polynomial 
of order 2. As the density and metric of an axially symmetric, stationary 
system remain symmetric under the reversal in the direction of rotation 
similar to the case of time reversal symmetry, the expansion of the 
perturbation terms contains only even power's terms, while the angular 
momentum expansion retains only odd terms. The first-order angular perturbation
term, which is considerable here, is denoted as $\omega(r,\theta)$. It is the
angular velocity of the local inertial frame and is proportional to the star's 
angular velocity $\Omega$. Further, 
under the coordinate transformation of type $r \rightarrow f(r)$, wherein 
there is no change of the form of the metric \eqref{rot_metric}, an additional 
condition: $k_0(r)=0$ can be imposed. Moreover, for convenience, a new 
variable can be introduced as $v_2=h_2+k_2$. With all these considerations 
equation \eqref{rot_metric} can be rewritten as
\begin{align}
	ds^{2} =& -e^{\nu} \left[1+2(h_{0}+h_{2}P_{2})\right] dt^{2} + \frac{\left[1 + 2(m_{0}+m_{2}P_{2})/\big(r-2M(r)\big)\right]}{1-2M(r)/r}\, dr^{2} 
\nonumber\\[2pt]
	&+ r^{2} \left[1 + 2(v_{2}-h_{2})P_{2}\right]
	\left[d\theta^{2} + \sin^{2}\theta(d\phi-\omega dt)^{2}\right].
\end{align}
The components of the four-velocity vector $u^\varrho$ of the fluid of the 
interior of the star, which satisfy the normalization condition 
$u^\varrho u_\varrho = -1$, are given as
\begin{align}
u^{t} &= \left(-g_{tt} - 2 \,\Omega\, g_{t\phi} - \Omega^{2} g_{\phi\phi} \right)^{-1/2} \nonumber \\[3pt]
		&= e^{-\nu/2} \left[1 + \frac{1}{2}\, \bar{\omega}^{2} r^{2}\sin^{2}\theta\, e^{-\nu} - h_{0} - h_{2}P_{2} \right]\!,\\
		u^{\phi} &= \Omega\, u^{t}, \\
		u^{r} &= u^{\theta} = 0,
	\end{align}
where the quantity $\bar\omega = \Omega - \omega$ represents the angular 
velocity of fluid relative to the local inertial frame. It is used to 
determine the magnitude of centrifugal force of a rotating star, which remains
in equilibrium by attaining a balance between pressure, gravitational and 
centrifugal forces. The equation to determine the value of $\bar\omega$ is 
given by
\begin{equation}\label{frame_dragging_diff_eqn}
	\frac{1}{r^{4}}\frac{d}{dr}\left(r^{4} \vartheta(r)\, \frac{d\bar{\omega}}{dr}\right) + \frac{4}{r}\frac{d\vartheta(r)}{dr}\, \bar{\omega} = 0,
\end{equation}
where
\begin{equation}\label{j_expression}
	\vartheta(r) = e^{-\nu(r)/2}\left(1 - \frac{2M(r)}{r}\right)^{\!1/2}\!\!\!\!\!.
\end{equation}
This equation \eqref{frame_dragging_diff_eqn} can be solved along 
with the TOV equation \eqref{TOV_equation} as coupled differential equations 
to find values of $\bar\omega$ considering an arbitrary value 
$\bar{\omega}_{c}$ and $d\bar{\omega}/dr = 0$ as initial conditions. Then we 
determine $\Omega$ and total angular momentum $J$ using the boundary 
conditions at $r = R$ with the equation of $J$ and $\Omega$ at $R$ as given by
\begin{align}
	J &= \frac{1}{6}\, R^{\,4}\! \left( \frac{d\,\bar{\omega}}{dr} \right)_{\!r\, =\, R}, \label{J} \\[5pt]
\Omega &= \bar{\omega}(R) + \frac{2J}{R^{3}}\,. \label{Omega}
\end{align}
This $\bar{\omega}$ term is related to the first order of star's angular 
velocity $\Omega$ ($\mathcal{O}(\Omega)$). For the second order of $\Omega$, 
i.e., $\mathcal{O}(\Omega^{2})$, there are two cases, $l = 0$ and $l = 2$.  
While $l = 0$ gives the spherical part of the rotational deformation, $l = 2$ 
gives the quadrupole part of the rotational deformation, as mentioned earlier. 
These terms used for calculating the deformation and redistribution of matter 
in rotating stars as discussed below.

\subsection{Spherical or Monopole $(l = 0)$ Deformation Terms}\label{secIIa}
The mass perturbation factor $m_{0}$ and pressure perturbation factor 
$p_{0}^{*}$ of the spherical part of rotational deformation are calculated 
using the following differential equations \cite{hartle_1968}: 
\begin{align}
	\frac{dm_{0}}{dr} & = 4\pi r^{2}\, \frac{d\epsilon}{dp}\,(\epsilon + p)\, p_{0}^{*} + \frac{1}{12}\, \vartheta^{2}(r)\, r^{4} \left(\frac{d\,\bar{\omega}}{dr}\right)^{\!2} - \frac{2}{3}\, r^{3} \vartheta(r)\, \frac{d\vartheta}{dr} \,\bar{\omega}^{2},\\[5pt]
	\frac{dp_{0}^{*}}{dr} & = -\frac{m_{0} (1 + 8\pi r^{2} p)}{\big[r - 2M(r)\big]^{2}} - \frac{4 \pi (\epsilon + p) r^{2}}{\big[r - 2M(r)\big]}\, p_{0}^{*} + \frac{1}{12} \frac{r^{4} \vartheta^{2}(r)}{\big[r-2M(r)\big]} \left(\frac{d\bar{\omega}}{dr}\right)^{\!2} + \frac{1}{3} \frac{d}{dr}\!\left[\frac{r^{3} \vartheta^{2}(r) \bar{\omega}^{2}}{r - 2M(r)}\right]\!.
\end{align}
These two equations can be solved along with the TOV equation 
\eqref{TOV_equation} as coupled differential equations by integrating
outward with the boundary conditions that both $m_{0}$ and $p_{0}^{*}$ vanish 
at the center of the star. Outside the star, $m_{0}$ is given as
\begin{equation}
	m_{0} = \delta{M} - \frac{J^{2}}{r^{3}},
\end{equation}
where $\delta{M}$ is an amount of perturbed mass. Thus, the total mass of the 
stars, including the deformed part due to the rotation is given by
\begin{equation}\label{Eqn:MPlusDelM}
	\mathfrak{M} \equiv M + \delta{M} = M + m_{0}(R) + \frac{J^{2}}{R^{3}}.
\end{equation}
Once $p_{0}^{*}$, $\delta M$, and $J$ are obtained, $h_{0}$ can be evaluated 
from the following relations:
\begin{align}
h_{0} & = -\,\frac{\delta{M}}{r - 2M} + \frac{J^{2}}{r^{2}(r - 2M)}\;\; 
(\text{outside the star}, r>R), \label{h_0_outside} \\
	h_{0} &= -\,p_{0}^{*} + \frac{1}{3}\, r^{2} e^{-\nu} \bar{\omega}^{2} + h_{0c} \;\; (\text{inside the star}), \label{h_0_inside}
\end{align}
where $h_{0c}$ is a constant, and it can be determined by equating equations 
\eqref{h_0_outside} and \eqref{h_0_inside} at the surface of the rotating star.

\subsection{Quadrupole $(l = 2)$ Deformation Terms} \label{secIIb}
The quadrupole part of rotation is calculated using a pair of differential 
equations that is solved along with other differential equations, starting 
from the TOV equation to spherical deformation as coupled differential 
equations. The differential equations to calculate the quadrupole part of the
deformation are given as \cite{hartle_1968}
\begin{align}
\frac{dv_{2}}{dr} &= h_2\,\frac{d\nu}{dr} - \left(\frac{1}{r} + \frac{1}{2} \frac{d\nu}{dr}\right)\left[-\frac{1}{3}\, r^{3} \frac{d\vartheta(r)^{2}}{dr}\,\bar{\omega}^{2} + \frac{1}{6}\, \vartheta(r)^{2} r^{4} \!\left(\frac{d\,\bar{\omega}}{dr}\right)^{\!2}\right]\label{eqn:v2_diff},\\[5pt]
\frac{dh_{2}}{dr} &= h_2\left[-\frac{d\nu}{dr} + \frac{r}{r - 2M(r)}\left(\frac{d\nu}{dr}\right)^{\!-1}\left(8 \pi (\epsilon + p) - \frac{4M(r)}{r^{3}}\right) \right] \nonumber \\[5pt]
	&- \frac{4v_{2}}{r\big(r - 2M(r)\big)}\left(\frac{d\nu}{dr}\right)^{\!-1} + 
\frac{1}{6}\! \left[\frac{1}{2} \frac{d\nu}{dr} r - \frac{1}{r - 2M(r)} \left(\frac{d\nu}{dr}\right)^{\!-1} \right] r^{3}\vartheta(r)^{2} \left(\frac{d\,\bar{\omega}}{dr}\right)^{\!2} \nonumber \\[5pt]
		 &- \frac{1}{3}\!\left[\frac{1}{2} \frac{d\nu}{dr}\, r + \frac{1}{r - 2M(r)}\left(\frac{d\nu}{dr}\right)^{\!-1} \right] r^{2} \frac{d\vartheta(r)^{2}}{dr}\, \bar{\omega}^{\,2}\label{eqn:h2_diff}.
\end{align}
The initial conditions for solving these coupled differential equations are 
$v_{2} = 0$ and $h_{2} = 0$, respectively, at the centre.
Outside the star $h_{2}$ and $v_{2}$ are given by \cite{hartle_1968}
\begin{align}
	h_{2} &= J^{2}\left(\frac{1}{M(R) r^{3}} + \frac{1}{r^{4}}\right) + K Q_{2}^{2}\left(\frac{r}{M} - 1\right)\label{eqn:h2_outside},\\[5pt]
	v_{2} &= -\,\frac{J^{2}}{r^{4}} + \frac{2KM(R)}{\left[r(r - 2M(R))\right]^{1/2}}\,Q_{2}^{1}\left(\frac{r}{M} - 1\right)\label{eqn:v2_outside},
\end{align}
where $K$ is constant and $Q_{n}^{m}$ is associated Legendre Polynomial of 
second kind.
After calculating $h_{2}$ and $v_{2}$, the non-radial mass and pressure 
perturbation factors $m_{2}$ and $p_{2}^{*}$ can be determined using the 
following equations \cite{hartle_1968}:
\begin{align}\label{eqn:m2_p2*}
	m_{2} & = (r - 2M(r)) \left[-h_{2} - \frac{1}{3}r^{3}\left(\frac{d\vartheta^{2}}{dr}\right)\bar{\omega}^{2} + \frac{1}{6} r^{4} \vartheta^{2} \left(\frac{d\bar{\omega}}{dr}\right)^{2}\right],\\[5pt]
	p_{2}^{*} &= -\, h_{2} - \frac{1}{3} r^{2} r^{-\nu} \bar{\omega}^{2}. 
  \label{eqn:p2*}
\end{align}

Due to rotation, the star is deformed and thus the star's number density,
energy density and pressure of the fluid are affected by the deformation. The 
pressure, energy density and number density at the interior of the deformed 
star at a given $(r,\theta)$, in a reference frame that is momentarily moving 
with the fluid can be expressed, respectively, as \cite{hartle_1968}
\begin{align}
\mathcal{P} &\equiv p + (\epsilon + p) (p_{0}^{*} + p_{2}^{*}\, P_{2}), \\[5pt]
\mathcal{E} &\equiv \epsilon + (\epsilon + p)\left(\frac{d\epsilon}{dp}\right)(p_{0}^{*} + p_{2}^{*}\, P_{2}), \\[5pt]
\mathcal{N} &\equiv n + (\epsilon + p)\left(\frac{dn}{dp}\right)(p_{0}^{*} + p_{2}^{*}\, P_{2}).
	\end{align}
where $p$ and $\epsilon$ are respectively the pressure and energy density for 
the non-rotating star, as mentioned earlier, and $n$ is the number density for 
such a star.

In the Hartle-Thorne slow-rotation formalism \cite{hartle_1967,
hartle_1968}, rotation causes the star to become oblate, with the equatorial 
radius exceeding the polar radius due to the centrifugal force. Let's consider 
a surface of constant density in a particular coordinate system, which is 
deformed due to rotation. Hence, the surface can be expressed as 
\cite{hartle_1968}
\begin{equation}
	r + \delta r_{0} + \delta r_{2} P_2\left(\cos \theta\right),
\end{equation}
where $\delta r_{0}$ and $\delta r_{2}$ can be expressed as
\begin{align}
	\delta r_{0} &= -\, p_{0}^{*} (\epsilon + p)/\left(\frac{dp}{dr}\right),	\\[5pt]
	\delta r_{2} &= -\, p_{2}^{*} (\epsilon + p)/\left(\frac{dp}{dr}\right).	
\end{align}
When the surface of a constant density is embedded in a $3$-D flat space, 
which has the same intrinsic geometry, we get a spheroid for a spherically 
symmetric system rotating with $\mathcal{O}\left(\Omega^{2}\right)$. The 
radius of the spheroid can be given by \cite{hartle_1968}
\begin{equation}
	R + \delta R = R + \delta r_{0}(R) + \left[\delta r_{2}(R) + R \left\{v_{2}(R) - h_{2}(R)\right\}\right]P_{2}(\cos\theta).
\end{equation}
The mean radius and the eccentricity of the system is given by 
\cite{hartle_1968}
\begin{align}
	\bar{R} &= R + \delta r_{0}(R), \\[5pt]
	e_{s} &= \left[-\,3\left(v_{2}(R) - h_{2}(R) + \frac{\delta r_{2}(R)}{R}\right)\right]^{1/2}\!\!\!\!.
\end{align}
The rotation causes the star to deform outward more significantly at 
the equator than at the poles, resulting in an oblate shape 
\cite{hartle_1967,hartle_1968}. Consequently, the stellar matter is 
redistributed, causing the mass distribution to deviate from spherical 
symmetry. This deviation is characterized by the mass quadrupole moment $Q$, 
which quantifies the rotational deformation of the star. For a rotating star 
with gravitational mass $\mathfrak{M}$ and angular momentum $J$, the mass 
quadrupole moment is given by
\begin{equation}
	Q = \frac{8}{5} K \mathfrak{M}^{3} + \frac{J^{2}}{\mathfrak{M}},
\end{equation}
where K is the same constant as in equations~\eqref{eqn:h2_outside} and 
\eqref{eqn:v2_outside}. The quantities $m_{2}$ and $p_{2}^{*}$ are 
introduced as auxiliary second-order perturbation quantities associated with 
the $l = 2$ terms. They are obtained algebraically using equations
~\eqref{eqn:m2_p2*} and \eqref{eqn:p2*}, respectively, once the $l = 2$ metric 
perturbation functions $h_{2}$ and $v_{2}$ have been determined. The 
quadrupole moment is obtained from the asymptotic behavior of the exterior 
$l = 2$ metric perturbation \cite{hartle_1968}. In particular, the constant 
$K$ is determined at the surface of the star using the equations
\eqref{eqn:v2_diff}, \eqref{eqn:h2_diff}, \eqref{eqn:h2_outside} and 
\eqref{eqn:v2_outside}, enters the expression for the quadrupole moment. 
Therefore, $m_{2}$ and $p_{2}^{*}$ do not appear explicitly in the final 
expression for $Q$. However, the total mass of the star $M$ can be replaced by 
$\mathfrak{M}$ without affecting the line element for 
$\mathcal{O}(\Omega^{2})$. The quantity $\mathfrak{M}$, $J$, and $Q$ 
characterize the exterior metric of a slowly rotating relativistic star 
\cite{hartle_1968}.

K.~Yagi and N.~Yunes proposed a relationship between the moment of 
inertia, the Love number and the quadrupole moment for spinning NSs and QSs 
\cite{yagi_2013}. This relation is independent of the NSs and QSs internal 
structure and breaks the degeneracies in GW detection to measure spin in 
binary inspiral and distinguishes NSs from QSs. We have already discussed the 
procedure for the calculation of rotationally modified mass and mean radius of 
the stars at different angular frequencies. We also have the surface values of 
angular momentum and angular frequencies for each stellar configuration. Using 
these quantities, the dimensionless moment of inertia $(\bar{I})$ and 
quadrupole moment $(\bar{Q})$ are defined as \cite{yagi_2013}
\begin{align}
	\bar{I} &= \frac{J}{\Omega M^{3}}\label{eq:barI}, \\[5pt]
	\bar{Q} &= -\, \frac{Q M}{J^{2}}\label{eq:barO}.
\end{align}
The dimensionless quantities given in equations~\eqref{eq:barI} and 
\eqref{eq:barO} depend only on observable macroscopic properties of rotating 
compact stars, namely the mass $(M)$, angular momentum $(J)$, angular 
frequency $(\Omega)$ and quadrupole momentum $(Q)$. Thus, it is useful for 
comparing theoretical predictions with astrophysical observations and can be 
used to constrain the underlying EoS of NSs and QSs.
\section{Equations of state}\label{secIII}
In this work, we employ the EoSs describing deconfined QM. These EoSs 
incorporate microscopic properties such as quark flavor number densities and 
stellar temperature. The details of the adopted EoSs, along with the procedure 
used to obtain the relations among pressure, energy density, and number 
density, are discussed in this section. As mentioned previously, we write the 
formalism in the natural unit system, and hence the thermodynamical quantities, 
like pressure, energy density, number density, etc., in EoSs are given in the natural unit system. We convert the values of the required physical quantities 
to the geometrized unit system while using them to solve the TOV equation. 

\subsection{Massless Quark Equation of State}
We take this EoS, denoted by EoS I, as given by Fraga, Pisarski and Bielich 
(FPB) for small, 
dense QSs \cite{fraga_2001}. In their work, FPB considered a strong first-order 
chiral transition, wherein they developed two models of QSs. In the first 
model, the QSs has a maximum mass $M_\text{max} \approx 2.14 M_{\odot}$ and 
radius $R_\text{max} \approx 12$ km with central quark density 
$\rho_\text{max} \approx 5.1\rho_{0}$. On the other hand, the second model of 
the QS has maximum mass $M_\text{max} \approx 1.05 M_{\odot}$ and radius 
$R_\text{max} \approx 5.81$ km with central quark density 
$\rho_\text{max} \approx 15\rho_{0}$. Here, $\rho_{0}$ is the density
of three quark flavors to remain at nuclear saturation density, and 
hence $\rho_{0} \sim 3 n_0$ is considered as the 
density of quarks in nuclear matter, where $n_0 = 0.17\, \text{fm}^{-3}$ is 
nuclear saturation density. In these models, FPB considered the 
system like NSs but denser and consisting of deconfined cold quarks. 
In such systems, individual quark masses are small relative to the quark 
chemical potential. Thus, FPB 
considered three flavors of massless quarks with equal chemical potential. 
The form of the thermodynamic potential of the plasma of massless quarks and 
gluons is given as
\begin{equation}\label{eos1}
	\mathcal{V}(\mu) = -\,\frac{N_{f}}{4\pi^{2}}\, a_\text{eff}\, \mu^{4} + 
B_\text{eff},
\end{equation}
where $\mu$ is the chemical potential of QM, $N_{f}$ is number of 
massless flavors, $a_\text{eff}$ measures the deviation from ideality and it 
can be given by $a_\text{eff} = 1 - 2\alpha_{s}/\pi$, with $\alpha_{s}$ being 
the strong coupling constant and $B_\text{eff}$ is an effective Bag constant.
From this thermodynamic potential, the pressure exerted by the quark matter
fluid can be obtained from the thermodynamic relation: 
$p(\mu) = -\, \mathcal{V}(\mu)$, and from this, the pressure and energy density 
relation, i.e., required EoS can be found from $\epsilon = -\,p + \mu n$.      

\subsection{Massive Quark Equation of State} \label{EOS_II}
In contrast to EoS I, in this EoS, denoted by EoS II, the quarks are considered 
to be massive.
This EoS is based on the fact that at very high densities, such as those 
expected in the cores of NSs, chiral symmetry is partially or fully restored, 
causing quarks to behave according to their deconfined rest masses rather than 
their dynamically generated constituent masses. This is because at such a very 
high density, the interquark interaction is screened in the medium and hence 
the interaction becomes weaker and shorter in range \cite{dey_1998}. Thus, 
deconfinement takes place at the high density medium. In such a situation, 
the Hamiltonian of the system contains both scalar and vector 
potentials \cite{dey_1998}. The scalar potential originates due to the mass 
of the quark $M_{i}$, which is density-dependent and is given as 
\cite{dey_1998, bagchi_2006}
\begin{equation}\label{den_dependent_q_mass_eqn}
 M_{i} = m_{i} + M_{Q} \sech \left(\frac{n_{B}}{N n_{0}}\right)\!,
\end{equation}
where index $i$ represents the quark flavors ($u,d,s$), $n_{B}$ is the baryon 
number density, $n_{0} = 0.17\, \text{fm}^{-3}$ is the nuclear saturation 
density, and $M_{Q}$ and $N$ are two adjustable parameters which ensure 
that the minimum value of energy per particle for QM is less than 
the energy per nucleon of iron to have stable configuration of the star 
\cite{farhi_1984, dey_1998, bagchi_2006}. As mentioned, at very high densities, the quark masses $M_{i}$ fall from their constituent masses to the deconfined 
rest masses ($m_i$), whose current values are: $m_{u} = 2.16$ MeV, 
$m_{d} = 4.7$ MeV,  and $m_{s} = 92.9$ MeV \cite{takahashi_2026}.
	
To describe the interaction between two quarks and the corresponding 
interquark potential originated from the gluon exchange, in 
Ref.~\cite{dey_1998}, the author used the Richardson potential \cite{richardson_1979}. This
potential incorporates the two crucial concepts: asymptotic freedom and 
linear quark confinement, which are required for systems like the very 
high-density cores of NSs, where quark deconfinement may take place. The 
Richardson potential is given as \cite{dey_1998, richardson_1979}
\begin{equation}\label{eq:richardsonPot}
	V_{ij} = \frac{12 \pi}{27}\, \frac{1}{\ln \Big[1 + \big(\vec{k}_{i} - \vec{k}_{j}\big)^{2}/\Lambda^{2}\Big]}\, \frac{1}{\big(\vec{k}_{i} - \vec{k}_{j}\big)^{\!2}},
\end{equation}
where $\vec{k}_{i}$, $\vec{k}_{j}$ are momenta of $i$th, $j$th flavors
of quark, respectively, and $\Lambda$ is a scale parameter. In a high-density 
medium, this potential will be screened due to pair creation and 
infrared divergence, and the inverse of this screening distance, $D^{-1}$, 
is given by
\begin{equation}
  (D^{-1})^{2} = \frac{2 \alpha_{0}}{\pi}\!\!\! \sum_{i\, =\, u,d,s}\!\!\! k_{i}^{f} \sqrt{(k_{i}^{f})^{2} + m_{i}^{2}},
\end{equation}
where $\alpha_{0}$ is the quark-gluon perturbative coupling parameter, and 
the Fermi momentum for each flavor is calculated as
\begin{equation}
	k_{i}^{f} = \left(n_{i}\, \pi^{2}\right)^{1/3},
\end{equation}
with $n_{i}$ as the number density of $i$th flavor for a given baryon 
number density, under the baryon number conservation, $\beta$-equilibrium and 
charge neutrality conditions. The respective conditions are given as
\begin{align}
	&n_{B} = \big(n_{u} + n_{d} + n_{s}\big)/3,\label{cond1}	\\
	&\mu_{d} = \mu_{s}, \,\,\, \mu_{d} = \mu_{u}+\mu_{e},\label{cond2}	\\
	&2 \left( k_{u}^{f} \right)^{3} - \left( k_{d}^{f} \right)^{3} - \left( k_{s}^{f} \right)^{3} - \left( k_{e}^{f} \right)^{3} = 0, \label{cond3}
\end{align}
where $\mu_{u}$, $\mu_{d}$, $\mu_{s}$ and $\mu_{e}$ are chemical potentials for 
up, down, strange quarks and electrons, respectively. The chemical potential 
for quark flavors is calculated from the kinetic and potential parts of the 
energy of quarks at a given $n_{B}$. At high densities, along with quarks, 
there is a degenerate, free electron gas to maintain charge neutrality and 
$\beta$-equilibrium in the medium. The $\mu_{e}$ is calculated using 
$\beta$-equilibrium and then $k_{e}^{f}$ is calculated using the relation, 
	\begin{equation}
		k_{e}^{f} = \sqrt{\mu_{e}^{2} - m_{e}}.
	\end{equation}
The number density of free electrons $n_{e}$ can be calculated using the 
relation $k_{e}^{f} = \left( 3\, n_{e} \pi^{2}\right)^{1/3}\!$. So the 
pressure and energy density inside the star due to the degenerate, free 
electron gas are given as
\begin{align}
	p_{e} &= \frac{\pi m_{e}^{4}}{3} \left[ x (2 x^{2} - 3) (x^{2} + 1)^{1/2} + 3 \sinh^{-1}x \right], \label{free_e_press}\\[5pt]
	\epsilon_{e} &= m_{e} \left[ \left( 1 + x^{2} \right)^{\!1/2} - 1 \right], \label{free_e_eneden}
\end{align}
where $x = k_{e}^{f}/m_{e}$. 
Finally, using the thermodynamic relation $\epsilon = -\, p + \mu n $, the 
pressure for each quark flavor can be calculated, and these pressures can 
be summed along with the free electron part to get the total pressure of the 
system at a given $n_{B}$. As such, a table correlating pressure, energy 
density, baryon number density, quark flavor number density 
and chemical potential can be generated, which can be used to find the 
mass-radius relation of the star using the TOV equation \eqref{TOV_equation}.
	
\subsection{Temperature-Dependent Massive Quark Equation of State}
\label{EOS_III}
This EoS, denoted by EoS III, is the modified form of EoS II obtained by
incorporating the temperature-dependent effect. In EoS II, the two main 
properties of quark-quark interactions, asymptotic 
freedom and confinement, were described by a single parameter $\Lambda$ in 
the Richardson potential \cite{dey_1998}. Only for a small value of $\Lambda$, 
around $100$ MeV, asymptotic freedom and confinement plays a role in 
determining the static properties of hadrons, according to perturbative QCD. 
In the paper \cite{bagchi_2006}, the authors extended the Richardson potential 
so that it embeds these two properties, i.e., asymptotic freedom and quark 
confinement, separately. This modified Richardson potential is given as
\begin{equation}\label{eq:modRichardsonPot}
 V_{ij} = \frac{12 \pi}{27} \left[\frac{1}{\ln\Big(1 + \big(\vec{k}_{i}-\vec{k}_{j}\big)^{2}/\Lambda^{2}  \Big)} - \frac{\Lambda^{2}}{\big(\vec{k}_{i} - \vec{k}_{j}\big)^{2}} + \frac{{\Lambda'}^{2}}{\big(\vec{k}_{i} - \vec{k}_{j}\big)^{2}} \right] \times \frac{1}{\big(\vec{k}_{i} - \vec{k}_{j}\big)^{2}},
\end{equation}
where $\Lambda'$ is the scale parameter for the confinement property 
and $\Lambda$ is that for the asymptotic freedom.

Moreover, in EoS II, the gluon mass was considered temperature independent and 
hence all the calculations were done for absolute zero temperature. Whereas 
in this EoS, the temperature dependence of the gluon mass is considered, and 
accordingly, the screening length is modified to show the temperature 
dependence of the gluon mass. The modified inverse screening length is given 
as \cite{bagchi_2006} 
\begin{equation}\label{eq:tempDependentScreeningLength}
	\left(D^{-1}\right)^{2} = \frac{2\alpha_{0}}{\pi}\!\!\! \sum_{i\, =\, u,d,s}\!\!\! k_{i}^{f} \sqrt{ (k_{i}^{f})^{2} + m_{i}^2 } + 7.152\, \alpha_{0} T.
\end{equation}
For a temperature-dependent system, to get the correlation between pressure, 
energy density, number density and chemical potential, one has to incorporate 
the Fermi-Dirac distribution function, which is given as
\begin{equation}\label{fermi_distribution_func}
	\mathcal{F}(k,T) = \frac{1}{e^{(\epsilon - \mu)/T} + 1}.
\end{equation}
Here, $\epsilon$ is the energy of a quark flavor at finite temperature $T$, 
whose expression for a flavor of quark is given as
\begin{equation}
	\epsilon_{i} = \sqrt{k^{2} + M_{i}^{2}} + U_{i}(k),
\end{equation}
where $M_{i}$ is the density-dependent quark flavor mass given in equation 
\eqref{den_dependent_q_mass_eqn}, $k$ is the momentum of that quark flavor. 
The quantity $U_{i}(k)$ is the single particle potential, which is calculated 
using the following expression:
\begin{equation}
	U_{i} = -\, \frac{1}{3 \pi}\!\! \sum_{j\,=\,u,d,s}\! \int_{-1}^{+1}\!\! dx \int_{0}^{k_{j}^{f}} \!\!f_{ij} \times V_{ij}\, dk_{j}.
\end{equation}
Here, $k_{j}^{f}$ is the Fermi momentum for the $j$th flavor of quark for a 
given value of $n_{B}$ at absolute zero temperature, $V_{ij}$ is the 
Richardson potential given in equation \eqref{eq:modRichardsonPot}, 
where $\big(\vec{k}_{i} - \vec{k}_{j}\big)^{2}$ is replaced by 
$\big[\big(\vec{k}_{i} - \vec{k}_{j} \big)^{2}  - D^{-2}\big]$, here $D$ is 
the modified screening length given in equation \eqref{eq:tempDependentScreeningLength}. 
The expression for $f_{ij}$ is given as
\begin{equation}
    f_{ij} = \left(e_{i} e_{j} + 2 k_{i} k_{j} x + \frac{ k_{i}^{2} k_{j}^{2} }{e_{i} e_{j}}\right) \frac{1}{( e_{i} - M_{i} )(e_{j} - M_{j})},
\end{equation}
where
\begin{equation}
	e_{i} = \sqrt{k_{i}^{2} + M_{i}^{2}} + M_{i}.
\end{equation}
The chemical potential $\mu$ in the Fermi-Dirac distribution function for 
finite temperature can be calculated using the chemical potential that is 
calculated for absolute zero temperature for a given value of $n_{B}$ as an 
initial guess. The values of $n_{i}$, $i = u, d, s, e$, can be obtained using 
the following expression after determining the maximum momentum $k$ at finite 
temperature for each particle:
\begin{equation}
	I = \frac{g}{2 \pi^{2}} \int_{0}^{\infty}\!\! \phi(\epsilon)\, k^{2} \mathcal{F}(k, T)\, dk,
\end{equation}
where $g$ is the spin-colour degeneracy, which is equal to $2$ for electrons 
and $6$ for quarks flavors. $I$ is basically the number density for 
$\phi(\epsilon) = 1$ and energy density for $\phi(\epsilon) = \epsilon$. While 
calculating $\mu_{i}$ and $n_{i}$, one has to make sure that they follow the 
conditions given in equations \eqref{cond1} - \eqref{cond3}. The calculated 
values of $\mu_{i}$, $n_{i}$ and $\epsilon_{i}$, $i = u, d, s, e$, for a given 
value of $n_{B}$ at finite temperature are used to calculate the entropy 
density and free energy density using the following respective equations:
\begin{align}
	& s(T) = -\,\frac{3}{\pi^{2}} \int_{0}^{\infty}\!\! k^{2} \big[\mathcal{F}(k,T)\ln\!\big(\mathcal{F}(k,T)\big) + \big(1-\mathcal{F}(k,T)\big)\ln\!\big(1-\mathcal{F}(k,T)\big)\big] dk,\\[5pt]
       & f(\epsilon,T) = \epsilon - Ts(T).
\end{align}
Further, the pressure can be calculated using the thermodynamic relation 
given as
\begin{equation}
	p = \sum_{i} \left( n_{i} \frac{\partial f_{i}(\epsilon,T)}{\partial n_{i}} - f_{i}(\epsilon,T) \right),\;\; i = u,d,s,e.
\end{equation}
With all these required calculations, as in the case of the previous EoSs, 
the mass-radius relation of the star can be obtained from the TOV 
equation \eqref{TOV_equation} for this temperature dependent EoS.

\section{Radial and quasiradial oscillations}\label{secIV}
\subsection{Radial Oscillations}\label{secIV_a}
In $1964$, S.~Chandrasekhar studied the dynamical instability of gaseous 
masses in the general relativity framework \cite{chandrasekhar_1964}. From his
studies, he established the relations describing the infinitesimal, baryon 
number conserved and adiabatic radial oscillations of a gas sphere. The final 
expression for radial perturbation and its corresponding Lagrangian 
perturbation are given by the dimensionless quantities $\xi = \delta r/r$ 
and $\eta = \delta p/p$, respectively. These Chandrasekhar's equations of 
radial and pressure perturbations can be written as 
\cite{chanmugam_1977, vaeth_1992}
\begin{equation}\label{xi_eqn}
	\frac{d\xi}{dr} = - \frac{1}{r} \left( 3\xi + \frac{\eta}{\gamma} \right) - \frac{dp}{dr} \frac{\xi}{p+\epsilon},
\end{equation}
\begin{equation}\label{eta_eqn}
	\frac{d\eta}{dr} = \xi \left[\sigma^{2} e^{\lambda - \nu} \left( \frac{p+\epsilon}{p} \right)r - \frac{4}{p} \frac{dp}{dr} - 8 \pi e^{\lambda} (p + \epsilon) r + \left(\frac{dp}{dr} \right)^{\!2} \frac{r}{p(p + \epsilon)} \right] + \eta \left[ -\frac{dp}{dr} \frac{\epsilon}{p(p + \epsilon)} - 4 \pi (p + \epsilon)r e^{\lambda} \right],
\end{equation}
where $\sigma$ is the characteristic frequency of radial oscillations, 
$\gamma$ is the relativistic adiabatic index. Equations \eqref{xi_eqn} and 
\eqref{eta_eqn} are coupled first-order differential equations. These equations
can be solved from the center of the stars to the surface of the stars. At 
$r \rightarrow 0$, the coefficient of $1/r$ in equation \eqref{xi_eqn} should 
vanishes, i.e.,
\begin{equation}\label{rad_osc_center_constr}
	3 \xi \gamma + \eta = 0,
\end{equation}
which is the first boundary condition that must be satisfied at the center 
with $\xi = 1$.
At the surface of the gaseous sphere, the pressure become zero. In general, 
even the energy density should be zero at the surface. However, since in our 
work, we have considered self-bound QS EoSs, which results into non-zero 
energy density at the surface of the stars, even though it is very small in 
comparison to its center \cite{alcock_1986}. So at the surface the coefficient of 
$\epsilon/p$ in equation \eqref{eta_eqn} must vanish. Hence, at $r = R$, we 
can have
\begin{equation}\label{rad_osc_boun_constr_eta_R}
	\eta_{r\,=\, R} = \xi_{R} \left[ \left(1 - \frac{2 M}{R}\right)^{\!-1}\!\! \left(-\frac{M}{R} -\frac{\sigma^{2} R^{3}}{M}\right) - 4 \right].
\end{equation}

The coupled differential equations \eqref{xi_eqn} and \eqref{eta_eqn} are the 
Sturm-Liouville type two point boundary value problem, which gives real 
eigenvalues $\sigma_{0}^{2}<\sigma_{1}^{2}<\hdots<\sigma_{n}^{2}\hdots$, 
corresponding to the eigenfunctions $\xi_{0}, \xi_{1}, \hdots, \xi_{n}, 
\hdots$, where eigenfunction $\xi_{n}$ has $n$ nodes. The $n = 0$ mode is 
called fundamental-mode or $f$-mode \cite{vaeth_1992}. The given model is 
unstable if $\sigma^{2} < 0$. For convenience, we denote the eigenfrequencies 
of radial oscillations of the non-rotating configuration by $\sigma^{(0)}$ and 
use this notation throughout the subsequent discussion.

Further, Chandrasekhar also derived the following pulsation equation
for the non-rotating spherically symmetric star with pulsation frequency
$\sigma^{(0)}$ and fluid displacement $\delta r:=\xi^{r}$ inside the star in 
radial direction \cite{chandrasekhar_1964}:
\begin{align}
        (\sigma^{(0)})^{2} e^{\lambda - \nu} (p + \epsilon) \xi^{r} &= \frac{4 p^{\prime} \xi^{r}}{r} - e^{(\lambda + 2\nu)/2} \left[e^{(\lambda + 3\nu)/2} r^{-2} \gamma p \left(r^{2} e^{-\nu_0/2} \xi^{r}\right)^{\prime}\right]^{\prime}\nonumber \\[5pt]
        &\quad+ 8 \pi e^{\lambda} p (p+\epsilon) \xi^{r} - \frac{(p^{\prime})^{2}}{p+\epsilon} \xi^{r}.
\end{align}
Here, $\xi^{r}$ can be expressed as
\begin{align}
        \xi^{r} = r^{-2} e^{\nu/2} U(r),
\end{align}
$U(r)$ is a radial function. Using this expression in the above pulsation
equation and then after simplification, we arrived at an expression as given by
\begin{align}
        -\, \sigma^{2} e^{(3\lambda + \nu)/2} (p+\epsilon) r^{-2} U(r) &= \left[e^{(\lambda + 3\nu)/2} r^{-2} \gamma p U^{\prime}(r) \right]^{\prime}\nonumber \\[5pt]
        &\quad + e^{(\lambda+3\nu)/2}\left[-\frac{4 p^{\prime}}{r^{3}} - 8 \pi r^{-2} e^{\lambda} p (p+\epsilon) + \frac{(p^{\prime})^{2}}{r^{2} (p+\epsilon)} \right] U(r).
\end{align}
As $\xi = \delta r/r$, and hence $\xi^{r} = r \xi$. Thus, the
function $U(r)$ can be expressed as
\begin{align}
        U(r) = r^{3} e^{-\nu/2} \xi
\end{align}
which can be used to check the fluid displacement in radial direction
inside the stars for a given eigen frequency of radial oscillations
$\sigma^{(0)}$. Additionally, using the quantity $U(r)$ along with other
non-rotating and rotating configuration parameters, we can get the expression
to calculate the change in $\sigma^{(0)}$ due to rotation of stars, as 
discussed in the following.

\subsection{Quasiradial Oscillations}\label{secIV_b}
We intend to study the quasi-radial modes of oscillations of QSs using the 
framework established by HTC in 
1972 \cite{hartle_1972}. In this framework, they used the slowly rotating 
relativistic stars that we have discussed in Section \ref{secII} as an 
equilibrium configuration and incorporated the displacement perturbation 
\textbf{$\xi(x,t)$} to describe the small oscillations in that slowly rotating 
relativistic configurations.

Usually, those solutions of oscillations contain both incoming and outgoing 
GWs, as measured by the behavior of the metric at large 
distances from the star. However, the solutions with only outgoing 
GWs are only the physically acceptable solutions 
\cite{hartle_1972}. It is to be noted that there are no possible GWs for 
radial oscillations of a non-rotating star. Whereas, if a small 
angular velocity $\Omega$ is given to the star, then the frequency of each
mode of oscillations becomes a function of $\Omega$, although the number 
of modes of oscillations of the star will remain the same. This indicates that 
the frequencies of oscillations of such rotating stars become complex in the 
sense that there is a decay mode in frequency of each mode of oscillations 
represented by an imaginary part of each frequency. Thus, for a slow-rotating
star, a discrete value of complex angular frequency is given by
\begin{equation}
	\sigma = \sigma_{R} + i \sigma_{I},
\end{equation}
where $\sigma_{R}$ is the real part and $\sigma_{I}$ is the imaginary part of 
the frequency. If $\sigma_{I} > 0$, the oscillation of the star is stable with 
a damping of vibration energy, whose half-life is $2/\sigma_{I}$, and if 
$\sigma_{I} < 0$, the mode of oscillation is unstable \cite{hartle_1972}.

In our work, the change induced in pulsation frequency $\sigma^{(0)}$ by the 
rotation is denoted by $\sigma^{(2)}$. HTC established a expression 
for $\sigma^{(2)}$, where related quantities can be evaluated only using 
non-rotating equilibrium configuration quantities such as $p$, $\epsilon$, 
$\nu$, $\lambda$, $\sigma^{(0)}$, fluid radial perturbation $U(r)$ and the rotational correction terms that we have 
discussed in Section \ref{secII}. The expression for $\sigma^{(2)}$ is given as
\cite{hartle_1972}
\begin{equation}
   \left(\sigma^{2}\right)^{\!(2)} = \frac{ \int_{0}^{R} dr\, e^{(\nu + \lambda/2)}\, U(r)\, \mathfrak{D}(r) }{ \int_{0}^{R} dr\, U(r)^{2}\, \mathfrak{W}(r)},
\end{equation}
where $\mathfrak{D}$ and $\mathfrak{W}$ are known as the driving term and
weighting function \cite{hartle_1972}, respectively. The respective expressions
of these two parameters are given in the Appendix \ref{appen1}.

It needs to be mentioned here that due to pulsation, the fluid gets displaced 
from its rotational equilibrium position by an amount $\xi^{r}$ in the radial 
direction, and $\delta\theta:= \xi^{\theta}$ and $\delta\phi:= \xi^{\phi}$ in 
tangential directions. Under the simultaneous reversal of $\phi \rightarrow -\,\phi$, 
$\Omega \rightarrow -\,\Omega$, $\xi^{r}$ and $\xi^{\theta}$ contains only 
even terms in rotational frame and $\xi^{\phi}$ contains only odd terms 
\cite{hartle_1972}. Furthermore, in a rotating but non-pulsating star, the 
metric terms contain even terms for $O(\Omega^2)$ expansion with spherical 
harmonics 
of order $l = 0$ and $2$, and $l = 1$ for $O(\Omega)$ terms as discussed in 
Section~\ref{secII} with an even parity \cite{hartle_1967}. For radial 
pulsations of a non-rotating configuration, all metric perturbations 
correspond to the monopole mode $(l = 0)$ and possess even parity 
\cite{chandrasekhar_1964}. This reveals that $\xi^{r}$ is a scalar under 
rotation and it must have $l = 0$ and $2$ poles. While $\xi^{\theta}$ and 
$\xi^{\phi}$ are vectors and can have $l = 1$ and $2$ \cite{hartle_1972}. In 
this work for rotational configuration, we only discussed the fluid 
displacement in the radial direction, $\xi^{r}$.

\section{Numerical Results and Discussions}\label{secV}
In this section, we present the numerically calculated results and 
corresponding related discussions for our considered models of QSs as 
discussed in previous sections. Starting from the mass-radius relations, we 
present the radial and quasiradial oscillations of these stars in the 
following subsections.       
\subsection{Mass-Radius Relations and Other Internal Characteristics}
\begin{figure}[!h]
	\centerline{
		\includegraphics[scale = 0.3]{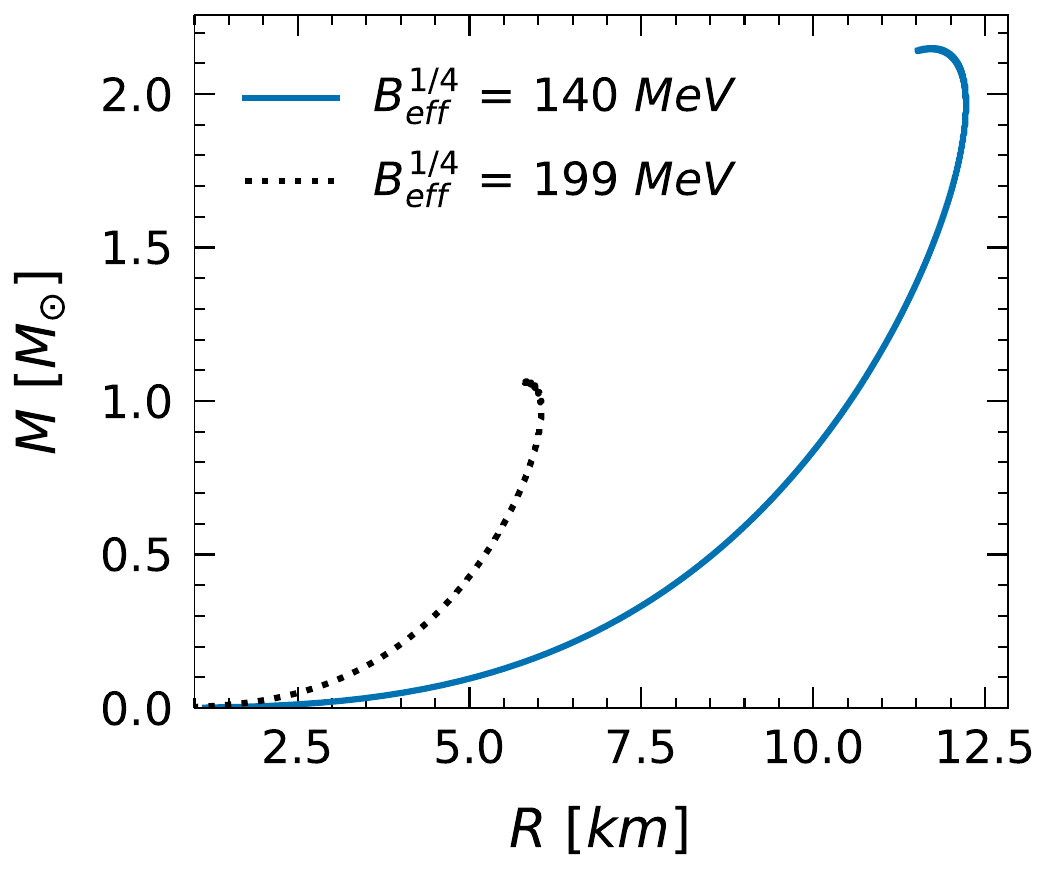}
		\includegraphics[scale = 0.3]{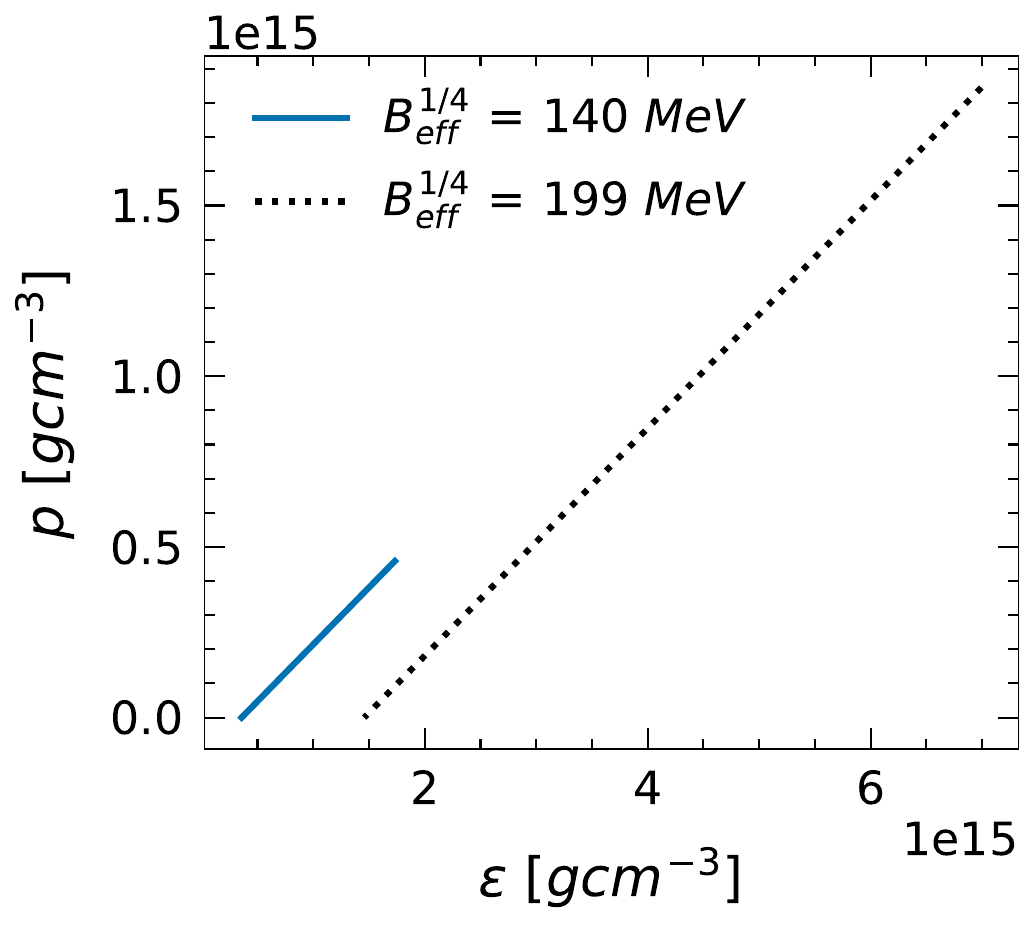}
		\includegraphics[scale = 0.3]{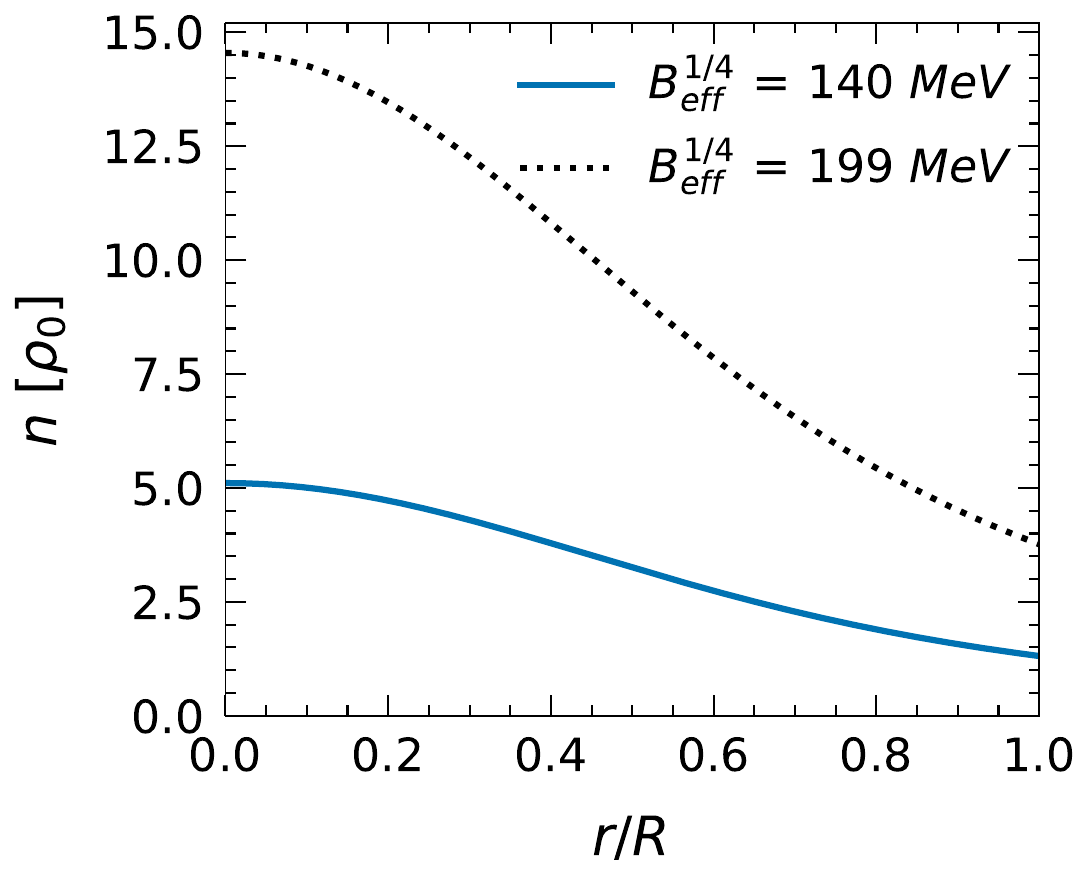}}
	\vspace{-0.3cm}
	\caption{Mass-radius relations (left), energy density vs pressure 
distribution inside the stars (middle) and number density distribution inside 
the stars (right) for different $B_\text{eff}$ values in EoS I.}
	\label{fig:eosIComparisonStudy}
\end{figure}

To compare the mass and radius of QSs for all three EoSs, we solve the TOV
equation \eqref{TOV_equation} along with equations \eqref{TOV_mass} and
\eqref{TOV_metric} numerically, as mentioned earlier, considering different
relevant parameter values of these EoSs. As EoS I is based on massless quarks,
the thermodynamic potential of massless quark-gluon plasma in equation 
\eqref{eos1} was calculated perturbatively to $\mathcal{O}(\alpha_s)$ using
the momentum-space subtraction scheme \cite{baluni1978,smith2000,blaizot2001} by Freedman and McLerran 
\cite{freedman1977_i,freedman1977_ii,freedman1977_iii,freedman1978}, and by Baluni \cite{baluni1978}. 
As mentioned earlier, FPB discussed two models in 
Ref.~\cite{fraga_2001}. The effective Bag constants used in these two models 
are $B_\text{eff} = (140$ MeV$)^4$ and $B_\text{eff} = (199$ MeV$)^4$. We 
plot the mass-radius relation, pressure-energy density and number density 
distribution for these two effective Bag constants as shown in 
Fig.~\ref{fig:eosIComparisonStudy}. The mass, radius and compactness of the 
stars with $B_\text{eff} = (199$ MeV$)^4$ are listed in the 
Table~\ref{table:MR}. For $B_\text{eff} = (140$ MeV$)^4$, the stars have 
mass $2.1497 M_{\odot}$, radius $11.7150$ km and compactness $0.271$.
On the basis of these calculations, for the EoS I, 
we take a star model (Model I) with the central value of chemical 
potential $\mu_c = 425 - 670$ MeV, and $a_\text{eff} = 0.628$ from  
Ref.~\cite{fraga_2001} with $B_\text{eff} = (199\,\text{MeV})^4$ and $N_f = 3$.
For $B_\text{eff} = (199$ MeV$)^4$, the model 
exhibits a strongly first-order chiral transition and produces small, dense 
QS configurations, in contrast to the relatively larger configurations 
obtained for $B_\text{eff} = (140$ MeV$)^4$. FPB showed that appropriate 
choices of the renormalization scale in perturbative QCD can lead to a strongly 
first-order chiral transition and consequently to QSs with masses and 
radii significantly smaller than those of ordinary NSs 
\cite{fraga_2001}. In the high-density central region, the quark chemical 
potential is sufficiently large that the effects of the strange-quark mass are 
relatively small, and the quarks can therefore be treated as approximately 
massless. Moving outward from the center, the quark chemical potential 
decreases. For $B_\text{eff} = (199$ MeV$)^4$, we obtain $\sim 668$ MeV at the 
center and $\sim 426$ MeV at the surface, whereas for 
$B_\text{eff} = (140$ MeV$)^4$ these values are approximately $\sim 472$ MeV 
and $\sim 300$ MeV, respectively. The occurrence of a strong first-order 
chiral transition is important because it permits a distinct high-density 
quark-matter branch. As the chemical potential decreases below the 
chiral-transition value ($\mu_\chi$), the system leaves the chirally restored 
phase and enters the chirally broken phase, where quarks acquire substantial 
masses and hadronic degrees of freedom become relevant \cite{fraga_2001}. 
Consequently, a stellar configuration whose central chemical potential is 
greater than $\mu_\chi$, but whose surface chemical potential falls below 
$\mu_\chi$, contains a chirally restored quark core surrounded by matter in 
the chirally broken phase and should therefore be regarded as a hybrid 
configuration rather than a pure QS. For $B_\text{eff} = (199$ MeV$)^4$, 
maintaining the chirally restored phase throughout the entire star requires, 
in particular, $\mu_\chi < 426$ MeV. Similarly, 
for $B_\text{eff} = (140$ MeV$)^4$, the 
corresponding condition is $\mu_\chi < 300$ MeV. Thus, the choice 
$B_\text{eff} = (199$ MeV$)^4$ is particularly useful for studying small, 
dense QS configurations associated with a strongly first-order chiral 
transition, provided that the adopted value of $\mu_\chi$ lies below the 
chemical-potential range realized inside the star. If $\mu_\chi$ lies within 
the range of chemical potentials sampled by the stellar profile, the outer 
layers undergo the chiral transition and the resulting object contains a 
massive-quark and hadronic component rather than being a pure QS.
The number density for this model is calculated as $n= \partial p/\partial 
\mu = N_f\pi^{-2}a_\text{eff}\, \mu^3$. Further, the value of $\mu$ at any 
radial distance from the core of the star is calculated as 
$\mu = 4\pi (p + B_\text{eff})^{1/4}/N_f\, a_\text{eff}$. 

\begin{figure}[!h]
	\centerline{
		\includegraphics[scale = 0.3]{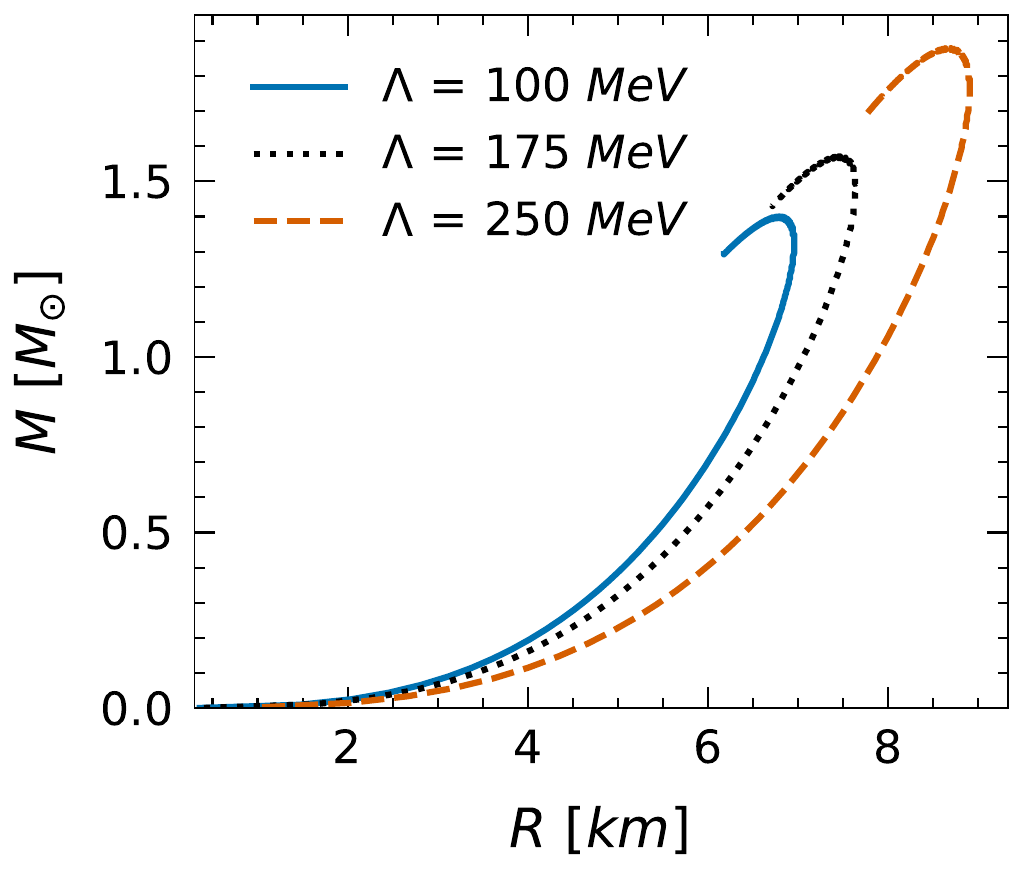}
		\includegraphics[scale = 0.3]{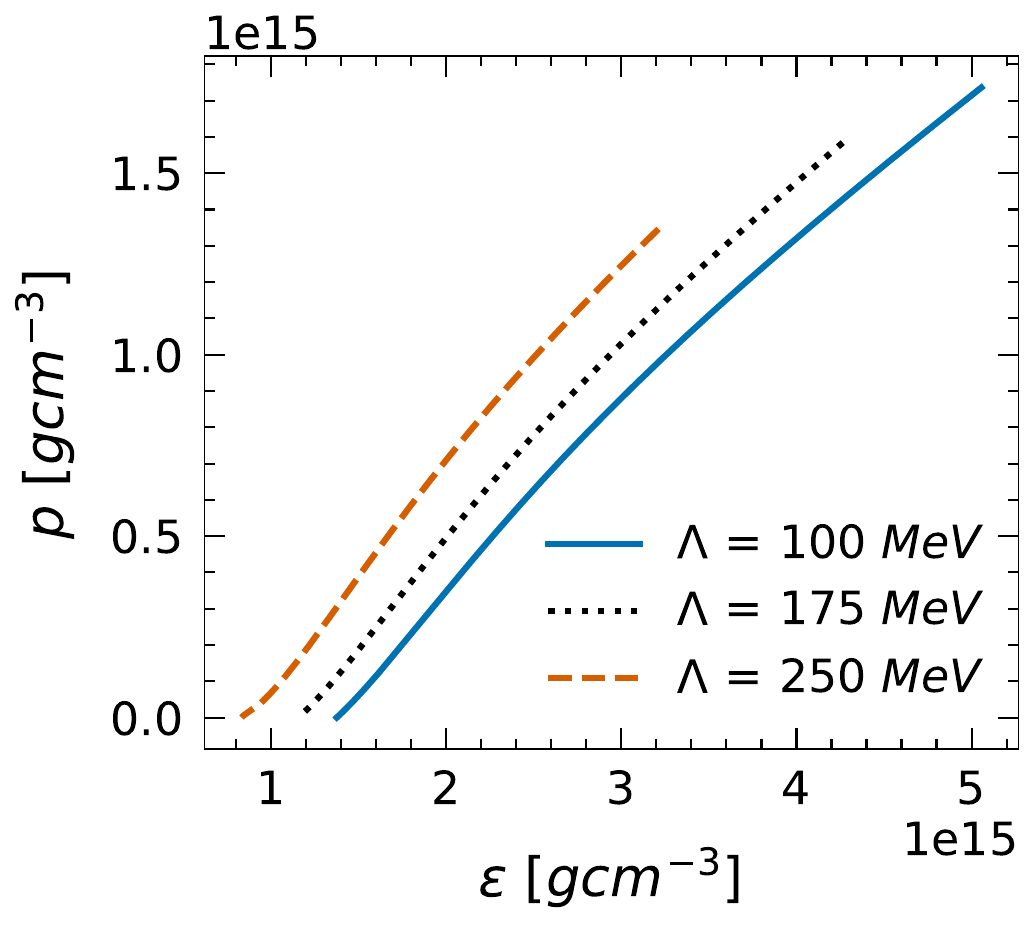}
		\includegraphics[scale = 0.3]{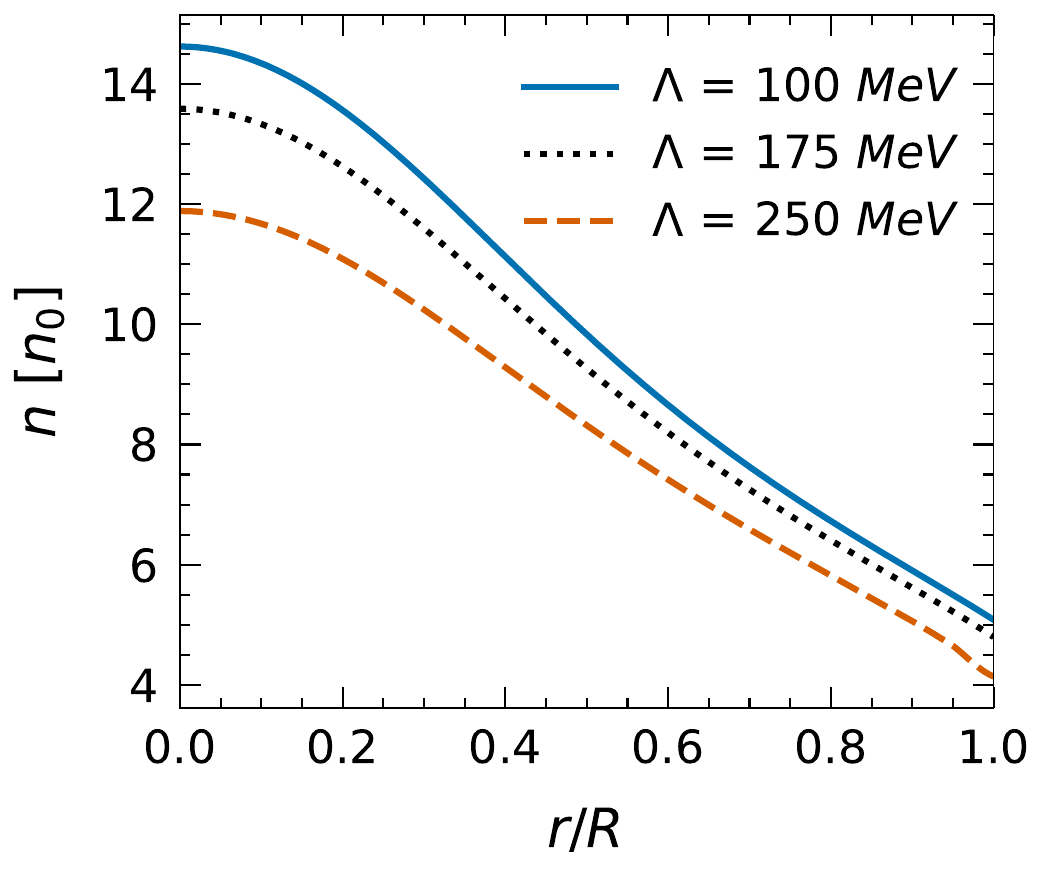}}
	\vspace{-0.3cm}
	\caption{Mass-radius relations (left), energy density vs pressure 
distribution inside the stars (middle) and number density distribution inside 
the stars (right) for different $\Lambda$ values in EoS II.}
	\label{fig:EOSIIComparsonStudy}
\end{figure}

For the temperature-independent massive quark EoS, i.e., for the EoS II, we 
develop a star model (Model II), in which we take $M_{Q} = 310$ MeV, 
$\alpha_{0} = 0.2$, $\Lambda = 100$ MeV and $N = 3.5$.
The parameter $\Lambda$ should have been $\sim 400$ MeV as
per Refs.~\cite{bagchi_2004,bagchi_2006} because the Richardson potential 
\eqref{eq:richardsonPot} has only the $\Lambda$ parameter to describe both 
confinement and asymptotic freedom of quark-quark interaction, where the 
confinement property is more important to determine the static properties of 
hadrons \cite{bagchi_2006, bagchi_2004}. However, studies of strange star 
properties in perturbative QCD show that asymptotic freedom is important to 
get a self-bound high-density star of strange quark matter. So based on 
perturbative QCD studies of deconfined strange quark matter, a smaller 
effective value of the Richardson scale parameter $\Lambda \simeq 100$ MeV 
was considered \cite{dey_1998,shifman_1979}. Here, for a comparison study, we 
plot the mass-radius relation, energy-pressure distribution and number 
density distribution for different values of $\Lambda$ as shown in 
Fig.~\ref{fig:EOSIIComparsonStudy}. The choice of $\Lambda$ was based on the 
microscopic properties of the stars. $\Lambda = 250$ MeV gives the maximum 
stable mass $1.8801M_{\odot}$ with radius $8.67$ km and compactness 
$0.3202$, while it gives less number density than $\Lambda = 100$ MeV at a 
given radial length of the stars.

\begin{figure}[!h]
	\centerline{
		\includegraphics[scale = 0.3]{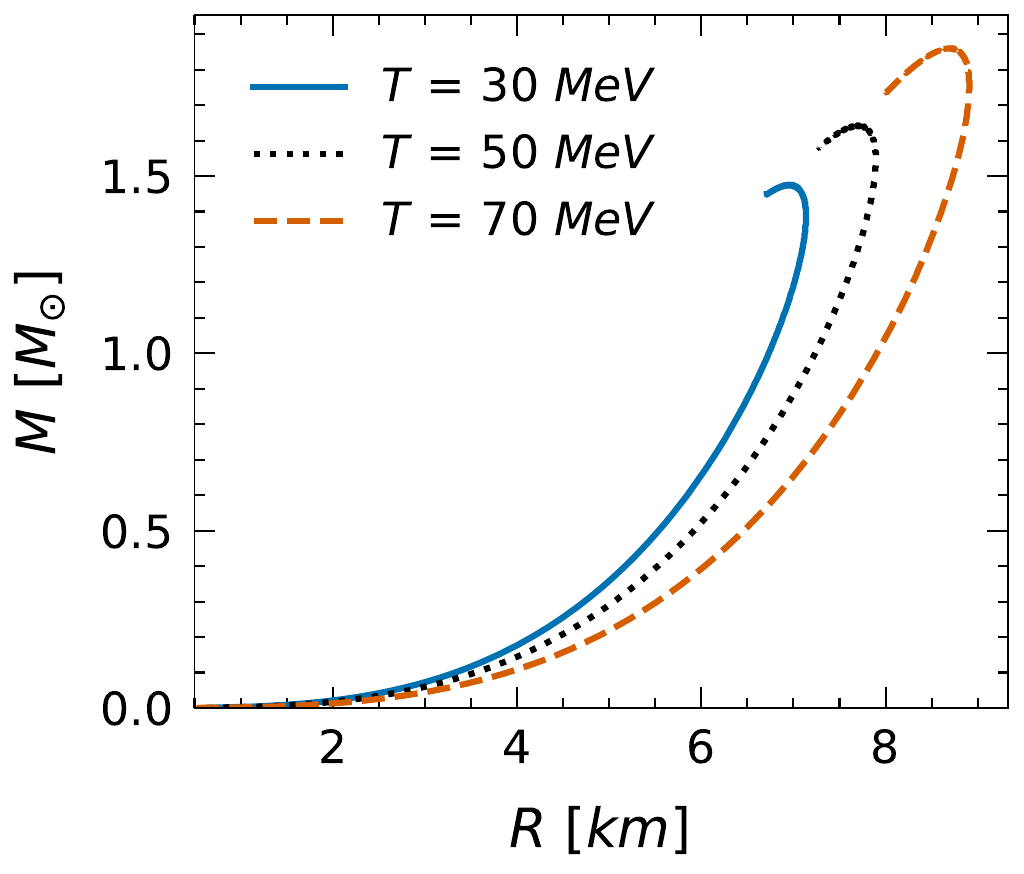}
		\includegraphics[scale = 0.3]{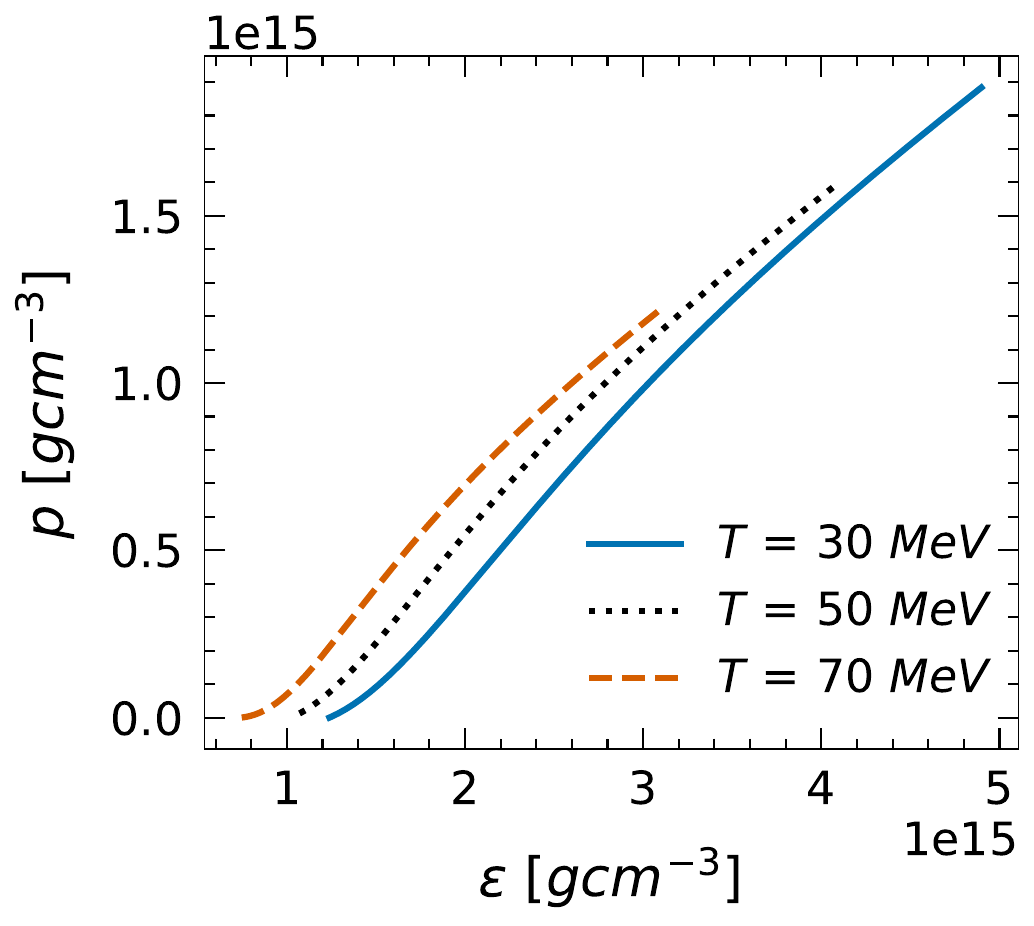}
		\includegraphics[scale = 0.3]{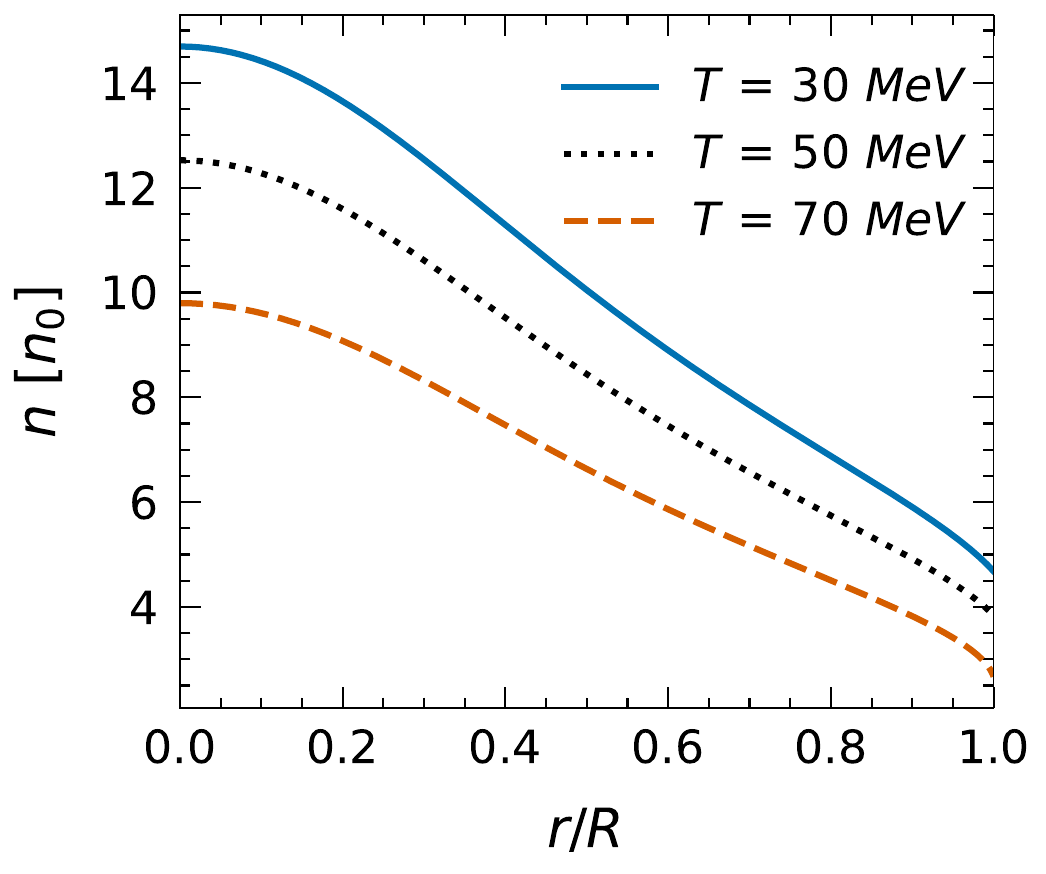}}
	\vspace{-0.3cm}
	\caption{Mass-radius relations (left), energy density vs pressure 
distribution inside the stars (middle) and number density distribution inside 
the stars (right) for different $T$ values in EoS III.}
	\label{fig:EOSIIIComparsonStudy}
\end{figure}

The stellar Model III and Model IV are based on the temperature-dependent 
massive quark EoS, i.e., 
on the EoS III. For the Model III, we take $T = 30$ MeV, and for the Model IV, 
we take $T=70$ MeV. Other free parameters that are considered for both Models 
III and IV are $M_{Q} = 345$ MeV, $\alpha_{0} = 0.65$, $\Lambda = 100$ MeV, and 
$\Lambda^{\prime} = 350$ MeV. However, for the Model III, $N = 4.4$ is 
considered, while for the Model IV, $N = 2.82$ is chosen to satisfy the 
condition: $\epsilon/n_B < 930$ MeV at the surface of the star, such that EoS 
III provides the $\epsilon/n_B$ value at the surface less than the binding 
energy per nucleon of $Fe^{56}$, as mentioned earlier. All these free 
parameters' values of Model II, Model III and Model IV are chosen from the 
range of their constrained values obtained from the asymptotic freedom and 
quark confinement properties in quark-quark interactions 
\cite{bagchi_2006,dey_1998,bagchi_2004}. We adopt the parameter set 
$\Lambda^{\prime} = 350$ MeV and $\Lambda = 100$ MeV proposed by Bagchi et 
al.~\cite{bagchi_2004}, who determined these values phenomenologically through 
relativistic Hartree–Fock calculations of the triple u-quark system 
($\Delta^{++}$) and triple s-quark system ($\Omega^{-}$) for the modified 
Richardson potential given in equation~\eqref{eq:modRichardsonPot}.
Additionally, the screening length in EoS III given by 
equation~\eqref{eq:tempDependentScreeningLength} has the temperature dependence 
of gluon mass, and in 1984 Witten had given the temperature of cosmic 
separation of gluon phases $\sim 100$ MeV \cite{witten_1984, bagchi_2006}. 
Here, we considered the temperature $T = 30$ MeV and $T = 70$ MeV based on the 
stability analysis of stellar structure by Bagchi et al.~\cite{bagchi_2006}. 
They found stable stellar structure up to a temperature of $80$ MeV. We have 
done a comparative study of different stable stellar structures at $T = 30$ MeV,
$50$ MeV and $70$ MeV as shown in Fig.~\ref{fig:EOSIIIComparsonStudy}. The 
maximum stable mass for $T = 30$ MeV and $70$ MeV, along with their respective 
radius and compactness have been listed in Table~\ref{table:MR}. The maximum 
stable mass for $T = 50$ MeV is $1.6437 M_{\odot}$ and its radius and 
compactness are $7.6960$ km and $0.3152$, respectively.
It is observed that the mass and radius of stable stars with maximum 
mass and their corresponding radius value increase with the increase in the 
temperature of the stars. However, the number density distribution shows that 
the stars with higher temperature have a lower number density at a given radial 
length inside the star in comparison to the stars with lower temperature. 
Which means the stars with higher temperature must have lower energy density 
while producing more pressure to balance the gravity. In 
Fig.~\ref{fig:EOSIIIComparsonStudy}, the middle plot shows that the stars with 
$T = 70$ MeV have a higher pressure value compared to $T = 30$ MeV and $50$ 
MeV at the same energy density value.
\begin{figure}[!h]
	\centerline{
		\includegraphics[scale=0.4]{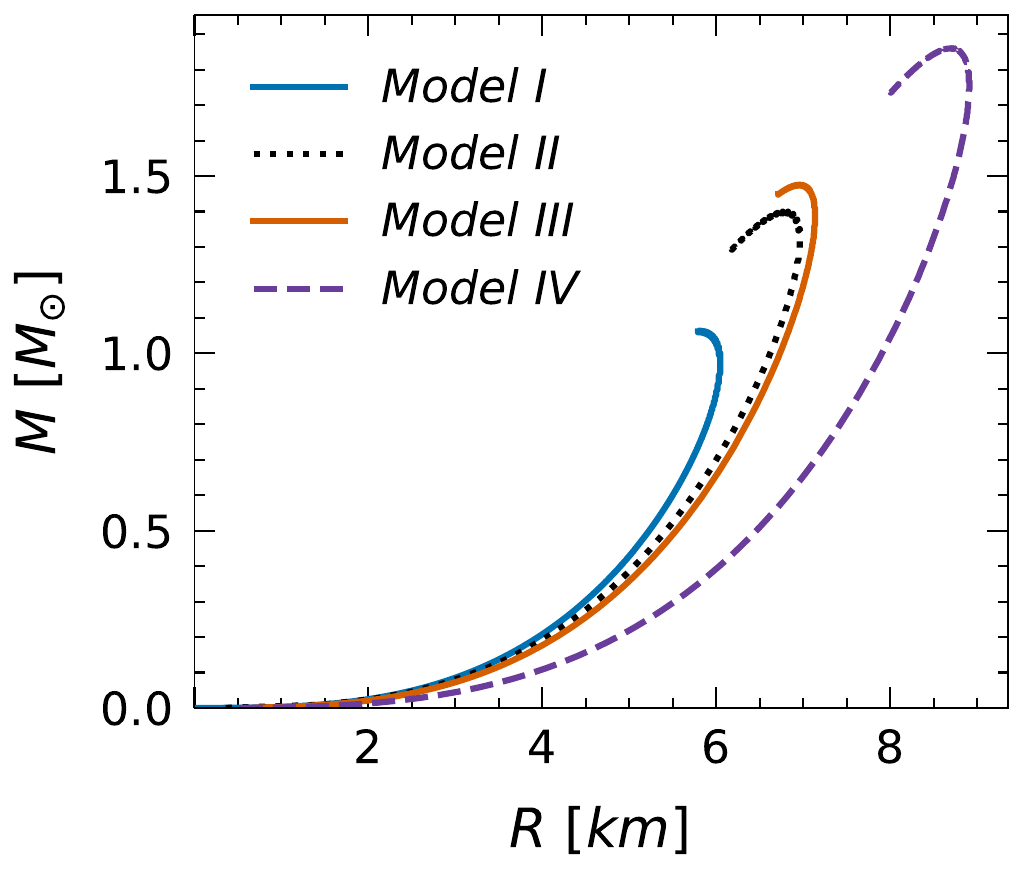}}
	\vspace{-0.2cm}
	\caption{Mass-radius relations for different QS models. Here, Model
		I is based on the EoS I (massless quark EoS), Model II is 
based on the EoS II (massive quark EoS), and Models III and IV are based on 
the EoS III (temperature-dependent massive quark EoS). For the Model III, 
$T=30$ MeV and for the  Model IV, $T=70$ MeV are used.}
	\label{fig:M_R_plot_comparison}
\end{figure}
\begin{table}[h!]
        \begin{center}
        \caption{Maximum mass and corresponding radius of stars for different
stellar models based on different EoSs.}\vspace{2pt}
        \setlength{\tabcolsep}{10pt}
        \begin{tabular}{ccccc}
                \hline \hline
                \rule{0pt}{10pt}Models & EoS & Mass ($M_\odot$) & Radius (km) & Compactness (M/R)\\[1pt]
                \hline \hline \\[-12pt]
                I   & EoS I & $1.0640$ & $5.8050$ & $0.2707$\\

                II  & EoS II & $1.4002$ & $6.7950$ & $0.3043$\\

                III & EoS III, $T = 30$ MeV & $1.4758$ & $6.9600$ & $0.3131$\\

                IV  & EoS III, $T = 70$ MeV & $1.8611$ & $8.7120$ & $0.3154$\\
                \hline
                \hline
        \end{tabular}
        \label{table:MR}
\end{center}
\end{table}

For a comparative understanding,
the mass-radius relations of stars for these stellar models are shown in 
Fig.~\ref{fig:M_R_plot_comparison}, and the maximum masses, maximum radii and 
corresponding compactness values of stars obtained from these relations for 
those different models are listed in Table~\ref{table:MR}. 
From this Fig.~\ref{fig:M_R_plot_comparison} as well as Table
\ref{table:MR}, it is seen that the Model I (EoS I) gives
the stars with the lowest maximum mass, maximum radius and corresponding 
compactness, whereas the Model IV (EoS III) with temperature $T = 70$ MeV 
gives the stars with the highest maximum mass, maximum radius and compactness. 
The maximum mass, maximum radius and compactness of stars given by Model II 
(EoS II) and Model III (EoS III) with $T = 30$ MeV are comparable. Thus, it is 
seen that the maximum mass, maximum radius and compactness of stars are higher 
for the massive quark EoS model. Further, these parameters of stars increase 
with increasing temperature of QM inside the stars.    

\begin{figure}[!h]
        \centerline{
                \includegraphics[scale = 0.3]{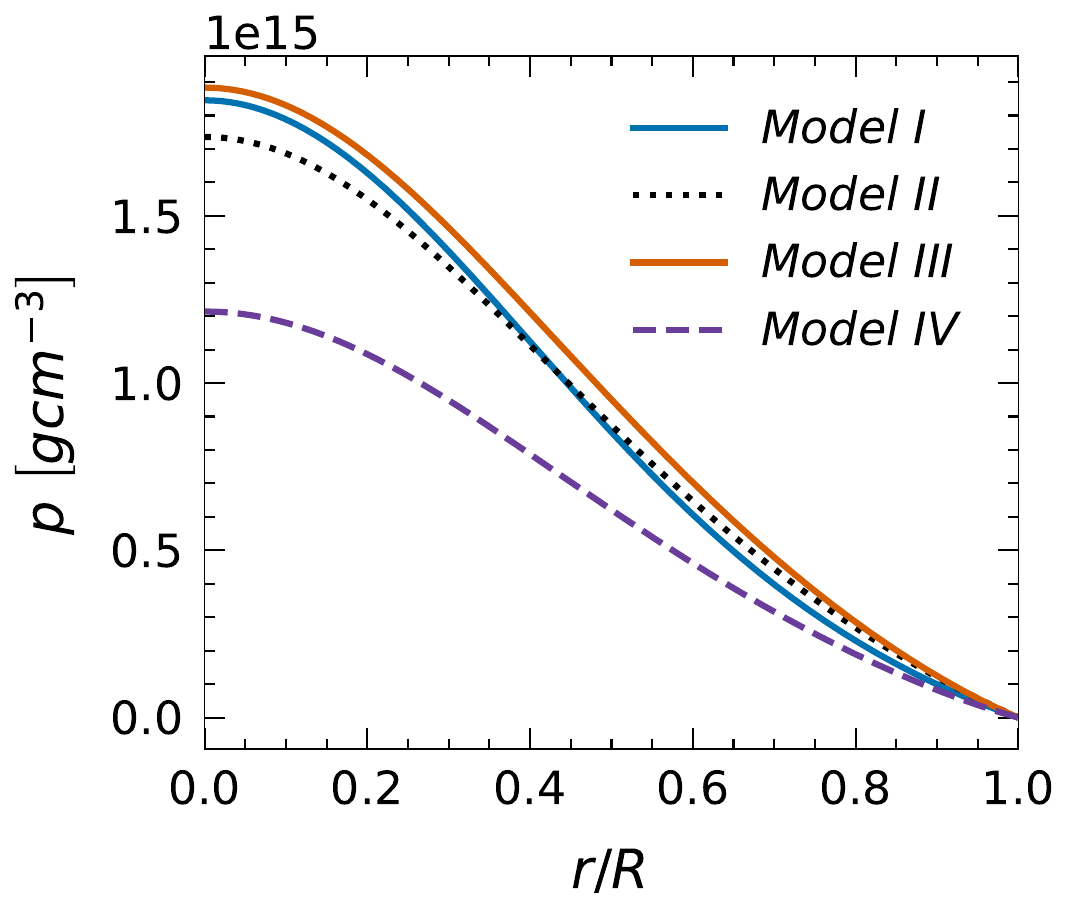}
                \includegraphics[scale = 0.3]{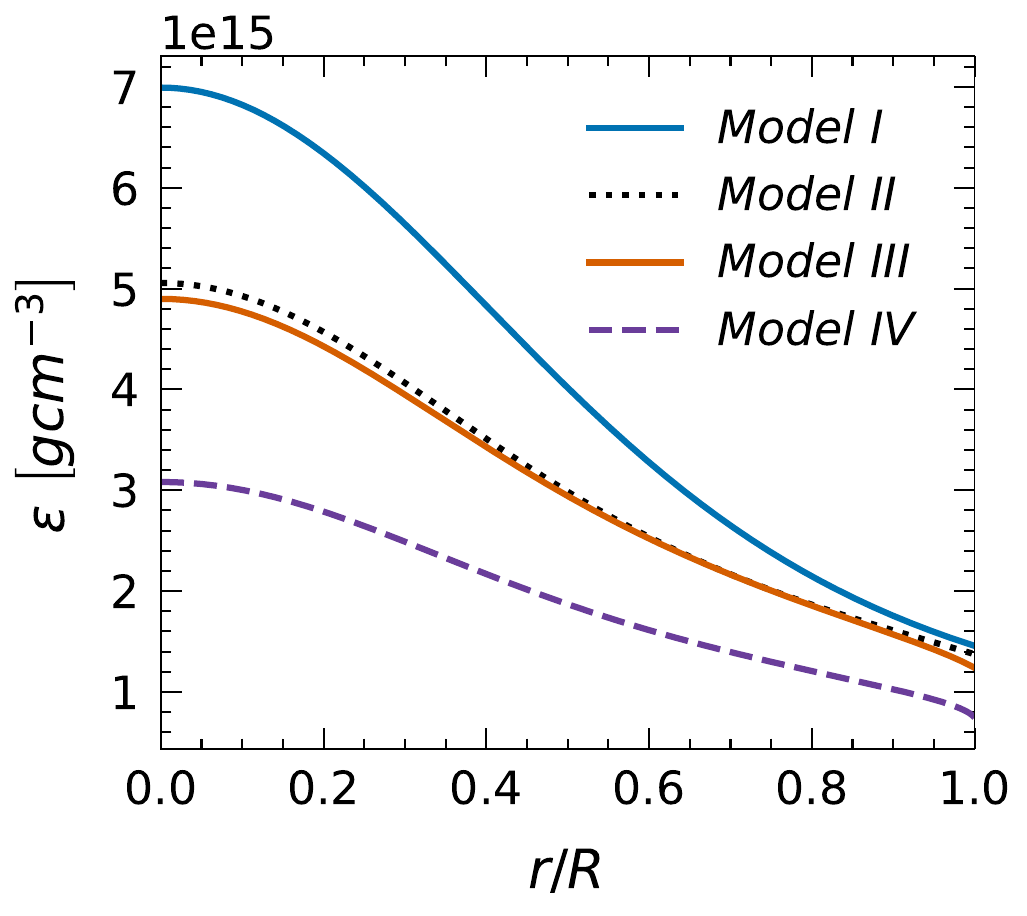}
                \includegraphics[scale = 0.3]{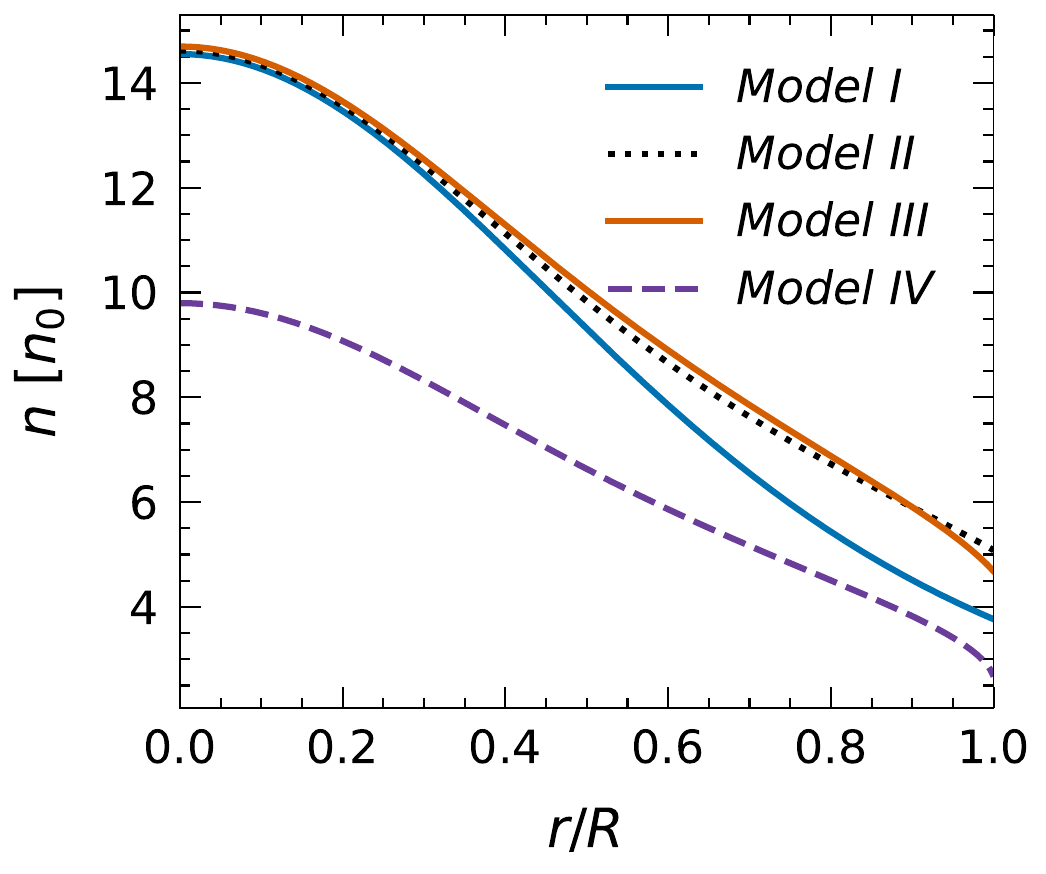}}
                \vspace{-0.3cm}
        \caption{Radial variations of pressure (left), energy density
(middle) and number density (right) for different stellar models.}
        \label{fig:pressure_energy_and_number_distribution}
\end{figure}
Moreover, the pressure, energy density and number density distribution
characteristics inside the stars obtained for these models are shown in
Fig.~\ref{fig:pressure_energy_and_number_distribution}. This figure shows 
that the Model III, based on the massive quark EoS at temperature $T=30$ MeV, 
gives stars of the highest central pressure, whereas the Model IV, based on the 
massive quark EoS at temperature $T=70$ MeV, gives stars of the lowest central
pressure. Thus, with increasing temperature of massive quark matter, pressure 
decreases inside the stars. The Model I, based on massless quark EoS, and the 
Model II, based on 
temperature-independent massive quark EoS, give stars of comparable central
pressure. Towards the surface of the stars, the pressures and differences in 
pressures for different models decrease. In the case of 
energy density, Model IV gives the lowest energy density from the center to 
the surface. While the Model I gives the highest central energy density, the 
Models II and III give comparable energy density throughout the interior of the
stars. However, towards the surface of the stars, energy densities for the
Models I, II, and III become nearly equal. For the case of the number density 
of stars, overall predictions of models are similar to the case of pressure;
however, apart from Models II and III, the number densities for the other two
models are different at surface. That is the number density of the stars
is lowest for Model IV throughout the stars.       

\subsection{Radial Oscillations}
Following the procedure as discussed in Section \ref{secIV_a}, we calculate 
the eigenfrequency spectrum of radial oscillations of non-rotating QSs 
for different stellar models developed based of EoS I, EoS II and EoS III, 
as mentioned already, and listed in Table~\ref{table:eigen-frequencies}. 
\begin{table}[!h]
	\begin{center}
        \caption{Eigenfrequencies of radial oscillations of stars for different
stellar models based on different EoSs. All frequencies are in kHz.}
        \vspace{3pt}
	\setlength{\tabcolsep}{15pt}
	\begin{tabular}{ccccc}
		\hline\hline\\[-8pt]
		\rule{0pt}{15pt} \shortstack{Modes\\[2pt] (Order $n$)} & 
\shortstack{Model I\\[2pt] (EoS I)} & \shortstack{Model II\\[2pt](EoS II)} & 
\shortstack{Model III \\[2pt] (EoS III, $T = 30$ MeV)} & \shortstack{Model IV 
\\[2pt] (EoS III, $T = 70$ MeV)} \\[2pt] \hline\hline\\[-12pt]
		$f (0)$		   & 0.25  & 0.08  & 0.21  & 0.52 \\
		$p_{1} (1) $   & 11.14 & 9.92  & 9.61  & 7.77 \\
		$p_{2} (2) $   & 18.29 & 16.12 & 15.49 & 12.47 \\
		$p_{3} (3) $   & 25.09 & 22.03 & 21.08 & 16.93 \\
		$p_{4} (4) $   & 31.77 & 27.83 & 26.58 & 21.30 \\
		$p_{5} (5) $   & 38.38 & 33.59 & 32.04 & 25.63 \\
		$p_{6} (6) $   & 44.96 & 39.32 & 37.48 & 29.94 \\
		$p_{7} (7) $   & 51.52 & 45.04 & 42.90 & 34.23 \\
		$p_{8} (8) $   & 58.07 & 50.74 & 48.31 & 38.52 \\
		$p_{9} (9) $   & 64.60 & 56.44 & 53.72 & 42.80 \\
		$p_{10} (10)$  & 71.13 & 62.13 & 59.12 & 47.80 \\
		$p_{11} (11)$  & 77.65 & 67.82 & 64.52 & 51.36 \\
		$p_{12} (12)$  & 84.17 & 73.51 & 69.92 & 55.64 \\
		$p_{13} (13)$  & 90.69 & 79.19 & 75.32 & 59.91 \\
		$p_{14} (14)$  & 97.20 & 84.87 & 80.71 & 64.18 \\
		$p_{15} (15)$  & $\_$  & 90.55 & 86.11 & 68.46 \\
		$p_{16} (16)$  & $\_$  & 96.23 & 91.50 & 72.73 \\
		$p_{17} (17)$  & $\_$  & $\_$  & 96.89 & 77.00 \\
		$p_{18} (18)$  & $\_$  & $\_$  & $\_$  & 81.27 \\
		$p_{19} (19)$  & $\_$  & $\_$  & $\_$  & 85.54 \\
		$p_{20} (20)$  & $\_$  & $\_$  & $\_$  & 89.81 \\
		$p_{21} (21)$  & $\_$  & $\_$  & $\_$  & 94.08 \\
		$p_{22} (22)$  & $\_$  & $\_$  & $\_$  & 98.35 \\[3pt]
		\hline
	\end{tabular}
	\label{table:eigen-frequencies}
        \end{center}
\end{table}
\begin{figure}[!h]
	\centerline{
		\includegraphics[scale = 0.35]{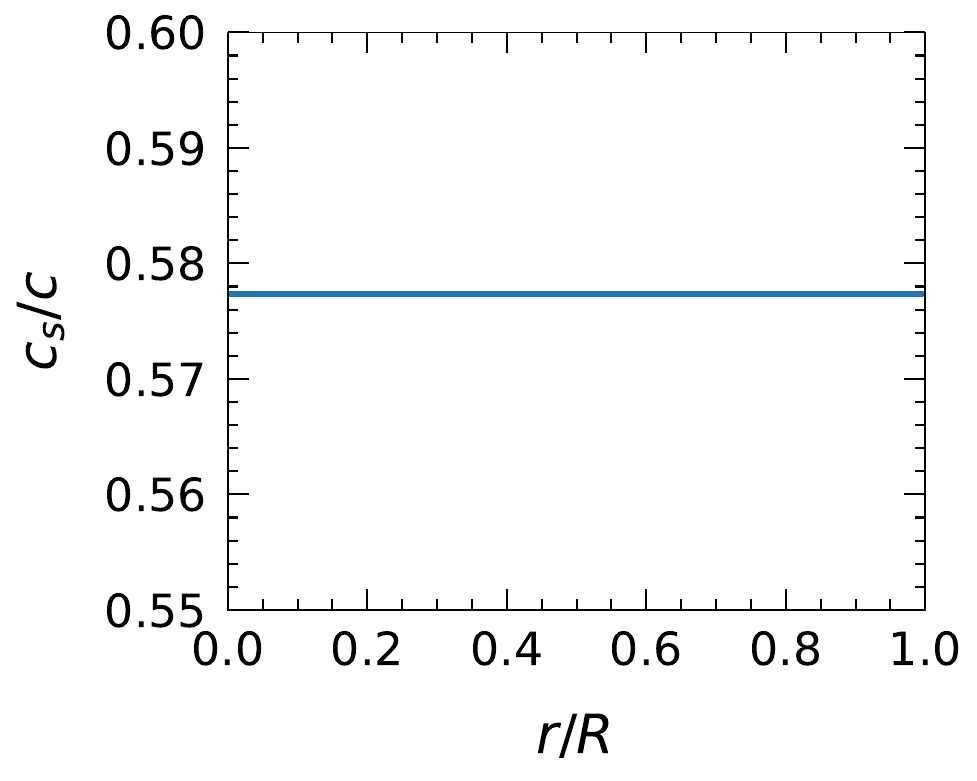}
                \hspace{0.5cm}
		\includegraphics[scale = 0.35]{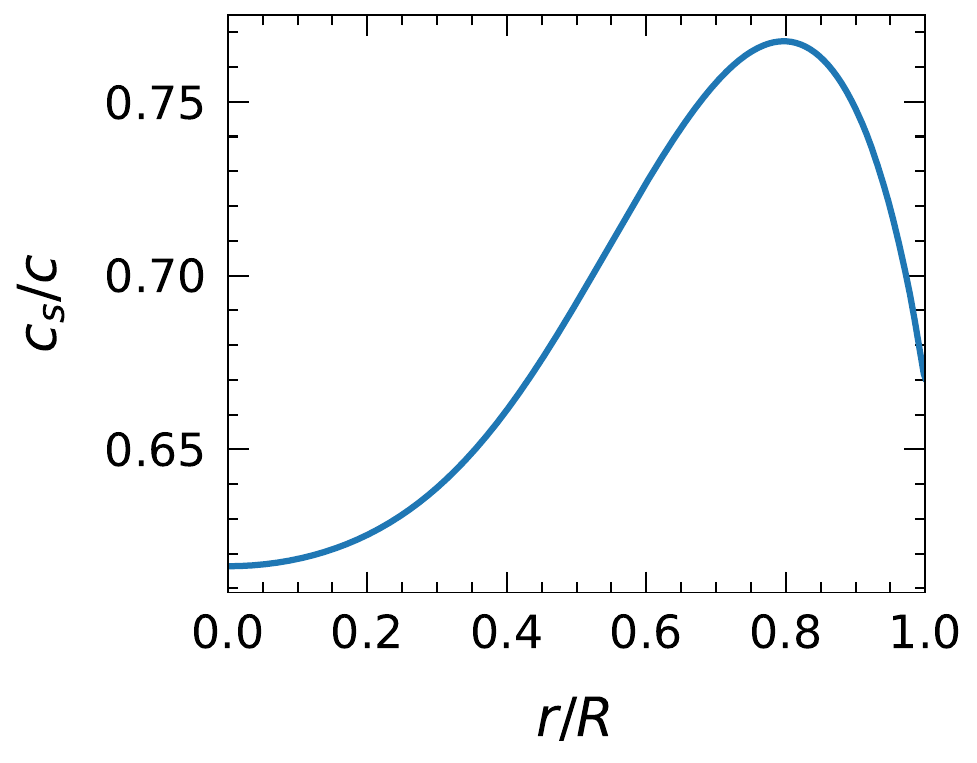}}
                \vspace{0.3cm}
        \centerline{
		\includegraphics[scale = 0.35]{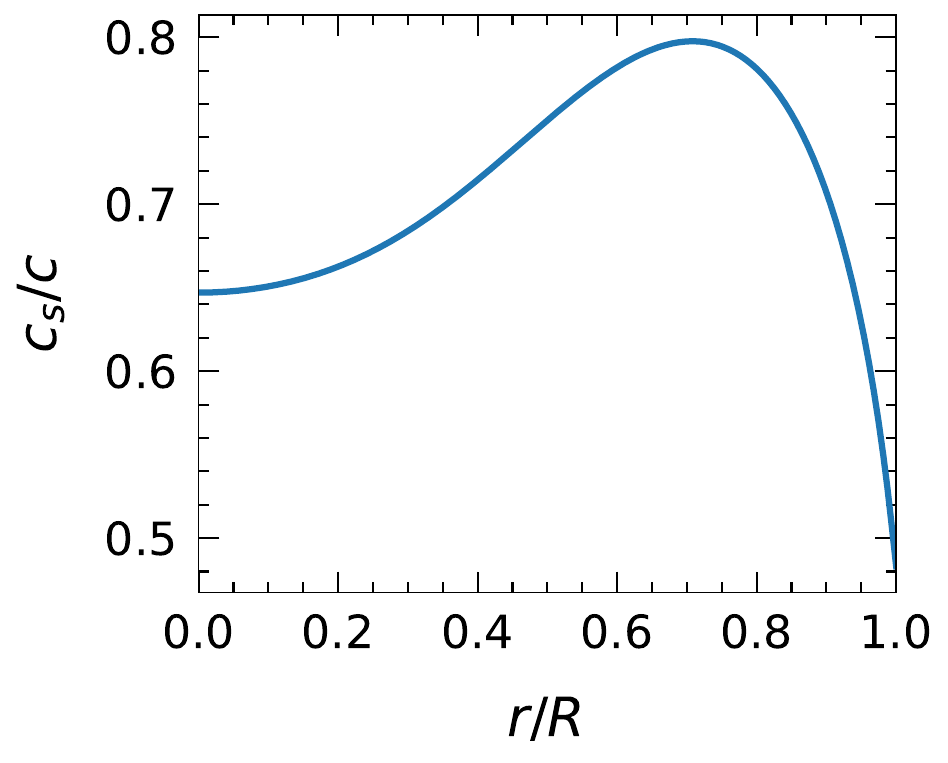}
                \hspace{0.5cm}
		\includegraphics[scale = 0.35]{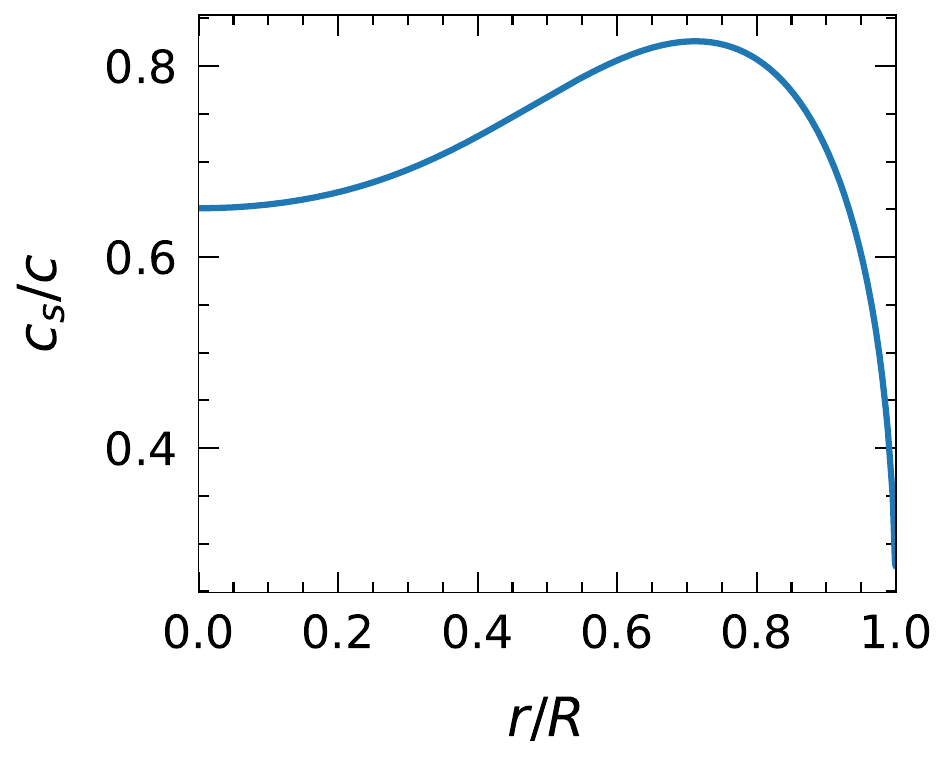}}
                \vspace{-0.3cm}
	\caption{Variation of relative acoustic wave velocity inside the 
non-rotating stars for Model I (top left), Model II (top right), 
Model III (bottom left), and Model IV (bottom right).}	
	\label{fig:acoustic_wave_non_rot}
\end{figure}
These oscillation frequencies are constrained up to $100$ kHz, within which 
different models exhibit different numbers of oscillation modes. To understand 
the origin of the variation in the number of modes among different models, we 
analyze the stellar mass, radius, compactness, and acoustic wave velocity 
profiles for these models. From Tables~\ref{table:MR} and 
\ref{table:eigen-frequencies}, one can observe that for models with higher 
stellar mass, radius and compactness tend to exhibit a greater number of 
oscillation modes. Further, in Models III and IV, temperature variations 
significantly affect the mass and radius of the stars, indicating a possible 
correlation between temperature and the number of oscillation modes. We also 
investigate the variation of acoustic wave velocity as a function of radial 
distance inside the stars. For Model I, the acoustic wave velocity remains 
nearly constant at $\sim \sqrt{1/3}$. In contrast, for the massive quark EoSs, 
Models II, III, and IV, the acoustic wave velocity varies significantly from 
the stellar center to the surface, as shown in 
Fig.~\ref{fig:acoustic_wave_non_rot}. In these three models, the acoustic wave 
velocity reaches a peak at approximately $70\%$ of the stellar radius and 
subsequently decreases toward the surface. This behavior indicates that the 
matter in Models II, III, and IV is relatively stiffer than that in Model I 
over a significant portion of the stellar interior, leading to stronger 
restoring forces. However, after the peak value, the acoustic wave velocity 
decreases rapidly, implying that the matter becomes softer in the outer layers 
of the star. Consequently, the local pressure restoring force and sound speed 
are reduced in these regions. The reduced sound speed increases the acoustic 
travel time through the outer layers, causing the radial overtone spectrum to 
become more densely packed. As a result, a larger number of eigenfrequency 
modes can appear within a fixed frequency range.

\begin{figure}[!h]
        \centerline{
               \includegraphics[scale = 0.3]{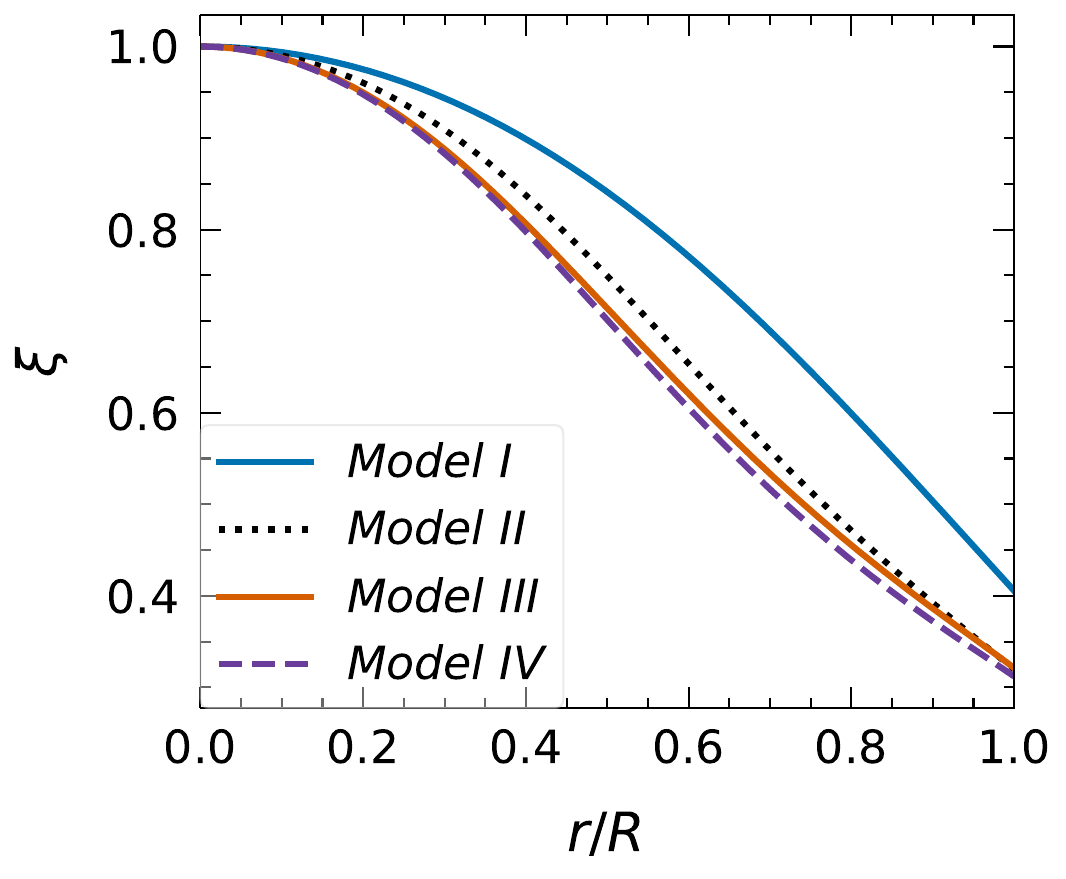}\hspace{0.3cm}
               \includegraphics[scale = 0.3]{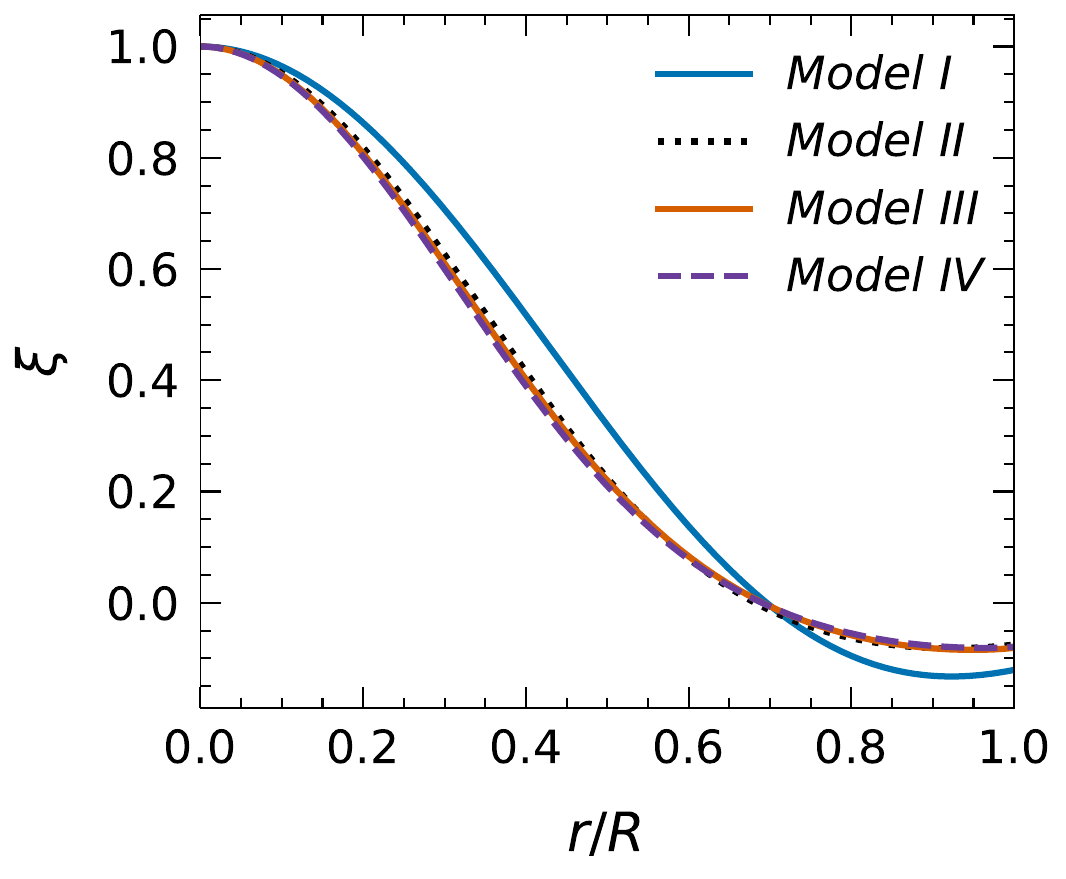}\hspace{0.3cm}
               \includegraphics[scale = 0.3]{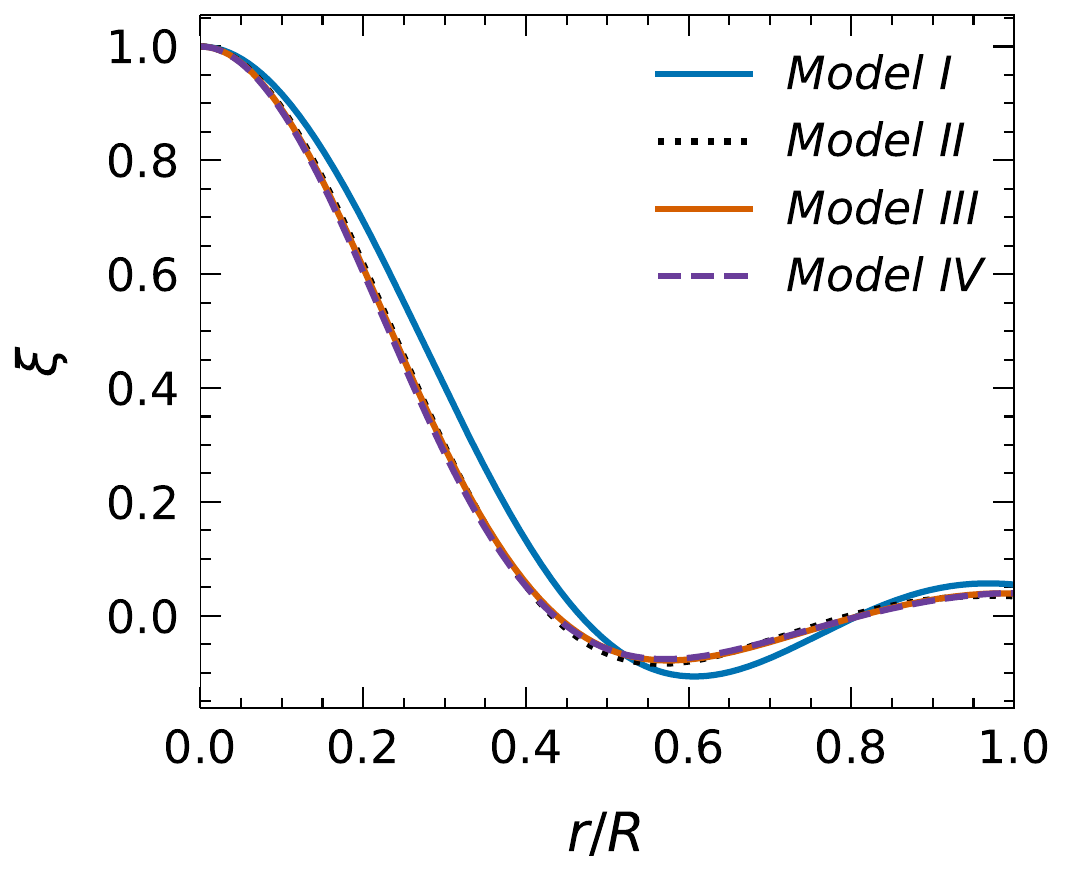}}\vspace{0.2cm}
        \centerline{
                \includegraphics[scale = 0.3]{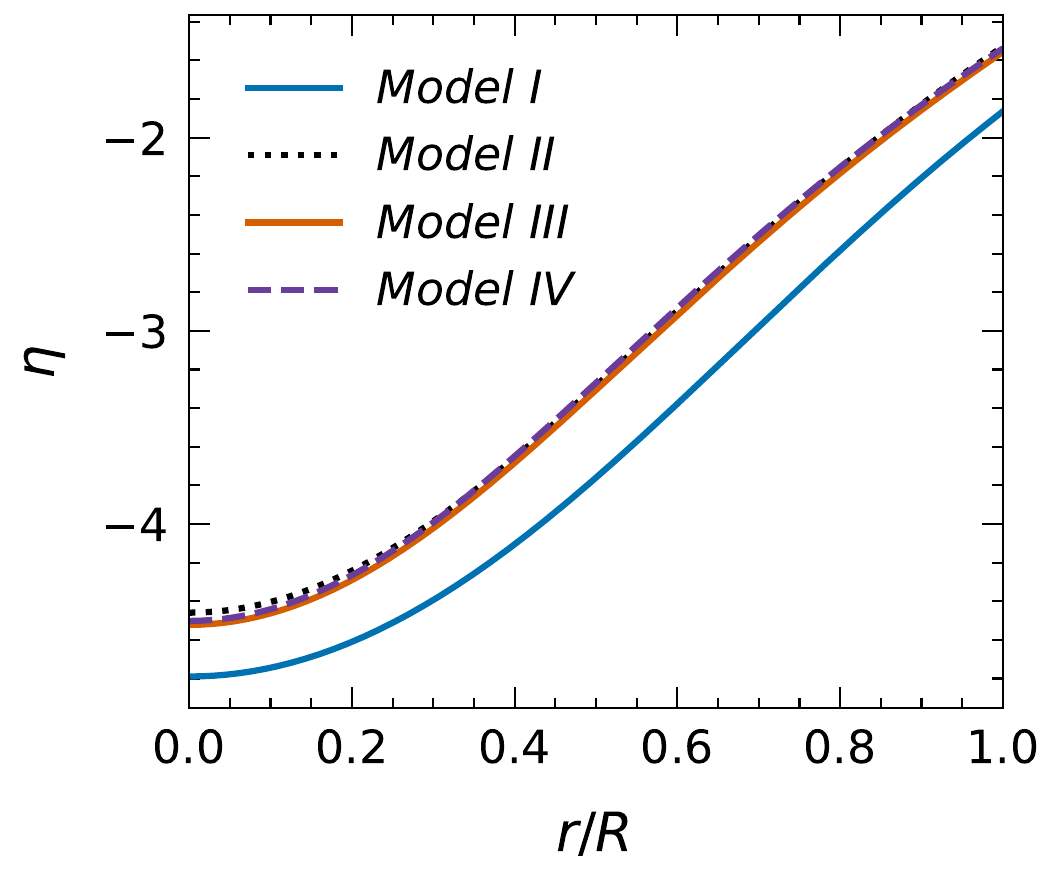}\hspace{0.3cm}
                \includegraphics[scale = 0.3]{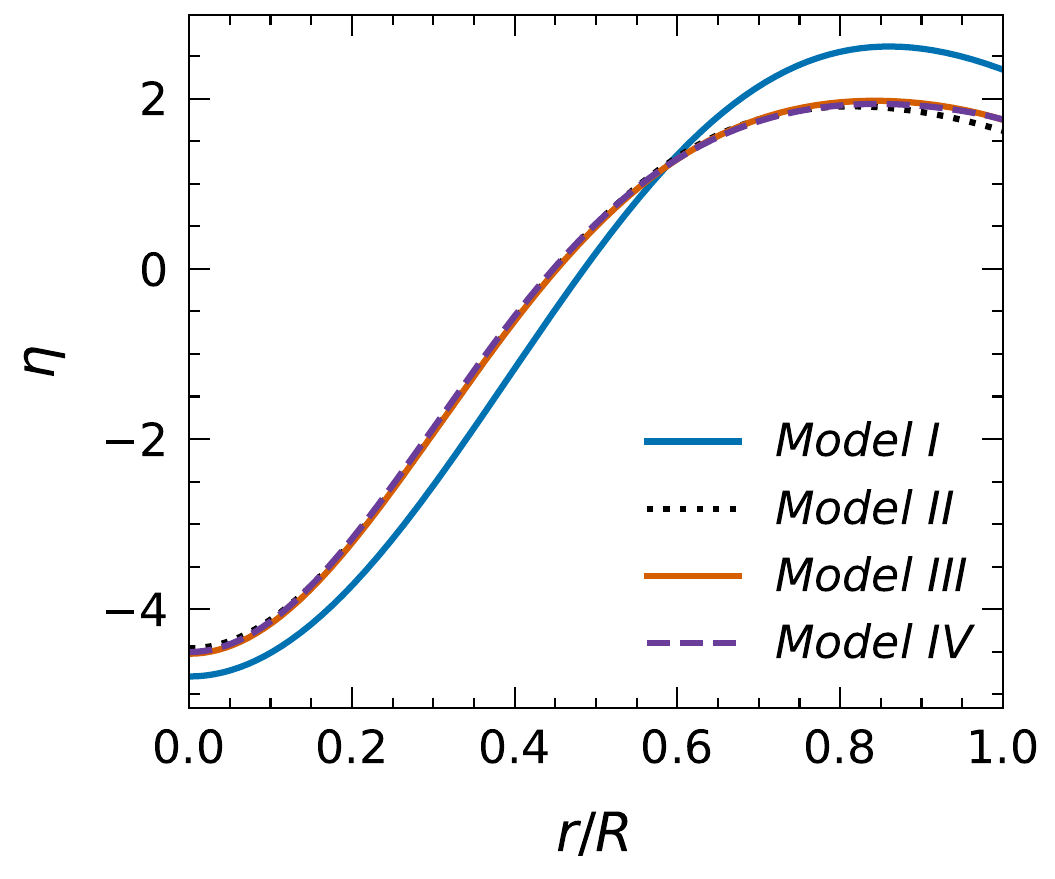}\hspace{0.3cm}
                \includegraphics[scale = 0.3]{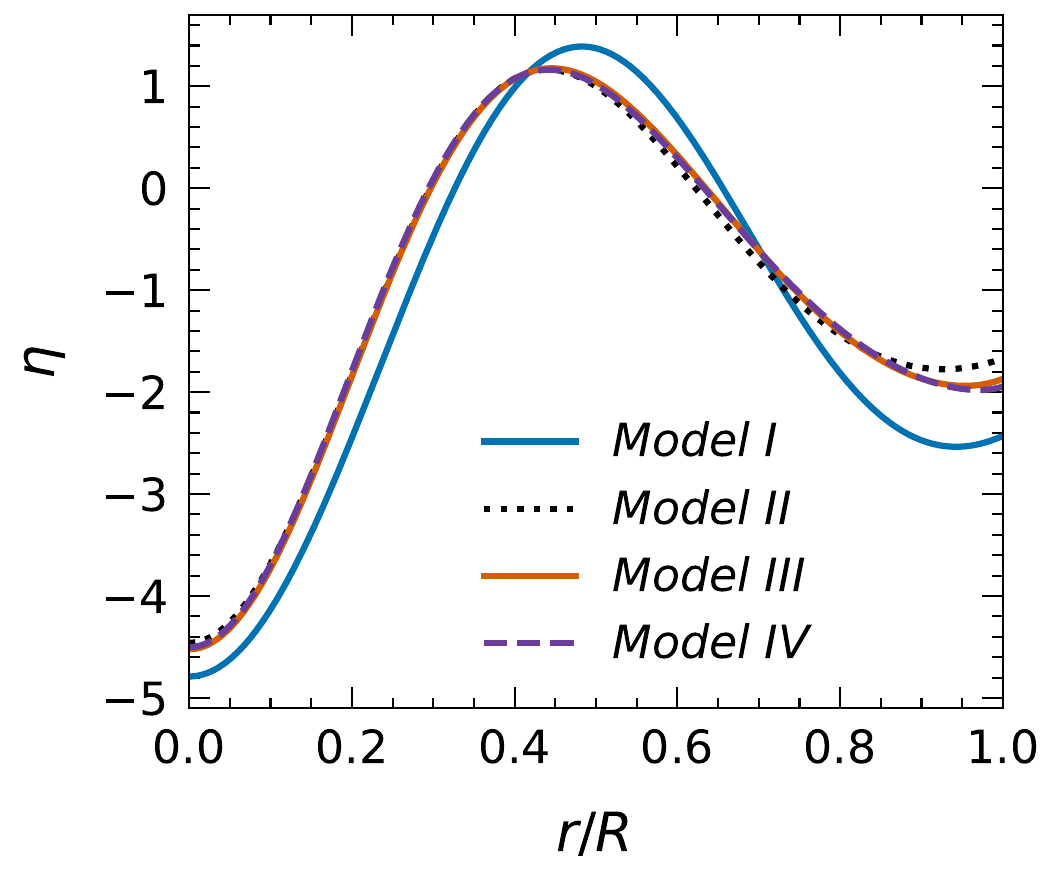}}
                \vspace{-0.3cm}
        \caption{Radial variations of $f$, $p_{1}$ and $p_{2}$ modes (from
left to right, respectively) of radial perturbation $\xi$ (top panels), and
corresponding modes of pressure perturbation $\eta$ for different stellar
models (bottom panels).}
\label{fig:f_p1_p2_mode_xi_and_eta}
\end{figure}
\begin{figure}[!h]
        \centerline{
                \includegraphics[scale = 0.3]{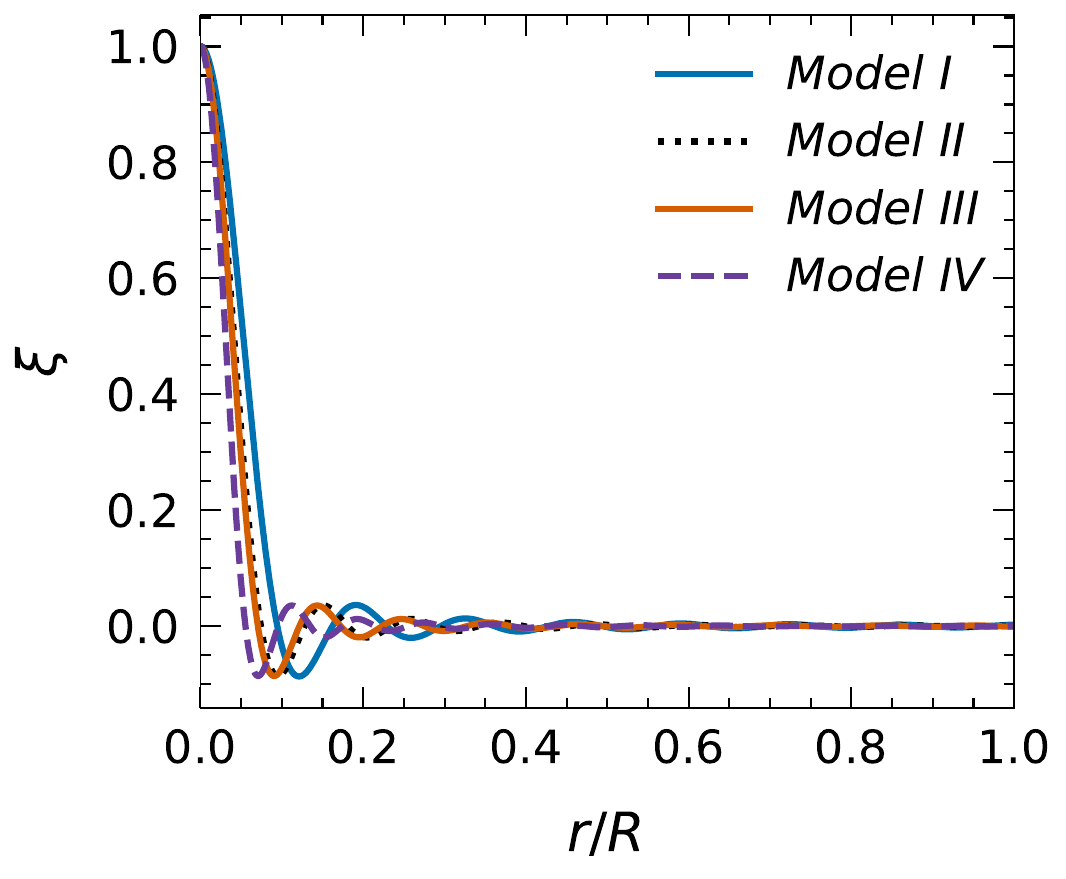}\hspace{0.3cm}
                \includegraphics[scale = 0.3]{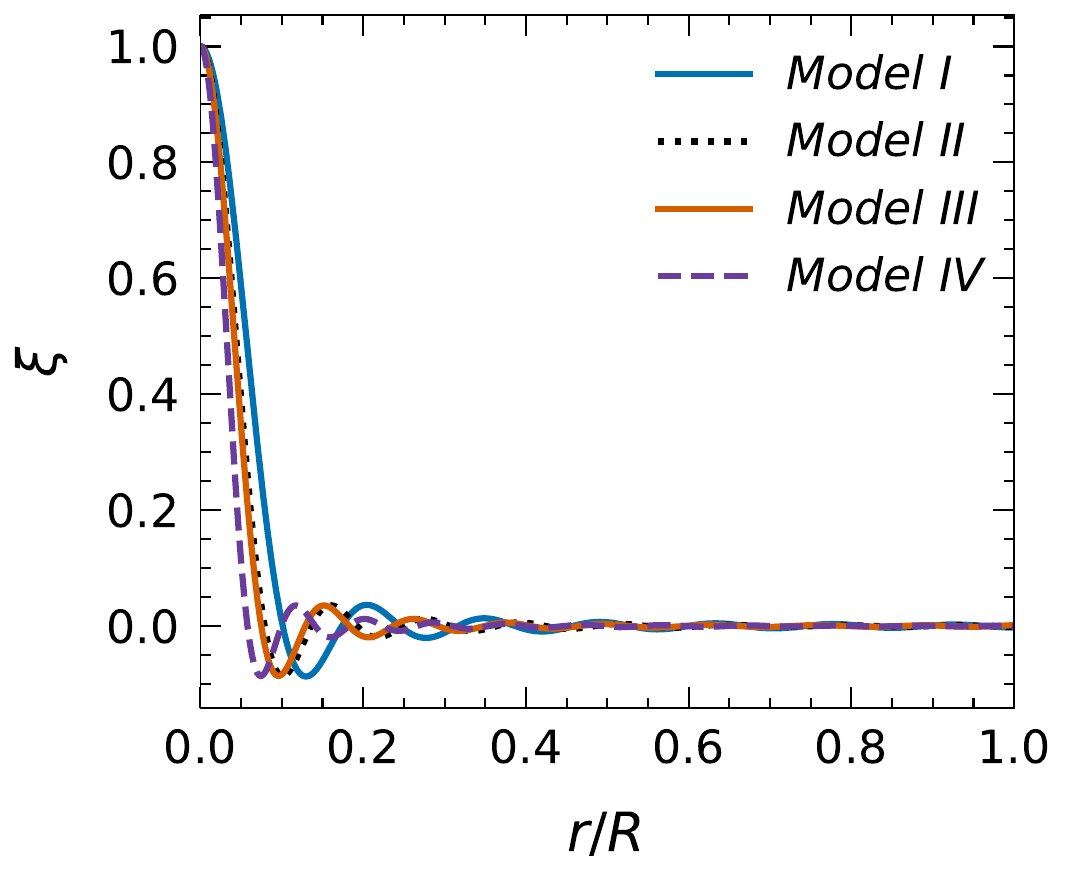}\hspace{0.3cm}
                \includegraphics[scale = 0.3]{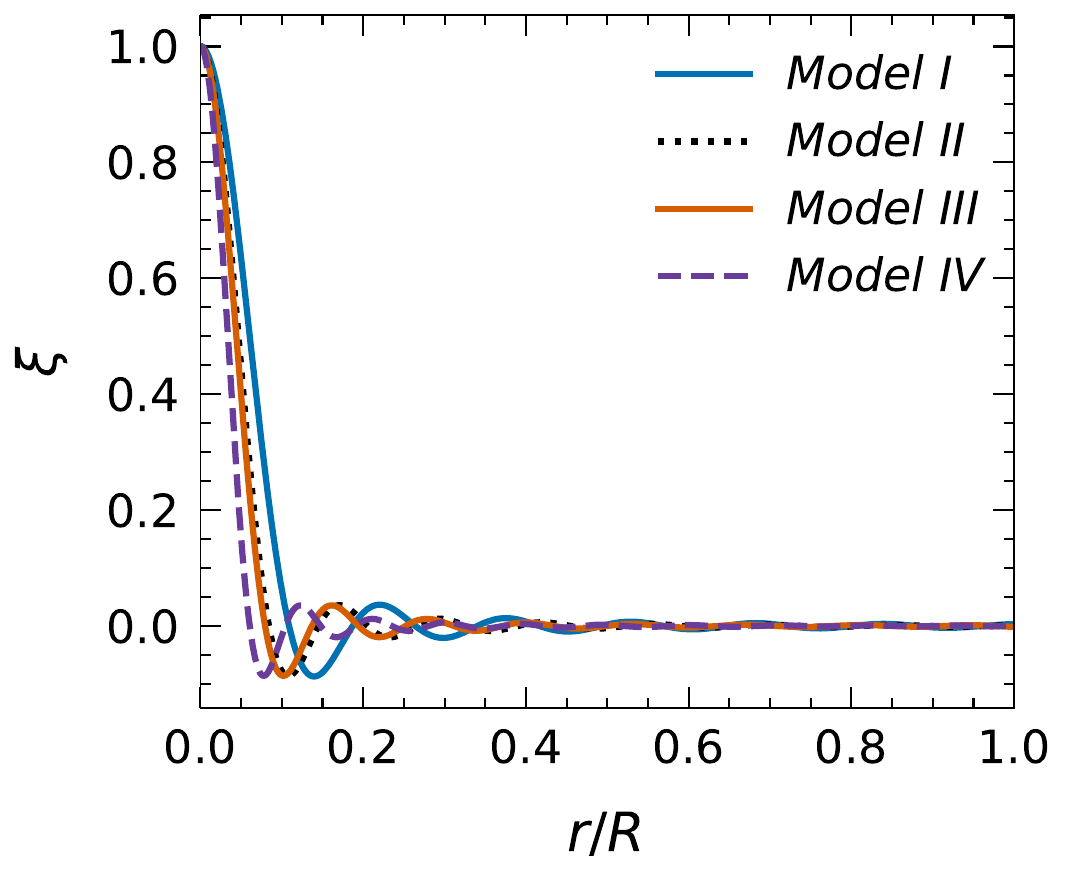}}\vspace{0.2cm}
        \centerline{
                \includegraphics[scale = 0.3]{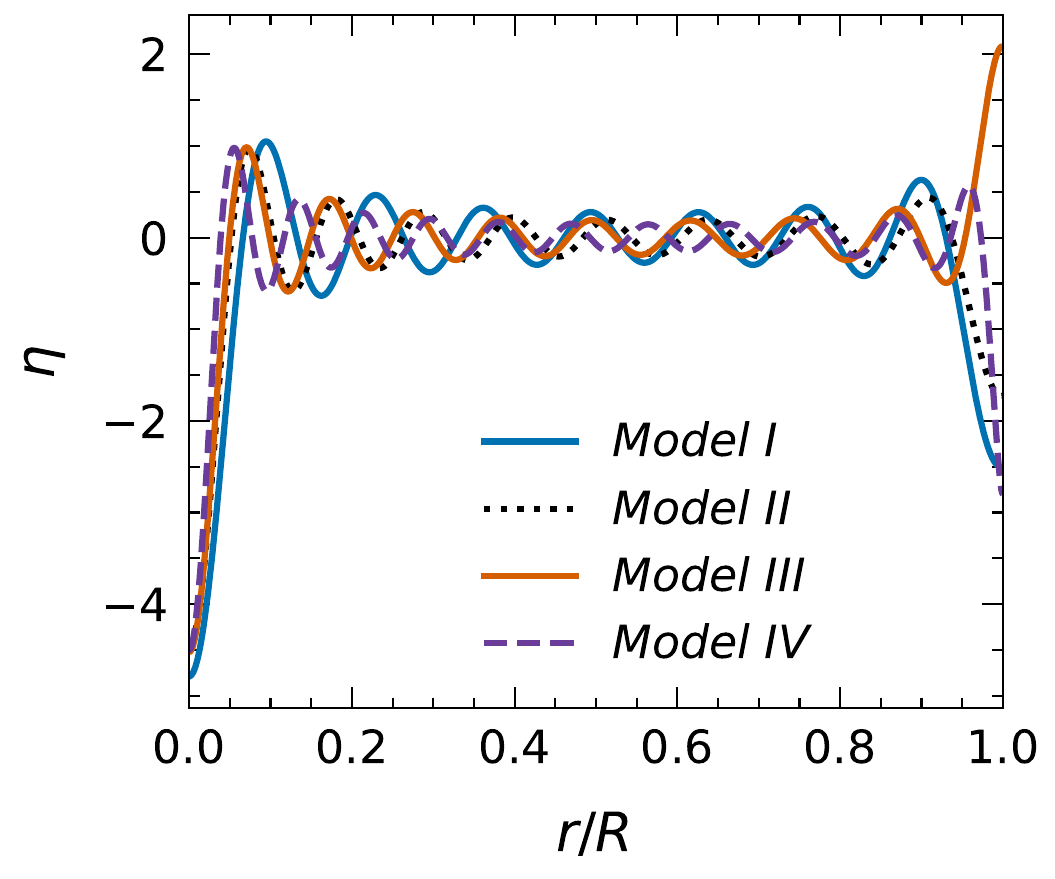}\hspace{0.3cm}
                \includegraphics[scale = 0.3]{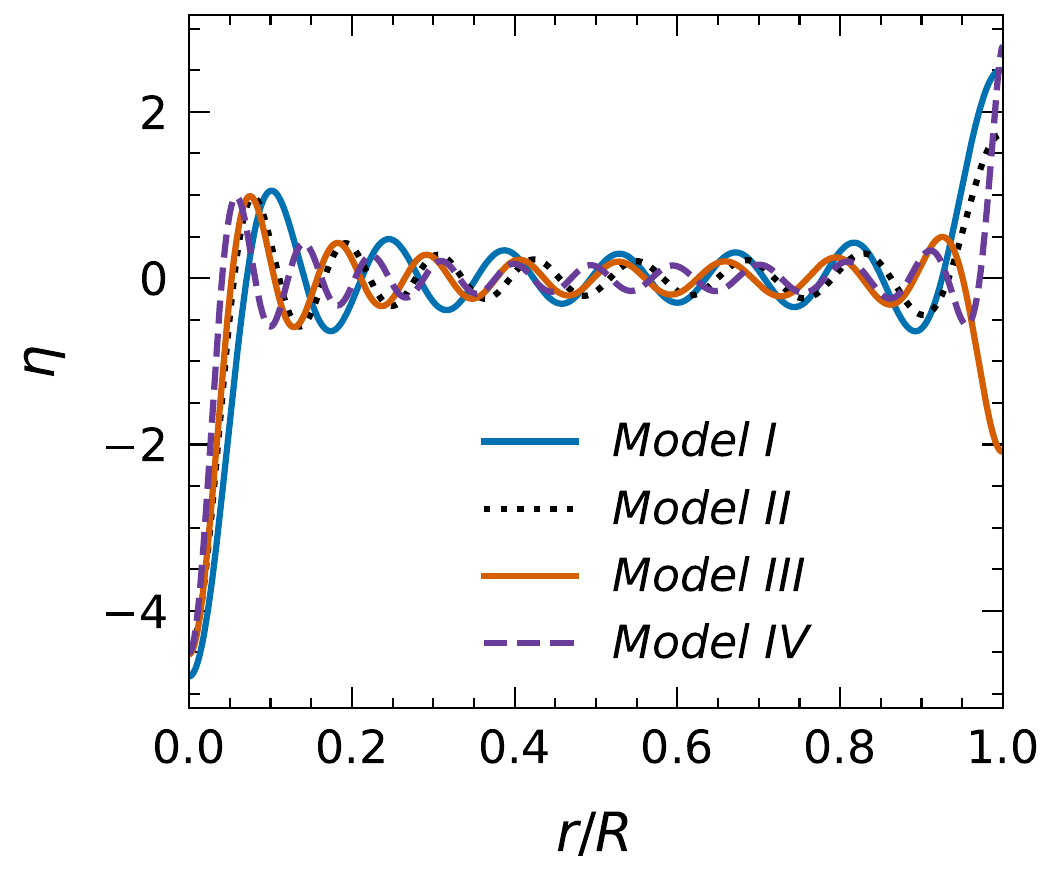}\hspace{0.3cm}
                \includegraphics[scale = 0.3]{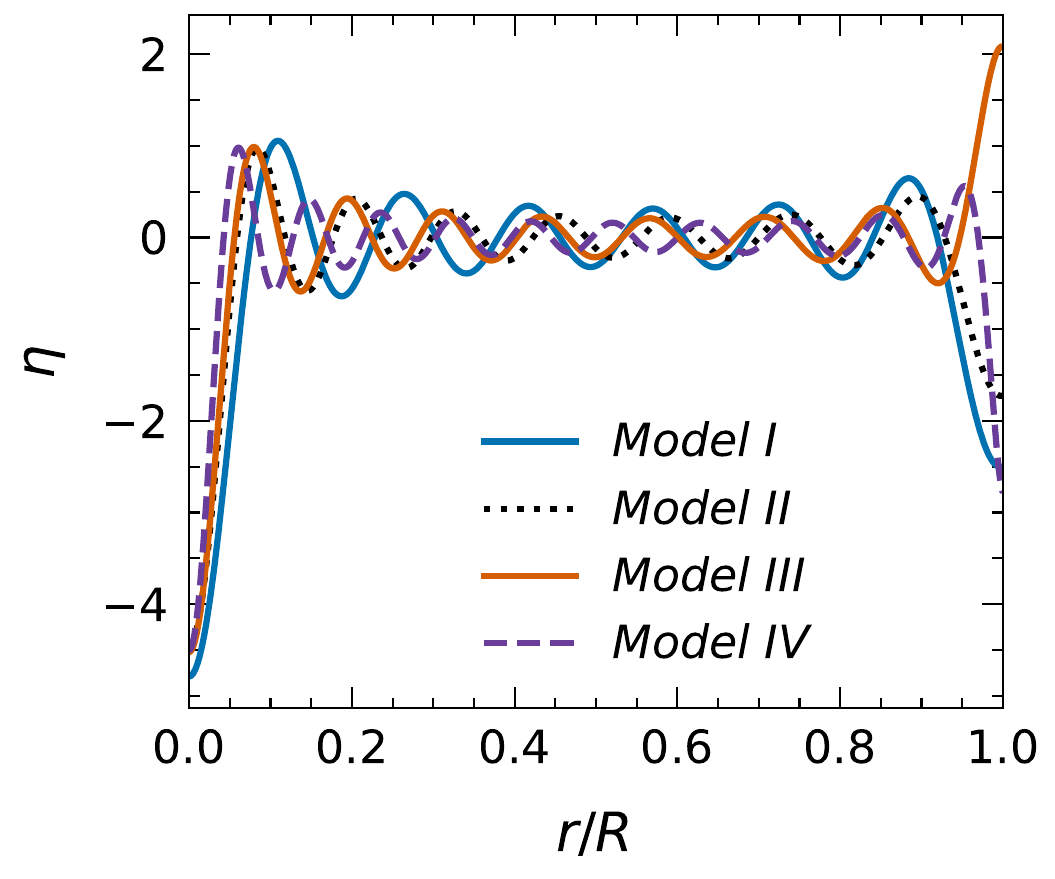}}
                \vspace{-0.3cm}
        \caption{Radial variations of the last three higher-order modes (from
left to right, respectively) of radial perturbation $\xi$ (top panels), and
corresponding modes of pressure perturbation $\eta$ for different stellar
models (bottom panels).}
\label{fig:(n-2)_(n-1)_n_mode_xi_and_eta}
\end{figure}
We also plot the $f$, $p_{1}$ and $p_{2}$ modes (three lower-order modes) 
and the last three modes (three higher-order modes) for each model in 
Fig.~\ref{fig:f_p1_p2_mode_xi_and_eta} and 
Fig.~\ref{fig:(n-2)_(n-1)_n_mode_xi_and_eta}, respectively. For the 
lower-order modes, the radial profiles of $\xi$ and $\eta$ exhibit only 
negligible variations with temperature, remaining nearly identical throughout 
the stellar interior. However, Model I shows notable deviation in radial
and pressure perturbation profiles within and on the surface of the stars, 
except at their centers, compared to the other three models. This deviation is 
specifically significant for the $f$-mode of oscillations, and it
decreases with increasing modes.  Whereas, for the higher-order modes, the 
effects of temperature become more apparent, leading to a visible separation 
of the $\xi$ and $\eta$ profiles, particularly in the central region of the 
stars. Nevertheless, towards the surface of the stars, while the behaviours of 
$\xi$ for all models become indistinguishable, the behaviours of $\eta$ remain 
distinguishable near the surface. 

\begin{figure}[!h]
	\centerline{
		\includegraphics[scale = 0.3]{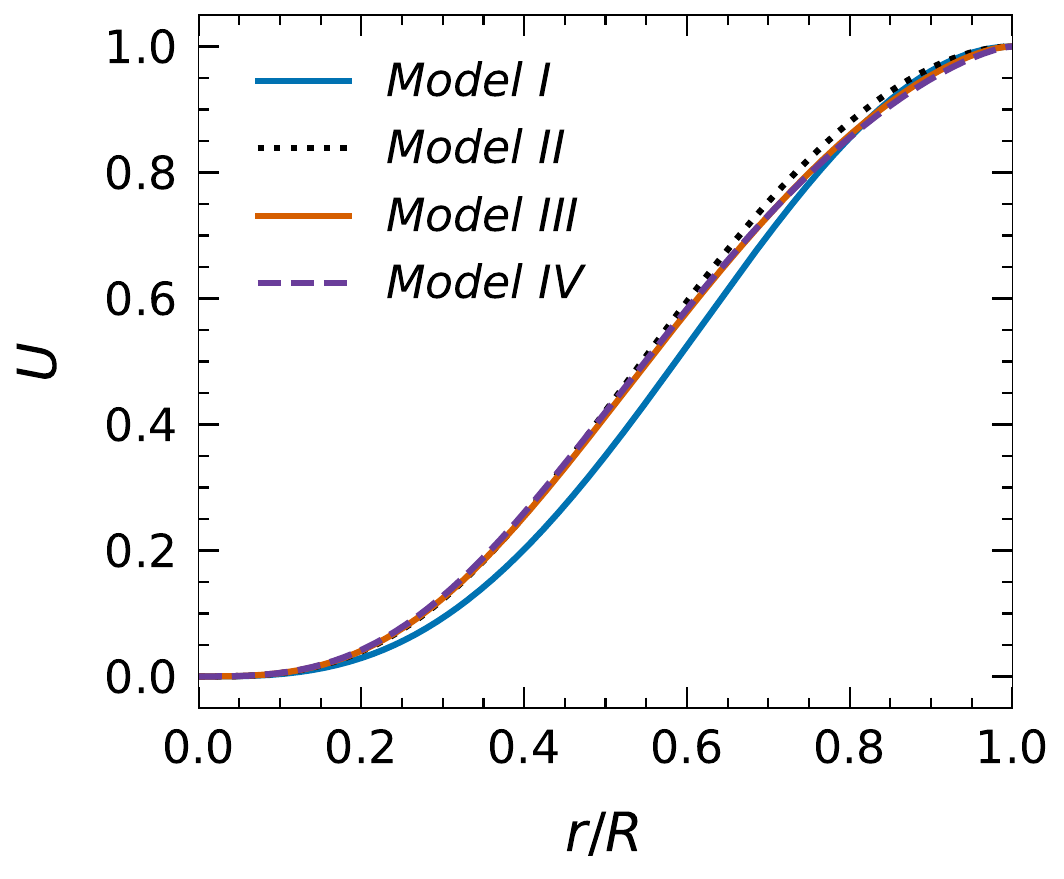}\hspace{0.3cm}
		\includegraphics[scale = 0.3]{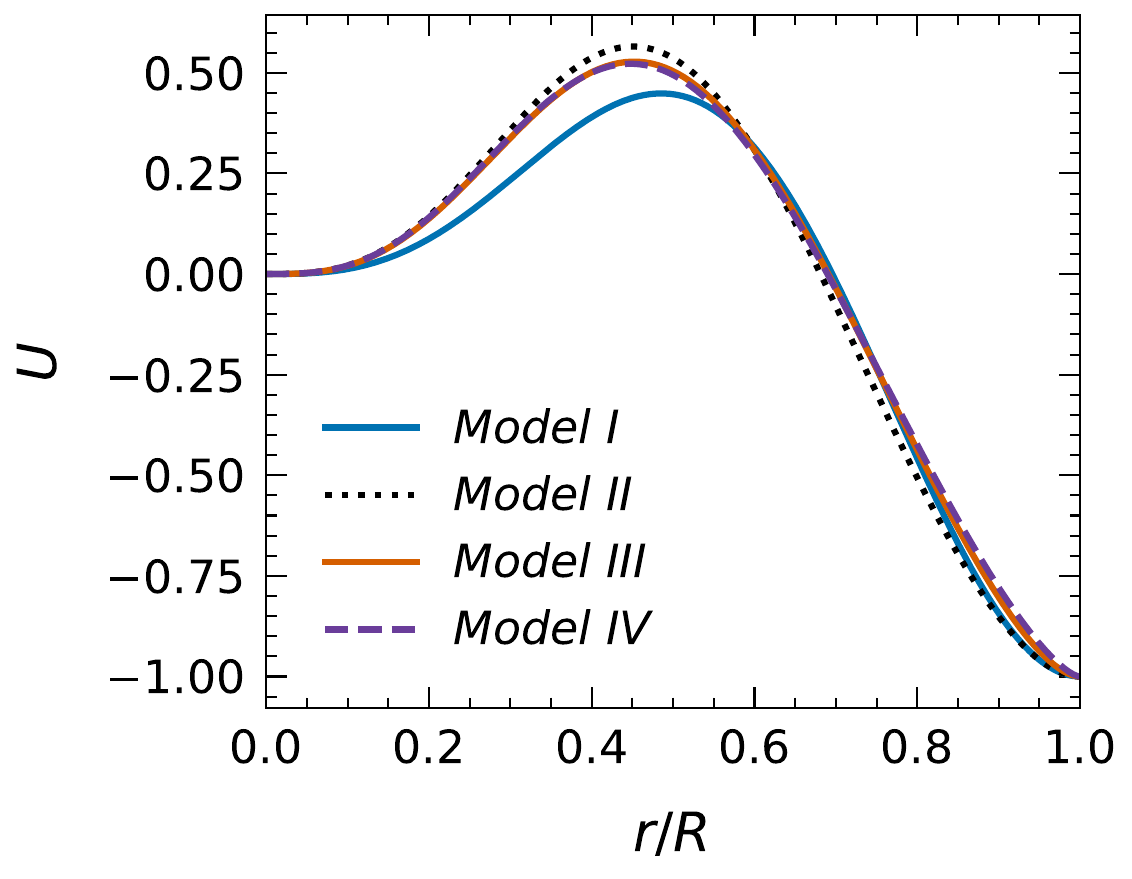}\hspace{0.3cm}
		\includegraphics[scale = 0.3]{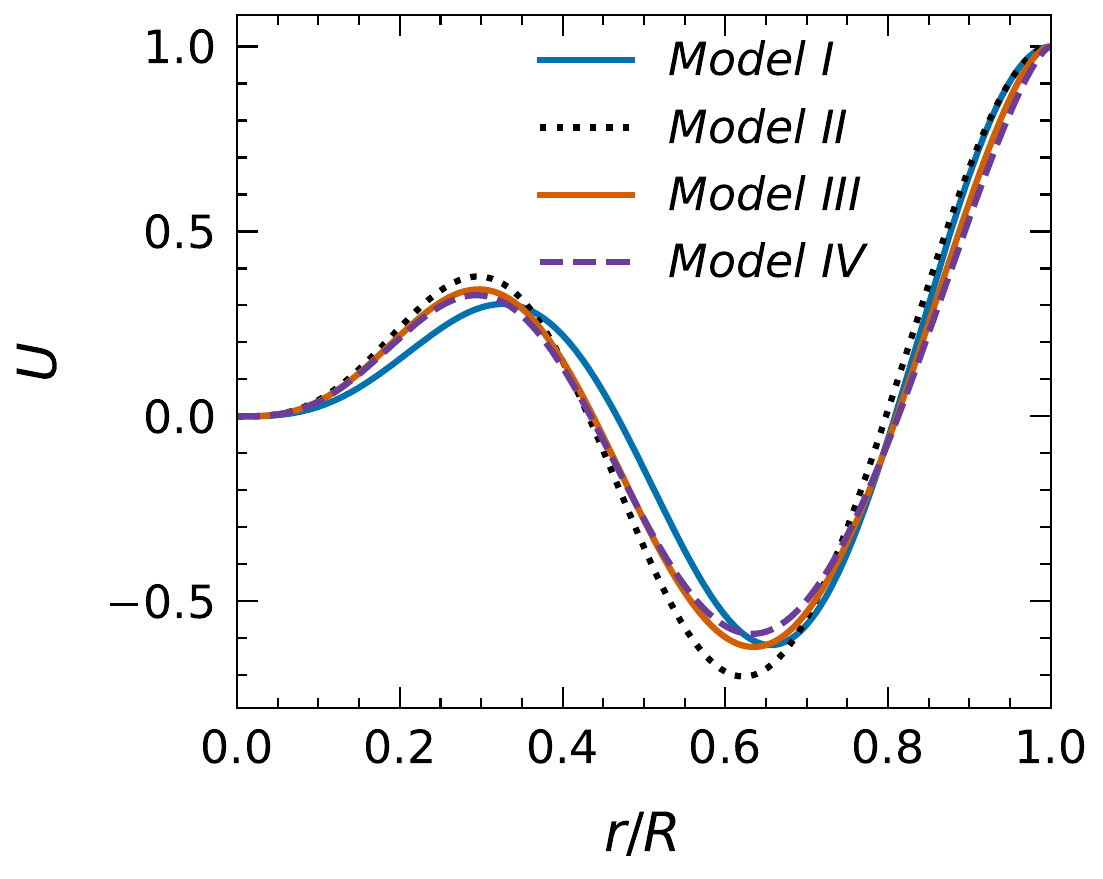}}\vspace{0.2cm}
	\centerline{
		\includegraphics[scale = 0.3]{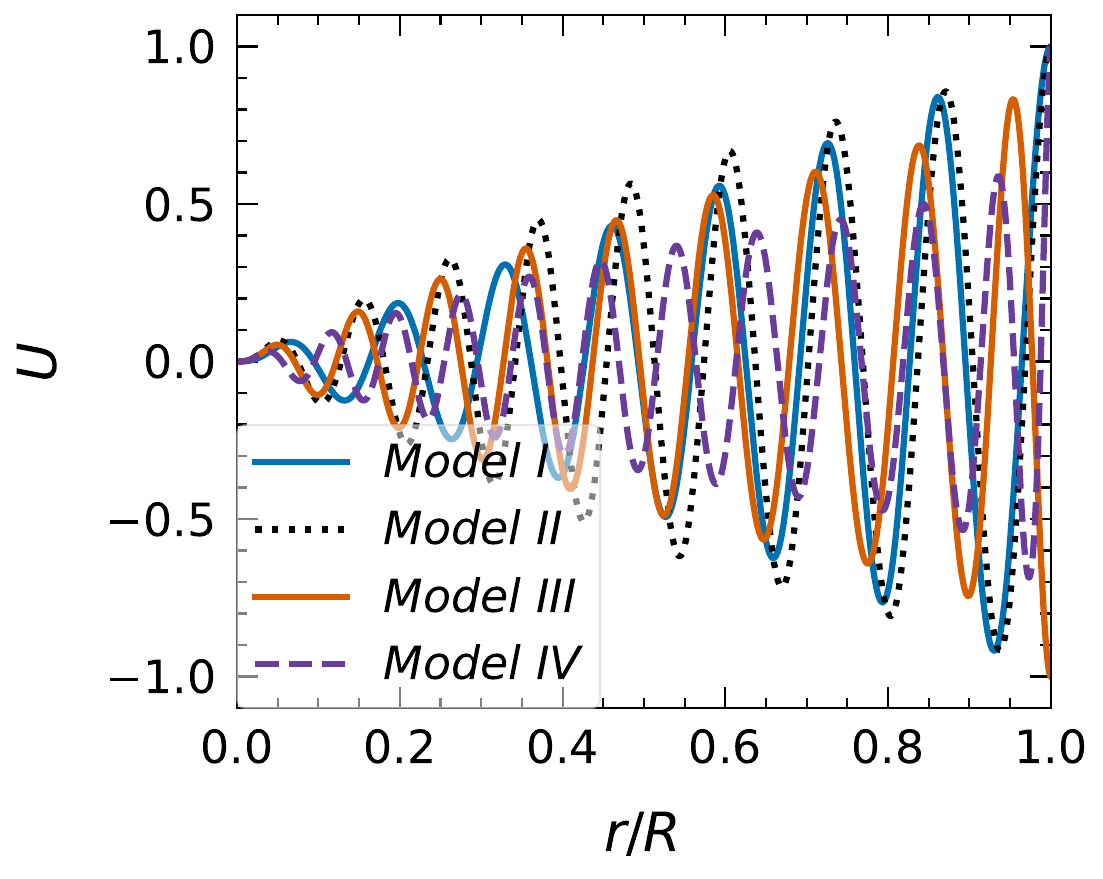}\hspace{0.3cm}
		\includegraphics[scale = 0.3]{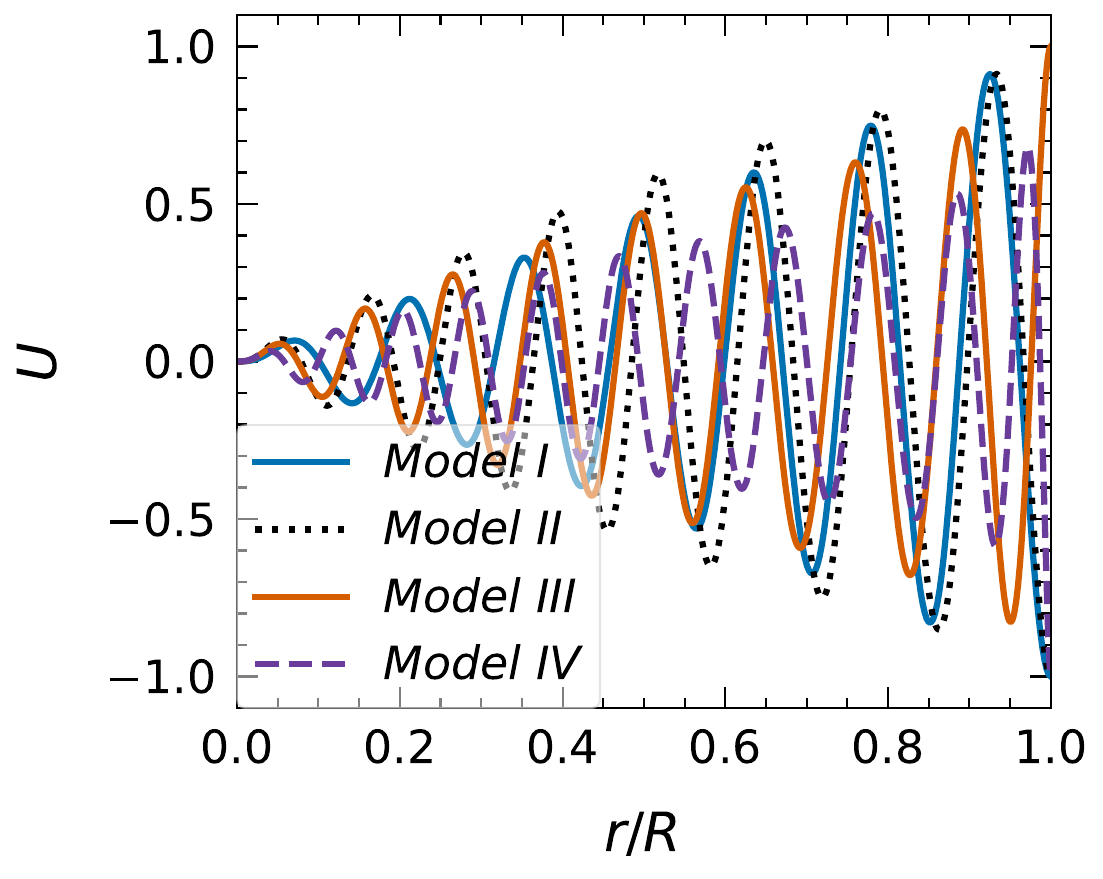}\hspace{0.3cm}
		\includegraphics[scale = 0.3]{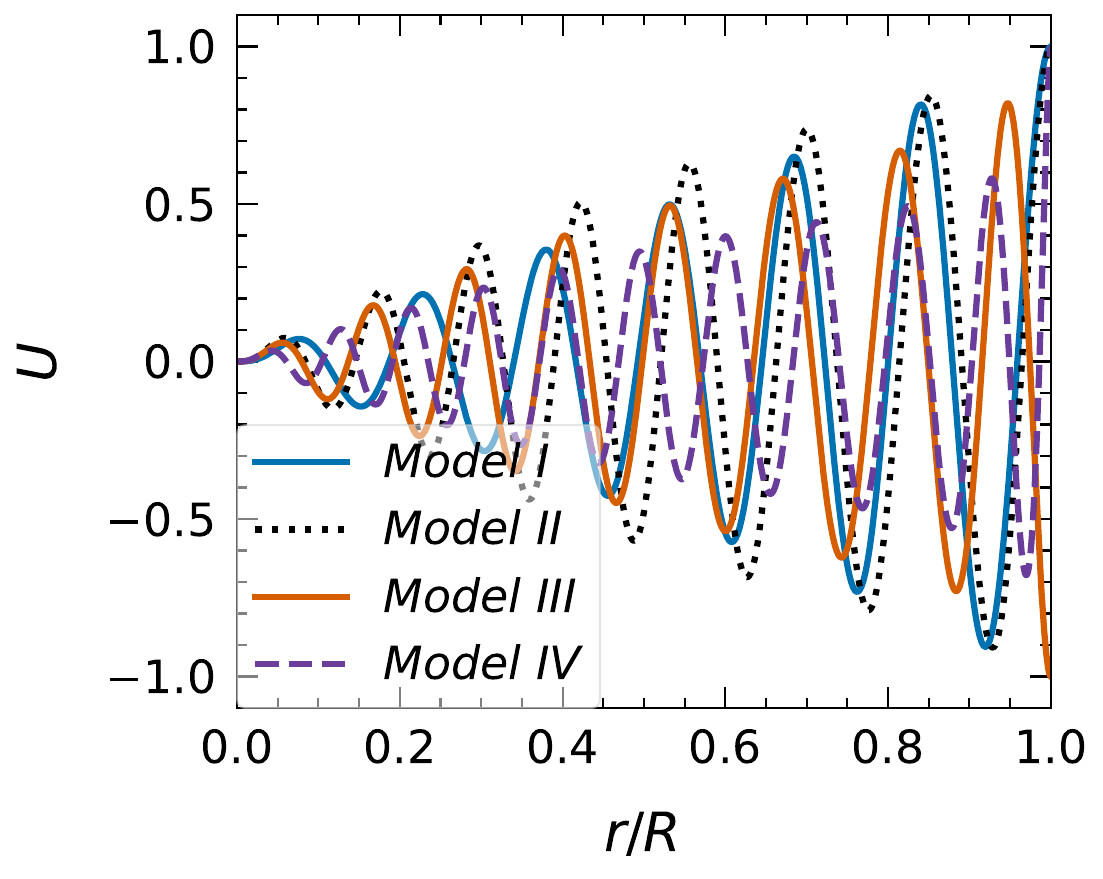}}
	\vspace{-0.3cm}
	\caption{Radial variations of the first three eigenfrequencies (first 
row from left to right, respectively) and the last three higher-order modes 
(second row from left to right, respectively) of fluid perturbation in radial 
direction given by $U(r)$. Here, we have normalized the $U(r)$ magnitude with 
respect to its absolute maximum value at each mode.}
	\label{fig:U_Plots_I}
\end{figure}

Further, the radial variations of $U(r)$ for the first three and last three 
modes of oscillations are shown in Fig.~\ref{fig:U_Plots_I}. Here, we plot the
normalized $U(r)$ values (denoted as $U$) with respect to their maximum value 
for each mode in every model. One can observe that for all the models, the 
radial variation of $U$ is almost the same for each mode. The magnitude of 
$U(r)$ starts at zero at the 
center and increases to its maximum value at the surface. The order of 
magnitude of $U(r)$ at the surface of stars for all four models is $10-10^{2}$ 
km$^{3}$ for the first three modes and $10^{-1}$ km$^{3}$ for Models I, II and 
III for the last three modes. For Model IV, the order of magnitude of $U(r)$ 
is $1$ km$^{3}$ for the last three modes. The higher overtones has lower value 
of $U(r)$ at the surface of the stars. The $f$-mode has the highest $U(r)$ 
values at the surface of the stars, which are $1.1711 \times 10^{2}$ km$^{3}$, 
$1.6040 \times 10^{2}$ km$^{3}$, $1.7703 \times 10^{2}$ km$^{3}$, and 
$3.3940 \times 10^{2}$ km$^{3}$ for Models I, II, III, and IV, respectively. 
While the magnitude of $U(r)$ at the surface of the stars for the last 
overtone for the respective models are $6.3832 \times 10^{-1}$ km$^{3}$, 
$5.4724 \times 10^{-1}$ km$^{3}$, $6.7189 \times 10^{-1}$ km$^{3}$, and 
$1.0976$ km$^{3}$.
%
\subsection{Rotational Corrections}
\begin{figure}[!h]
        \centerline{
           \includegraphics[scale = 0.3]{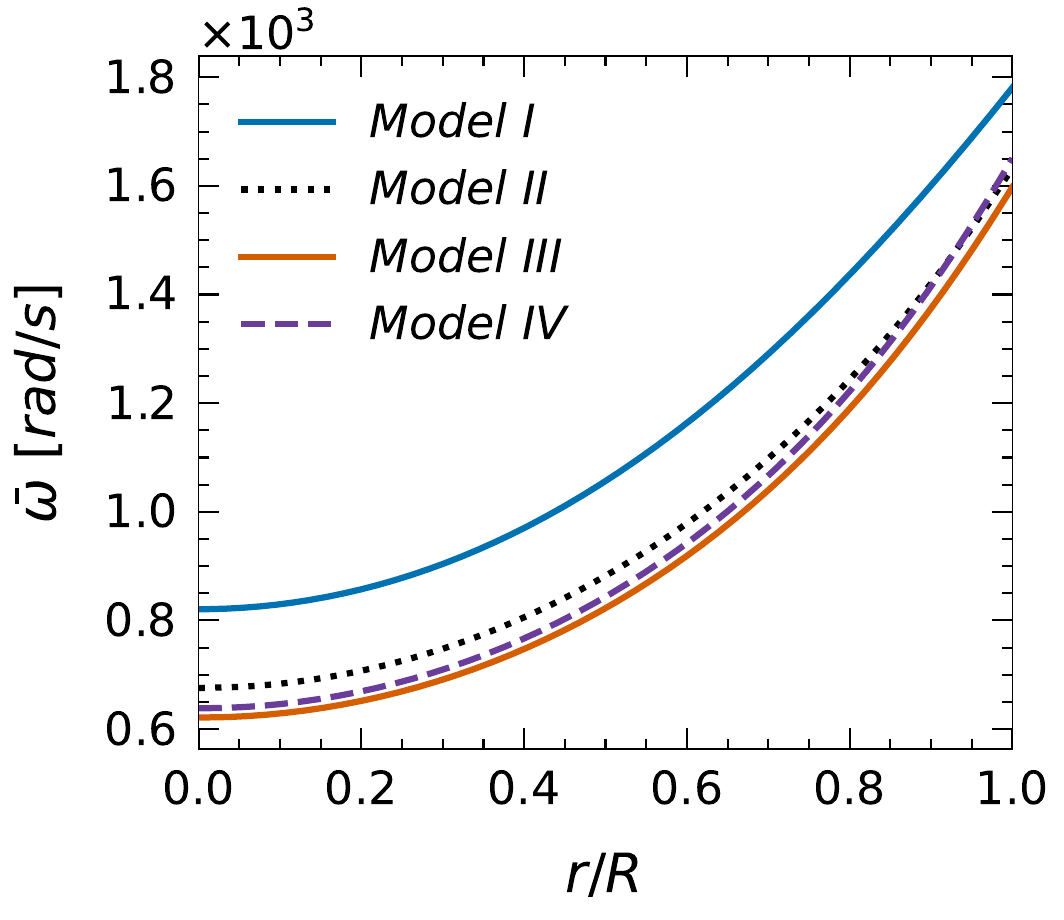}\hspace{0.3cm}
           \includegraphics[scale = 0.3]{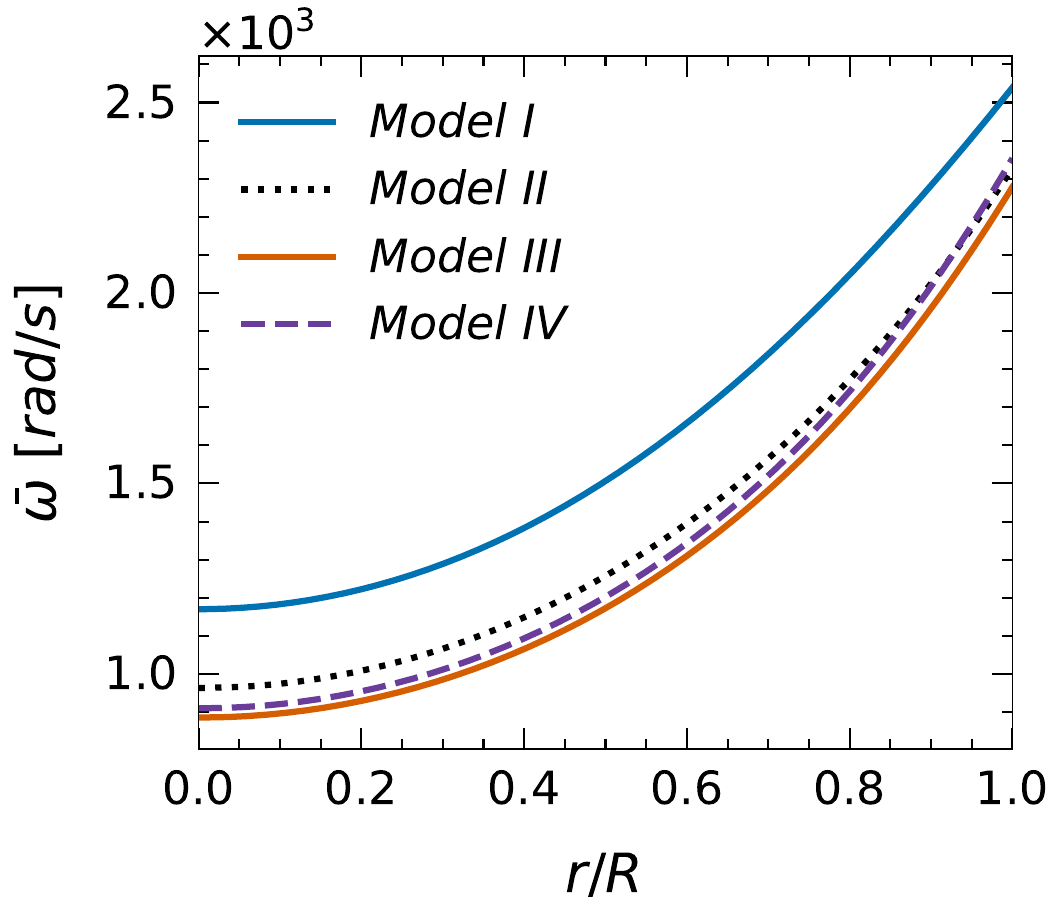}\hspace{0.3cm}
           \includegraphics[scale = 0.3]{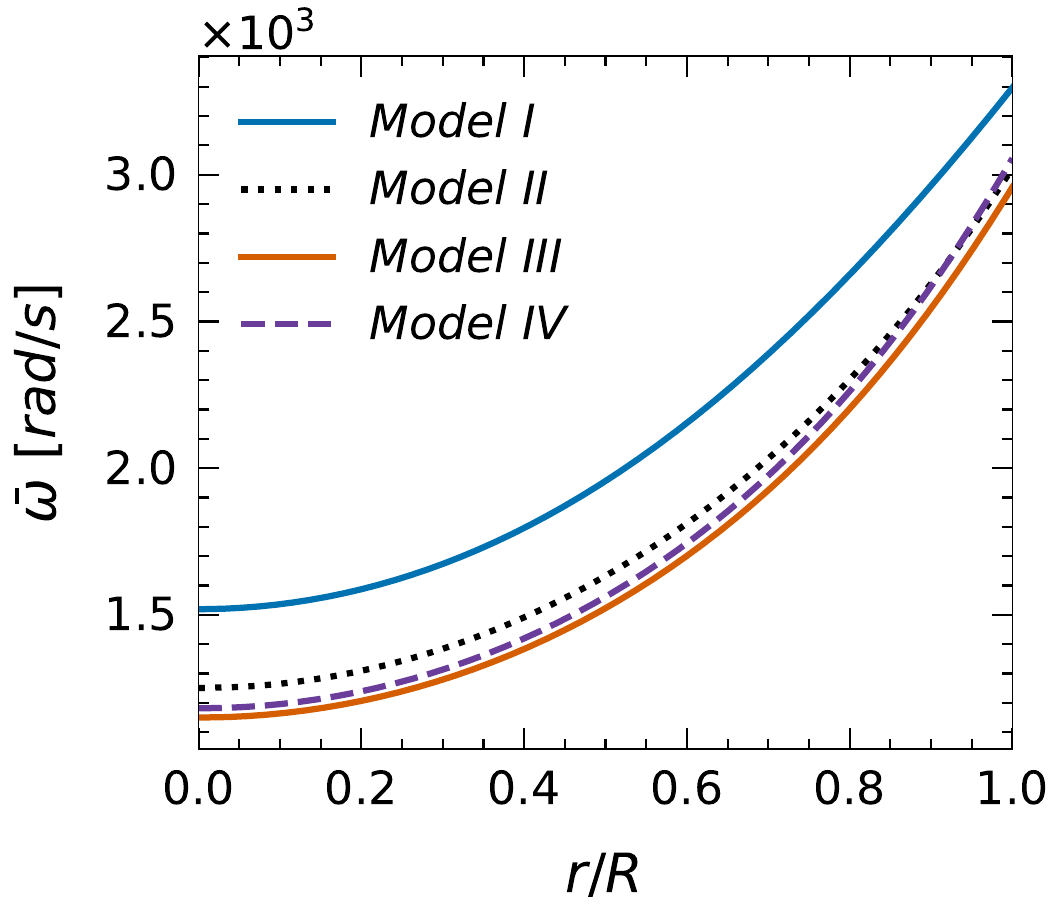}}
           \vspace{-0.3cm}
        \caption{Radial variation of frame-dragging parameter $\bar{\omega}$ 
of the slow-rotating stars with the angular frequencies of $383$ Hz (left 
plot), $546$ Hz (middle plot) and $709$ Hz (right plot) under different stellar
models.} 
\label{fig:frameDraggingForAllEOSForAllThreeFrequency}
\end{figure}

To study the oscillating behaviours of slowly rotating stars, one has to 
first implement the rotational corrections to different 
governing parameters of the stars due to the rotational perturbations of 
those parameters as discussed in Section \ref{secII}. In this context, 
we consider here three angular frequencies ($\Omega$) of stars, viz., 
$383$ Hz, $546$ Hz, and $709$ Hz based on the observational data 
\cite{sullivan_2024, romani_2026, smith_2023} of the pulsars, 
PSR J2215+5135, PSR J1311-3430, PSR J1810+1744 
and PSR J0952-0607, mentioned in the introduction section. As detailed
in Section \ref{secII}, the stellar configurations that satisfy the 
slow-rotation condition for relativistic stars can be described by the 
Hartle–Thorne formalism \cite{hartle_1967,hartle_1968}. From this formalism, we first 
calculate the frame-dragging angular frequency $\bar{\omega}$ and plot its 
radial variation from the center to the surface of the stars as shown in the 
Fig.~\ref{fig:frameDraggingForAllEOSForAllThreeFrequency}. It is obvious that 
with the change of angular frequency of the stars, there will be a change in  
the amount of frame-dragging. 
\begin{figure}[!h]
        \centerline{
           \includegraphics[scale = 0.3]{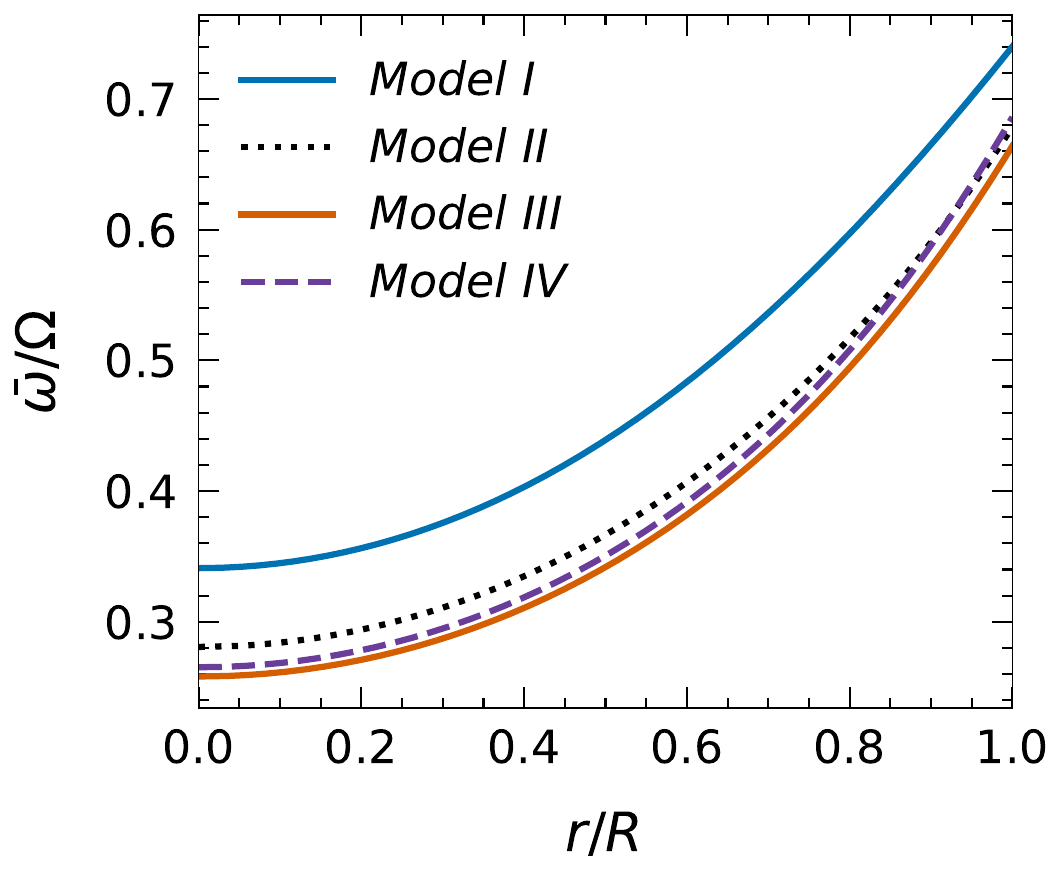}\hspace{0.3cm}
           \includegraphics[scale = 0.3]{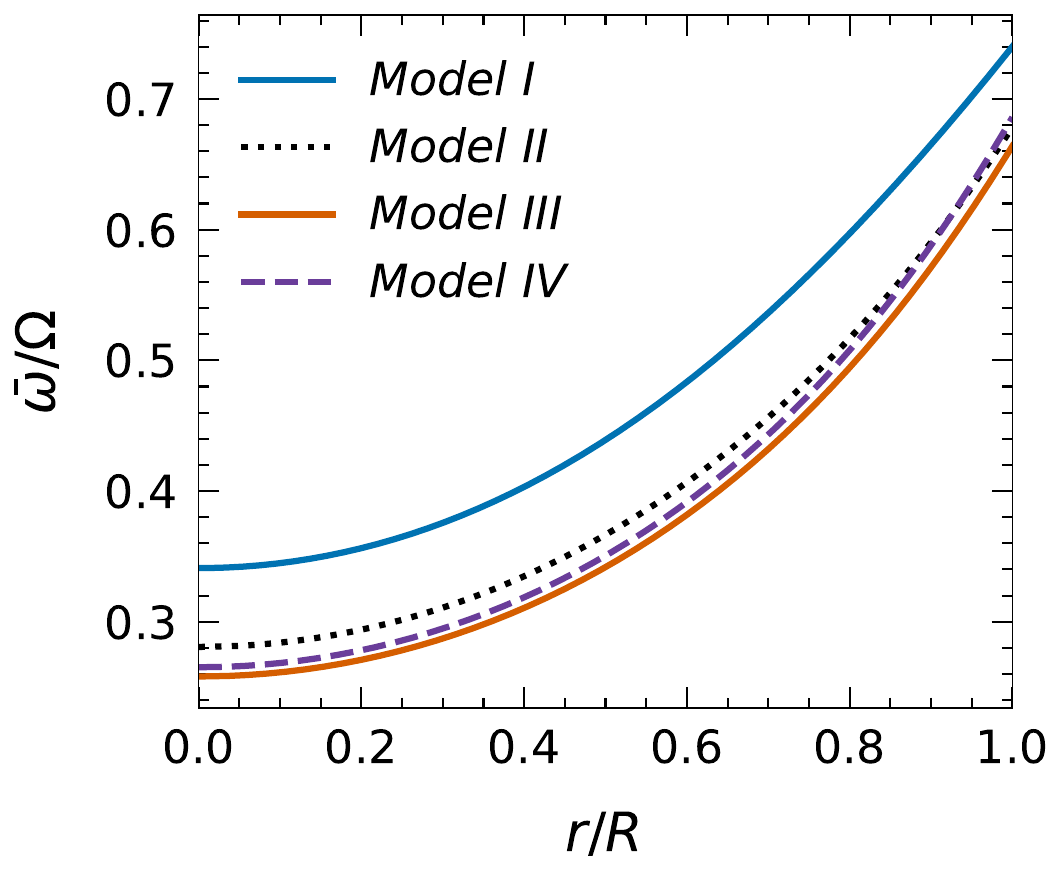}\hspace{0.3cm}
           \includegraphics[scale = 0.3]{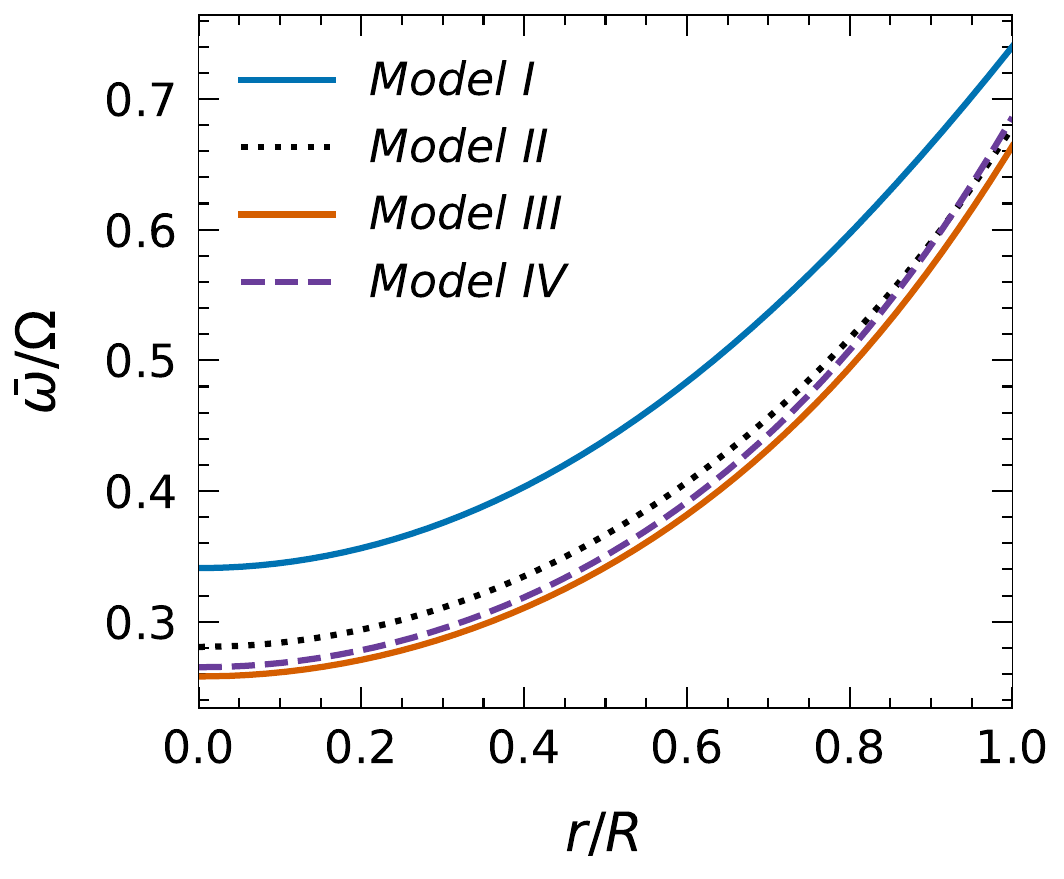}}
           \vspace{-0.3cm}
        \caption{Radial variation of the dimensionless frame dragging parameter
$\bar{\omega}/\Omega$ of the slow-rotating stars with the angular frequencies 
of $383$ Hz (left plot), $546$ Hz (middle plot) and $709$ Hz (right plot) 
under different stellar models.}
        \label{fig:dimensionlessBarOmegaVsr}
\end{figure}
\begin{figure}[!h]
        \centerline{
            \includegraphics[scale = 0.3]{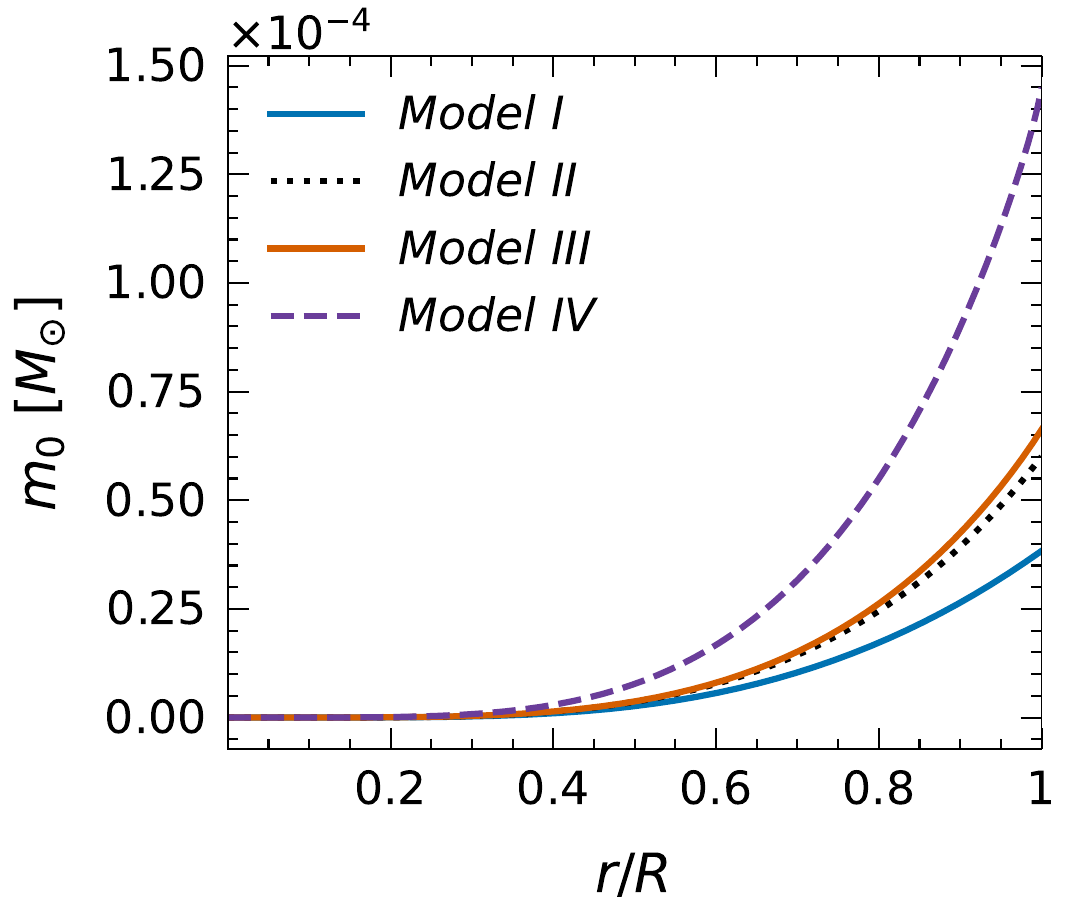}\hspace{0.3cm}
            \includegraphics[scale = 0.3]{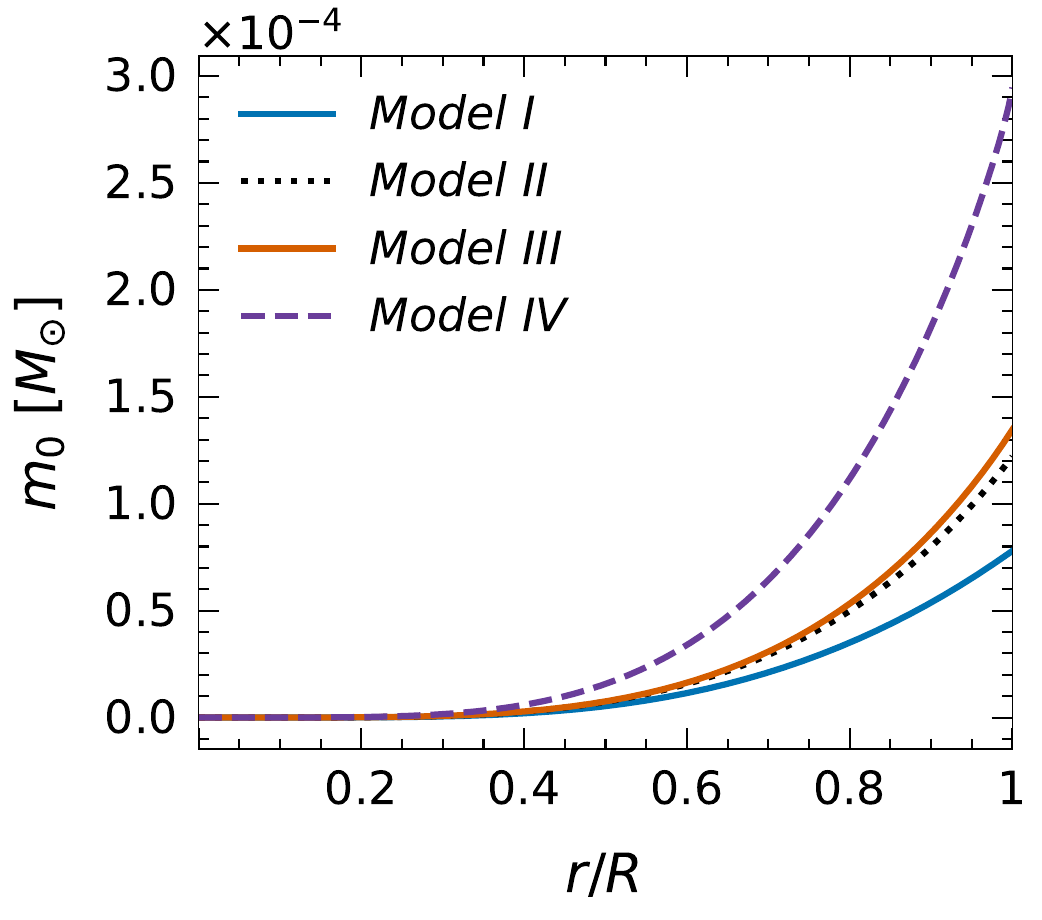}\hspace{0.3cm}
            \includegraphics[scale = 0.3]{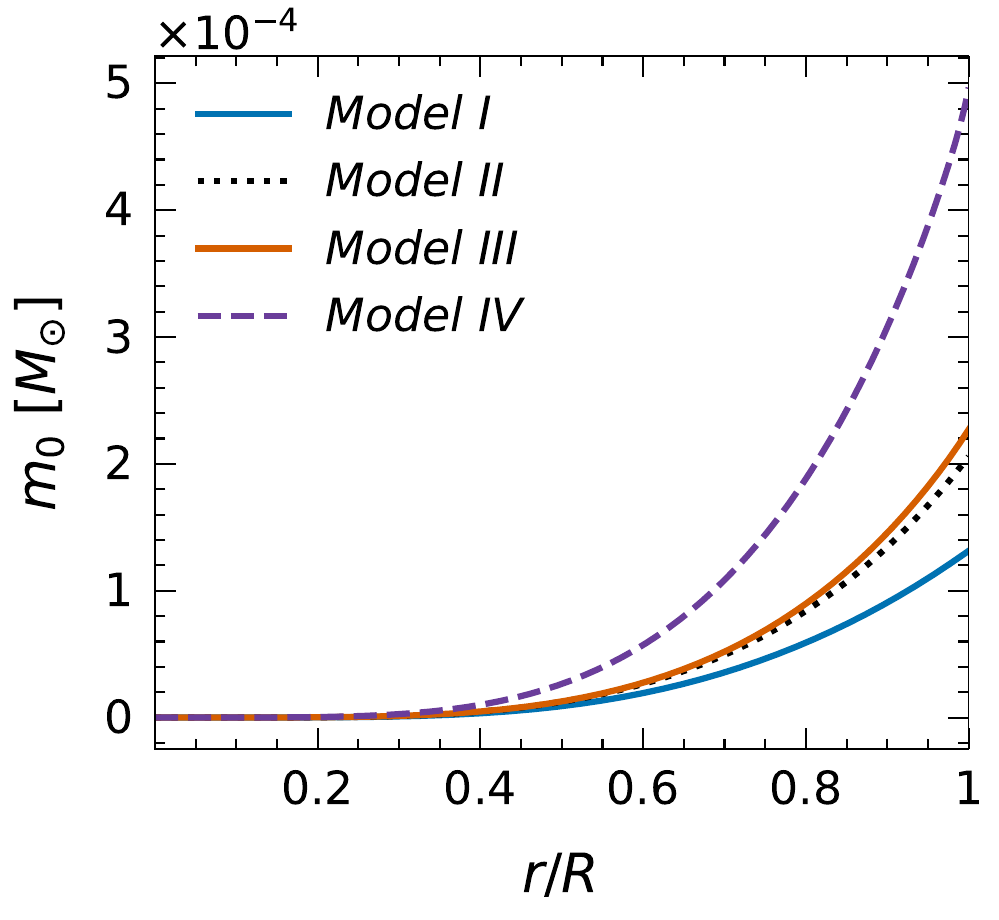}}
            \vspace{-0.3cm}
        \caption{Radial variation of the mass correction term $m_{0}$ of the
slow-rotating stars with the angular frequencies of $383$ Hz (left plot),
$546$ Hz (middle plot) and $709$ Hz (right plot) under different stellar
models.}
\label{fig:massCorrectionTerm_m0_variation}
\end{figure}
From Fig.~\ref{fig:frameDraggingForAllEOSForAllThreeFrequency} we observe that 
the background local inertial frame is twisting slowly from the center to the
surface of the stars. Also higher the compactness of the stars, the higher 
the twisting of the local inertial frame. To demonstrate these we plot the 
dimensionless quantity $\bar{\omega}/\Omega$ from the center to the surface 
of the stars as shown in Fig.~\ref{fig:dimensionlessBarOmegaVsr}. It 
shows that the ratio $\omega/\Omega$ must be the smallest for Model I than the 
other higher compact stellar models. That is, the angular velocity of the local 
inertial frame $\omega$ is the smallest for Model I for a given stellar 
angular velocity $\Omega$. For all three angular frequencies of the stars, 
for Model I, $\bar{\omega}/\Omega$ starts at $0.341$ near the center and 
ends at $0.7401$ at the surface. Similarly, for Model II, it starts at 
$0.2807$ and ends at $0.6785$, for Model III, from $0.2581$ to $0.664$, and 
for Model IV, from $0.2651$ to $0.686$. Thus, we have 
observed that the amount of twist produced at par angular frequency in the 
local inertial frame for stars with different angular frequencies is the same. 
Moreover, with the increase in compactness of the stars, $\bar{\omega}/\Omega$ 
decreases, which means, the local inertial frame is being twisted more for 
stars with higher compactness in comparison to the stars with lower 
compactness of the same angular frequency. Further, the effect of the 
temperature of QM is very minimal in the frame-dragging process.

\begin{table}[!h]
        \begin{center}
         \caption{Rotation corrected mass, and other parameters for
rotational deformation at the equatorial plane of the slow-rotating stars with
the angular frequencies of $383$ Hz, $546$ Hz and $709$ Hz under different
stellar models.}\vspace{3pt}
        \setlength{\tabcolsep}{8pt}
        \begin{tabular}{cccccccccc}
                \hline \hline\\[-12pt]
                \rule[24pt]{0pt}{0pt}
                \shortstack{Models \\[8pt]}
                & \shortstack{$\Omega$ \\[2pt] (Hz)}
                & \shortstack{$M+\delta M$ \\[2pt] ($M_{\odot}$)}
                & \shortstack{$J$ \\ ($\times 10^{40}$g cm$^{2}$s$^{-1}$)}
                & \shortstack{$R+\delta R$ \\ (km)}
                & \shortstack{$\bar{R}$ \\[2pt] (km)}
                & \shortstack{$e_{s}$\\[8pt]}
                & \shortstack{$Q$ \\[2pt] $(M_{\odot}^{3})$}
                & \shortstack{$\bar{Q}$\\[8pt]}
                & \shortstack{$\bar{I}$\\[8pt]}\\[3pt]
                \hline\hline
                \rule[10pt]{0pt}{0pt}
& 383 & 1.0640 & 1.3108 & 5.8054 & 5.8051 & 0.0169 & 5.1914e-04 & 2.4913 & 6.5528 \\[2pt]
                I & 546 & 1.0641 & 1.8687 & 5.8057 & 5.8052 & 0.0241 & 1.0552e-03 & 2.4914 & 6.5520 \\[2pt]
                & 709 & 1.0641 & 2.4267 & 5.8062 & 5.8053 & 0.0313 & 1.7793e-03 & 2.4915 & 6.5510 \\[2pt]

                \hline
                \rule[10pt]{0pt}{0pt}

                & 383 & 1.4002 & 2.6010 & 6.7955 & 6.7951 & 0.0185 & 1.174e-3 & 1.8839 & 5.7041 \\[2pt]
                II & 546 & 1.4003 & 3.7076 & 6.7960 & 6.7952 & 0.0264 & 2.3867e-03 & 1.8840 & 5.7033\\[2pt]
                & 709 & 1.4004 & 4.8143 & 6.7966 & 6.7953 & 0.0343 & 4.0240e-03 & 1.8841 & 5.7021\\[2pt]

                \hline
                \rule[10pt]{0pt}{0pt}

                & 383 & 1.4758 & 2.9209 & 6.9605 & 6.9601 & 0.0185 & 1.3052e-3 & 1.7496 & 5.4714 \\[2pt]
                III & 546 & 1.4759 & 4.1638 & 6.9610 & 6.9601 & 0.0264 & 2.6524e-03 & 1.7497 & 5.4705 \\[2pt]
                & 709 & 1.4760 & 5.4067 & 6.9616 & 6.9603 & 0.0343 & 4.4722-03 & 1.7498 & 5.4693 \\[2pt]

                \hline
                \rule[10pt]{0pt}{0pt}

                & 383 & 1.8612 & 6.0487  & 8.7130 & 8.7121 & 0.0240 & 4.3644e-03 & 1.7205 & 5.4278 \\[2pt]
                IV & 546 & 1.8614 & 8.6228  & 8.7140 & 8.7123 & 0.0343 & 8.8694e-03 & 1.7206 & 5.4263 \\[2pt]
                & 709 & 1.8616 & 11.197 & 8.7154 & 8.7125 & 0.0445 & 1.4955e-02 & 1.7208 & 5.4242 \\[2pt]

                \hline\hline
        \end{tabular}
        \label{table:rotCorrectionMassAndAngMom}
        \end{center}
\end{table}
\begin{figure}[!h]
	\centerline{
		\includegraphics[scale = 0.3]{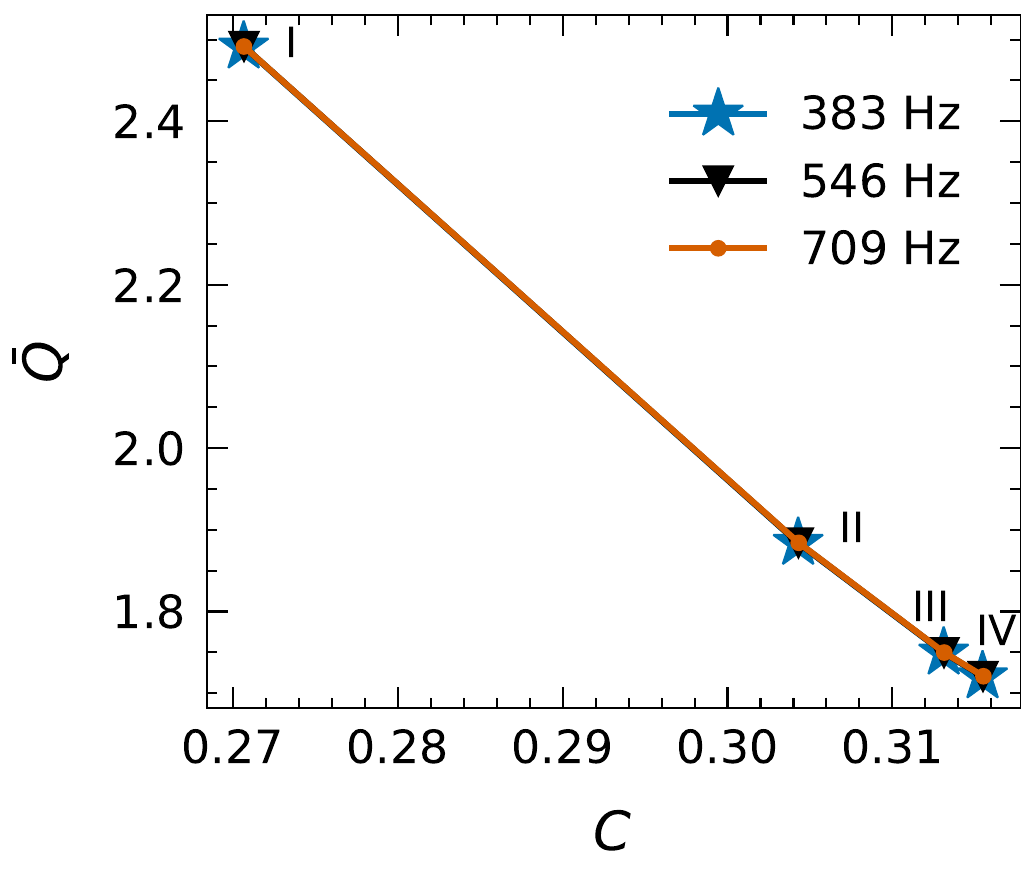}
		\hspace{0.3cm}
		\includegraphics[scale = 0.3]{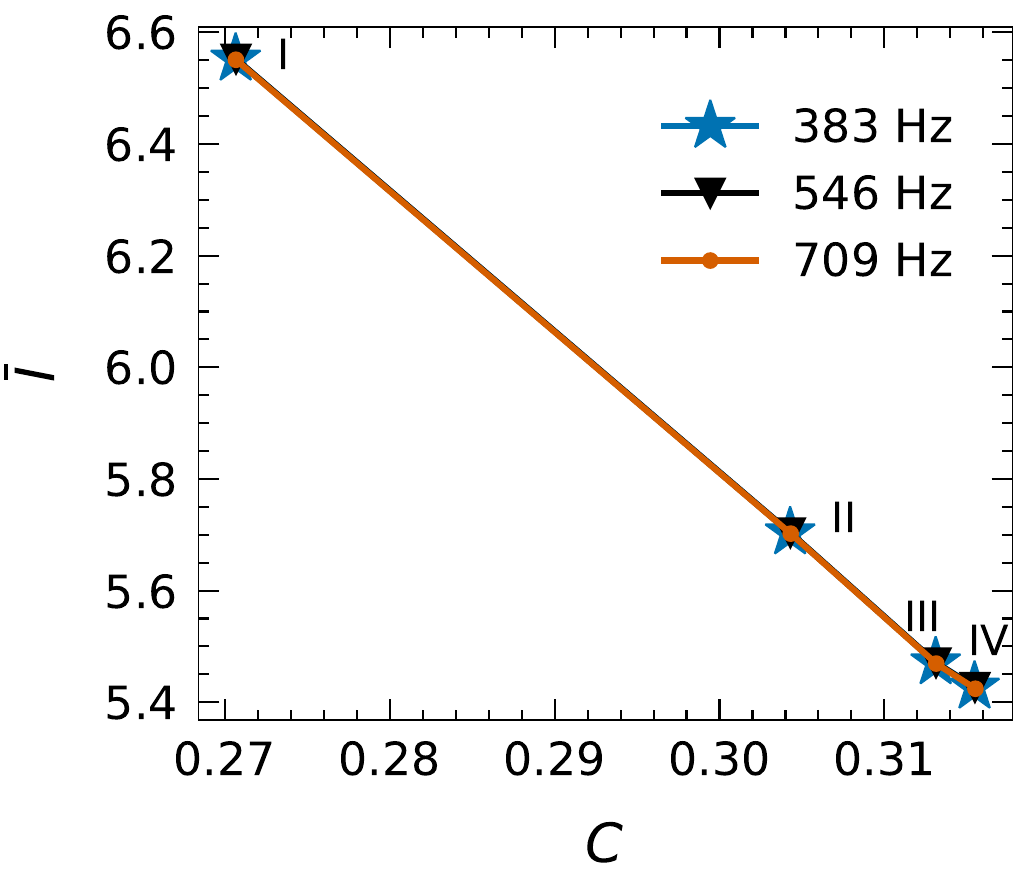}}
		\vspace{-0.3cm}
		\caption{Variation of dimensionless quantity $\bar{Q}$ (left) 
and $\bar{I}$ (right) against the compactness of the stars for different 
models at different angular frequencies. Here, the line joining two points has 
neither theoretical nor observational significance and is only used to join two 
theoretically predicted points.}
		\label{fig:dimenslessQandI}
\end{figure}
As we have discussed in Section \ref{secII}, for the $l = 0$ case of 
the rotation correction, the mass correction term is $m_{0}$. We plot the 
radial distribution of this $m_{0}$ term from the center to the surface of 
the stars in Fig.~\ref{fig:massCorrectionTerm_m0_variation} for all three 
considered angular frequencies and for different models of stars. The 
corresponding corrected masses ($M+\delta M$) of the stars are given in 
Table~\ref{table:rotCorrectionMassAndAngMom} with their surface angular 
momentum values ($J$), corrected radii ($R+\delta R$), mean corrected radii 
($\bar{R}$), eccentricities ($e_s$), quadrupole moments ($Q$), dimensionless
quadrupole moments ($\bar{Q}$), and dimensionless moment of inertias 
($\bar{I}$). From Fig.~\ref{fig:massCorrectionTerm_m0_variation}, it is
observed that $m_{0}$ is maximum at the surface and almost zero near the 
center of the stars. For the Model IV, $m_{0}$ term is maximum, whereas for 
the Model I it is minimum, and the Model II and  Model III have comparable 
values throuthout the stars from center to the surface. Thus, the 
temperature of quark matter has a dominating effect on the $m_{0}$ term.  
The variation of masses of stars due to the rotation from the 
non-rotating stars' masses is of the order of $\sim 10^{-5}\, M_{\odot}$ to 
$\sim 10^{-4}\, M_{\odot}$, for Models II, III and IV. Where the variation of 
masses increases with the increase in angular frequencies of the stars.
However, for the Model I, the variation of masses of the stars is of the order 
of $\sim 10^{-5}\, M_{\odot}$ for $546$ Hz and $709$ Hz, while there is no
variation for $383$ Hz frequency. As we have mentioned 
at the end of Section~\ref{secIIb}, the universal relationship between 
moment of inertia, Love number and quadrupole moment is independent of the 
internal structure of NSs and QSs \cite{yagi_2013}. We plot 
dimensionless quantities $\bar{Q}$ and $\bar{I}$ in 
Fig.~\ref{fig:dimenslessQandI} against the compactness of the stars using the 
rotation corrected mass and radius values for all models at different angular 
frequencies. We observe that $\bar{Q}$ and $\bar{I}$ both decrease with 
increasing compactness. Which suggests that higher compact stars are less 
susceptible to rotational deformation and their gravity is stronger. It is also 
observed that the plots for different angular frequencies overlapped almost 
completely. The stars defined by each model at different angular frequencies 
are only separated by their compactness, while there is negligible dependence 
of angular frequencies in the Hartle slow rotation approximation 
\cite{hartle_1967,hartle_1968}. These trends in Fig.~\ref{fig:dimenslessQandI} 
shows that the stars are consistent with the expected behavior of slowly 
rotating compact stars and suggests that the adopted EoSs satisfy the 
theoretical framework underlying the I--Q universality.  
\begin{figure}[!h]
	\centerline{
		\includegraphics[scale = 0.3]{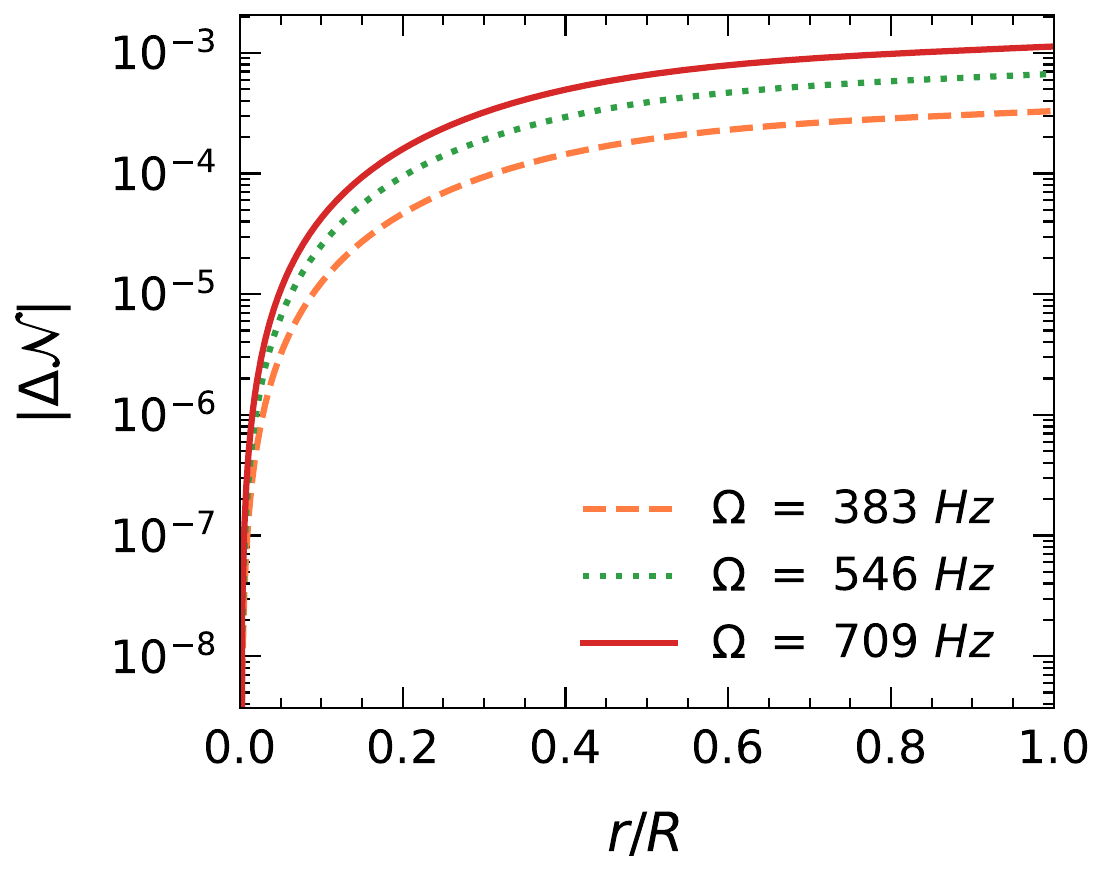}\hspace{0.5cm}
		\includegraphics[scale = 0.3]{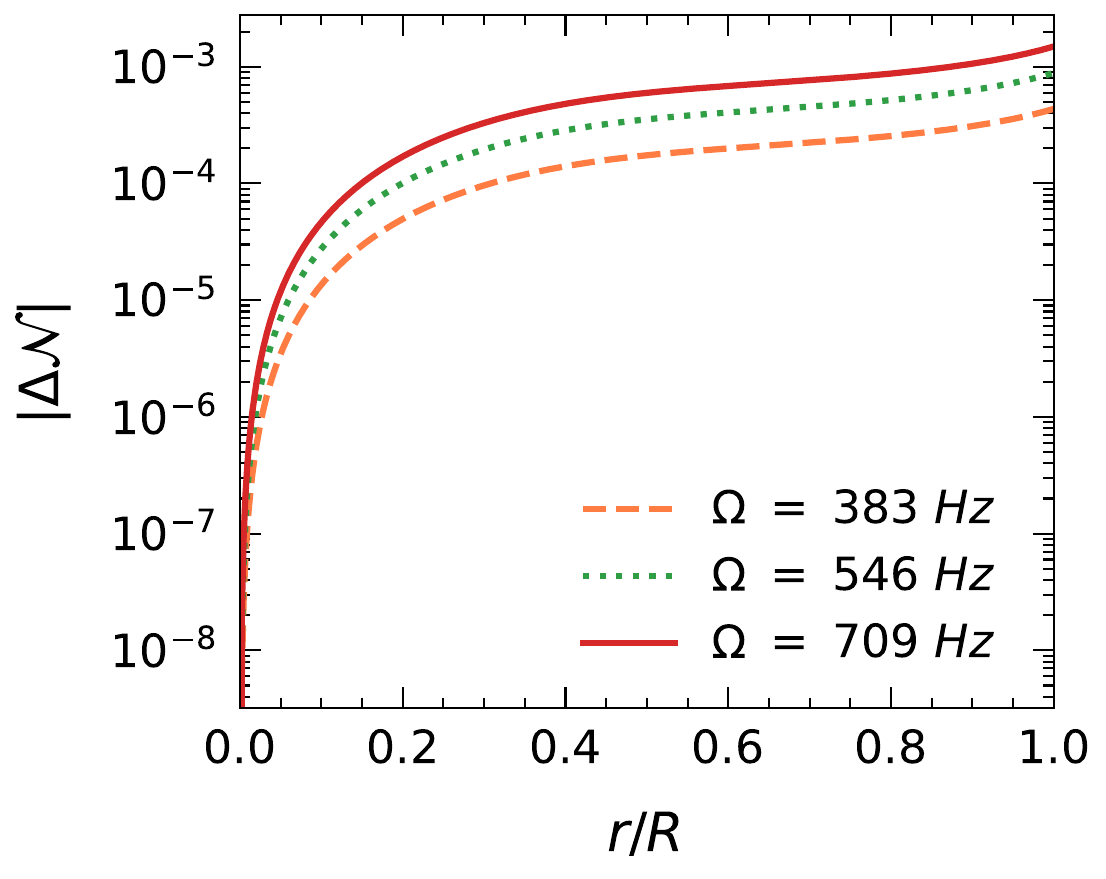}}
	\vspace{0.3cm}
	\centerline{
		\includegraphics[scale = 0.3]{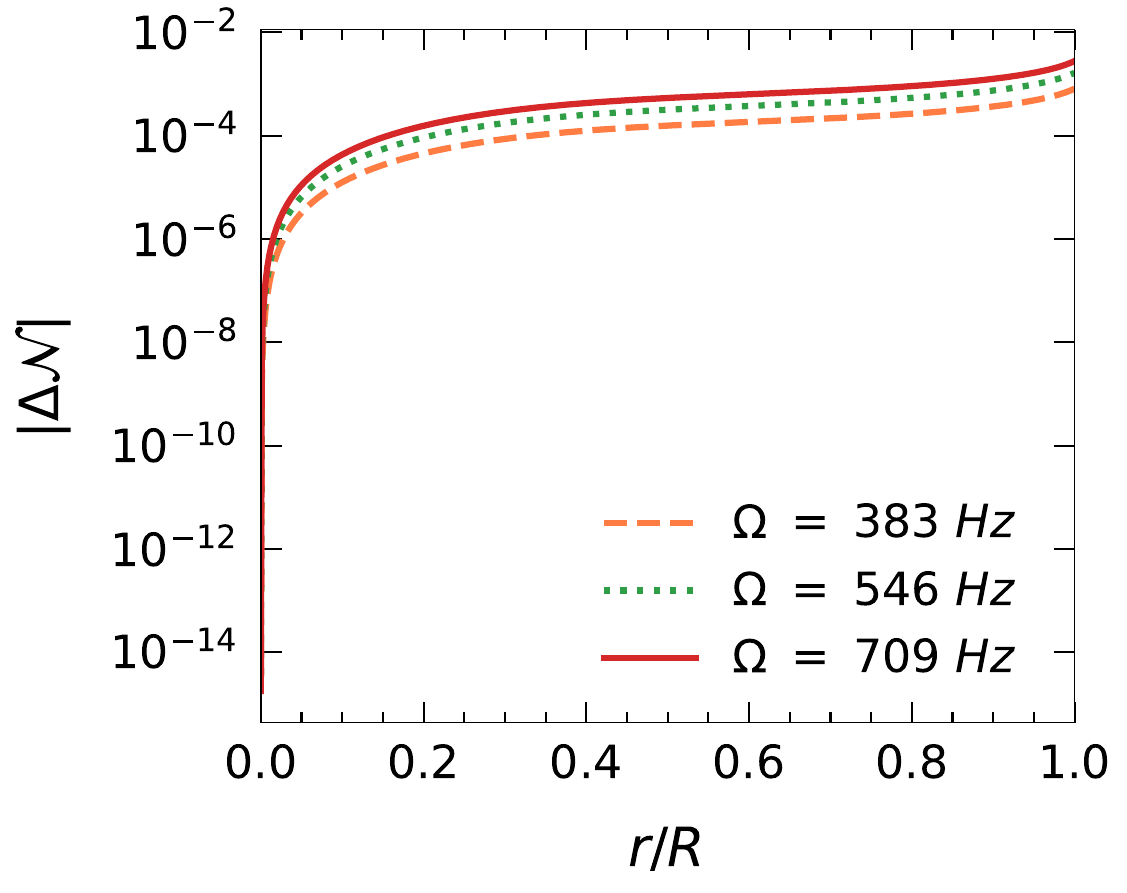}\hspace{0.5cm}
		\includegraphics[scale = 0.3]{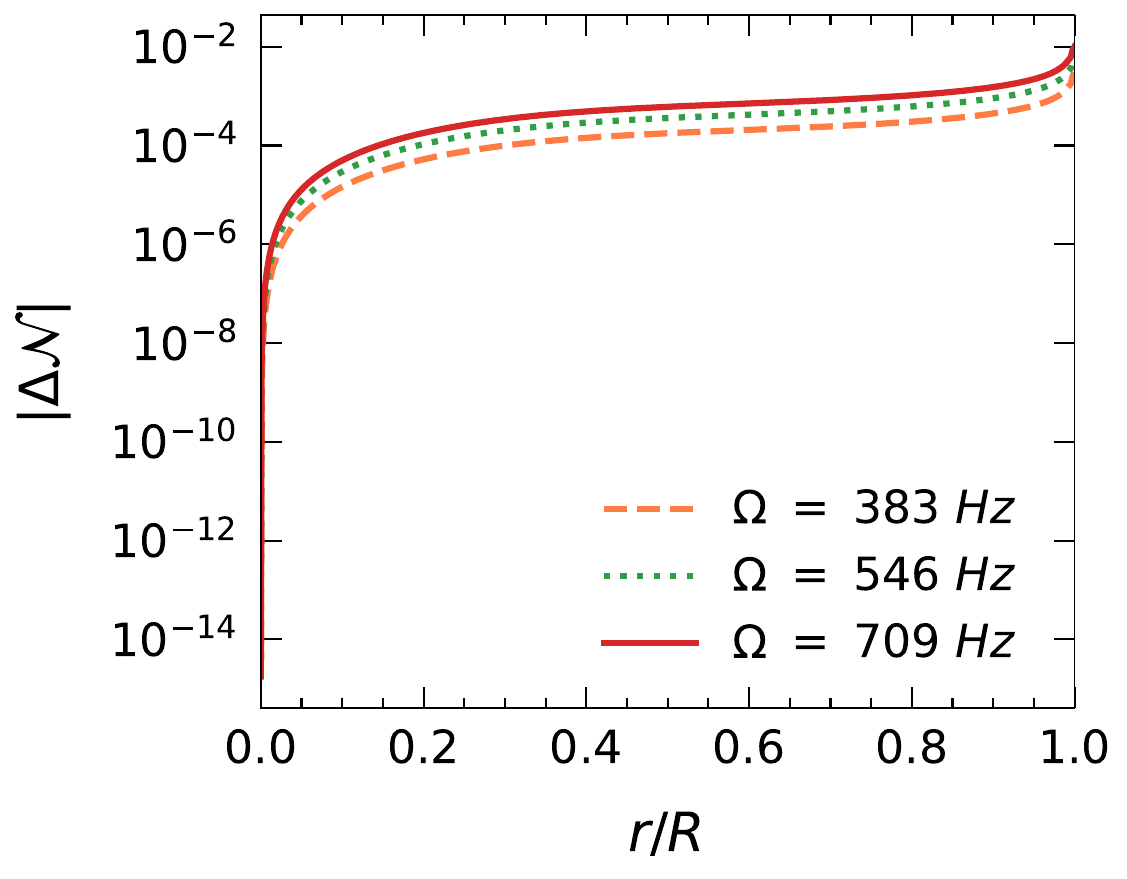}}
	\vspace{-0.3cm}
	\caption{Radial variation of the rotation corrected part of the number
		density of slow-rotating stars with the angular frequencies of 
$383$ Hz, $546$ Hz and $709$ Hz for the Model I (top left), Model II 
(top right), Model III (bottom left) and Model IV (bottom right).}
	\label{fig:comparison_num_den_with_non_rot_part}
\end{figure}
\begin{figure}[!h]
	\centerline{
		\includegraphics[scale = 0.3]{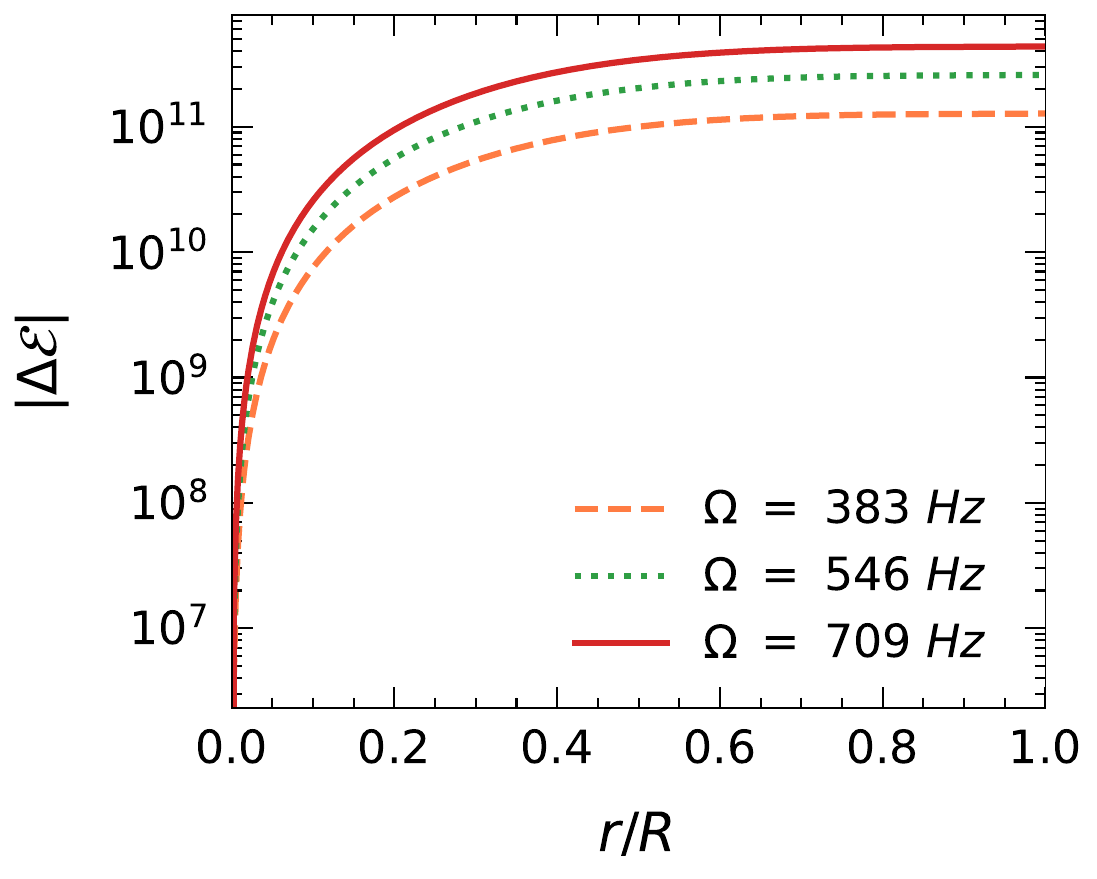}\hspace{0.5cm}
		\includegraphics[scale = 0.3]{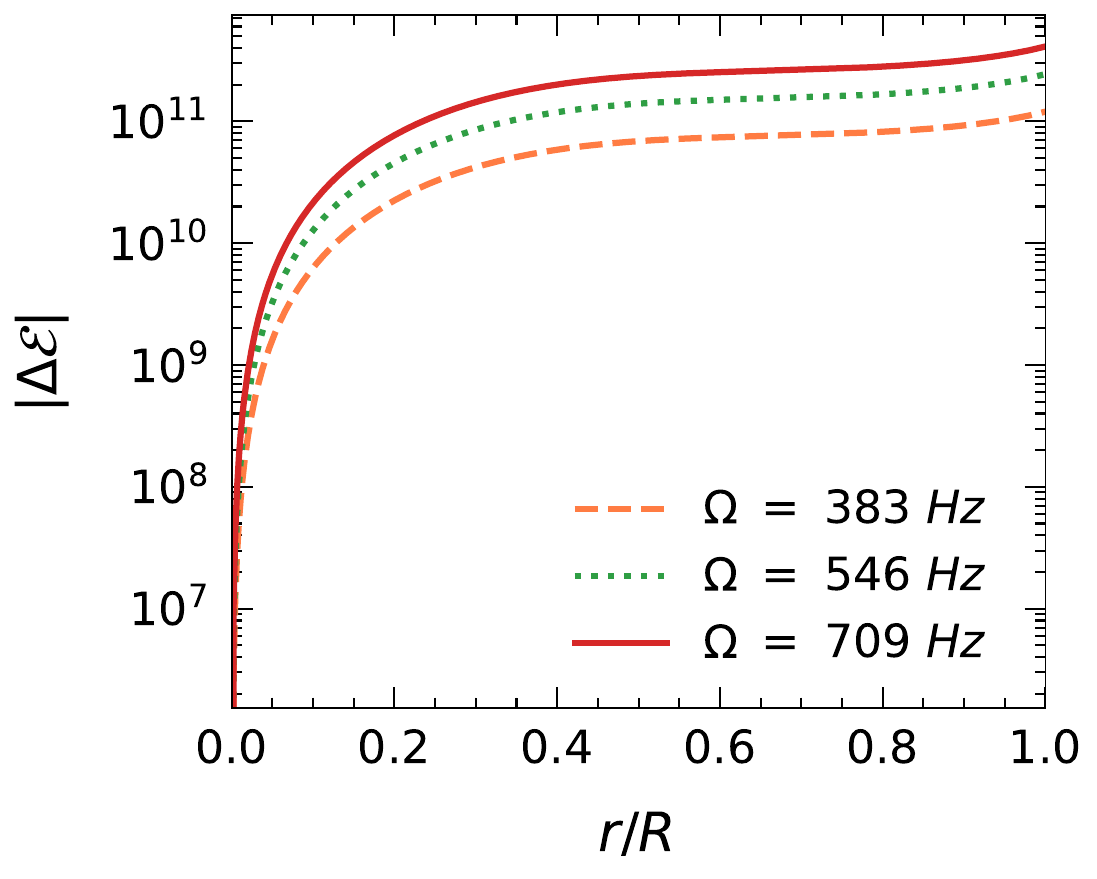}}
	\hspace{0.3cm}
	\centerline{
		\includegraphics[scale = 0.3]{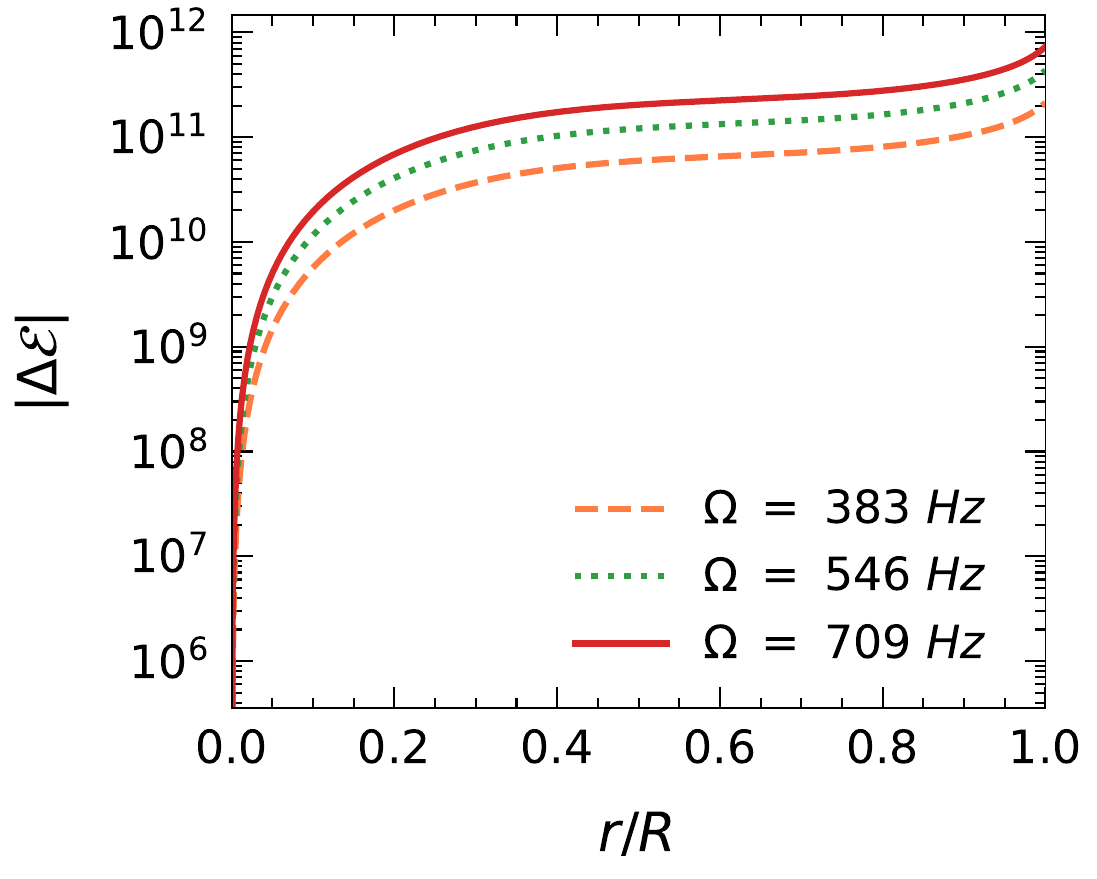}\hspace{0.5cm}
		\includegraphics[scale = 0.3]{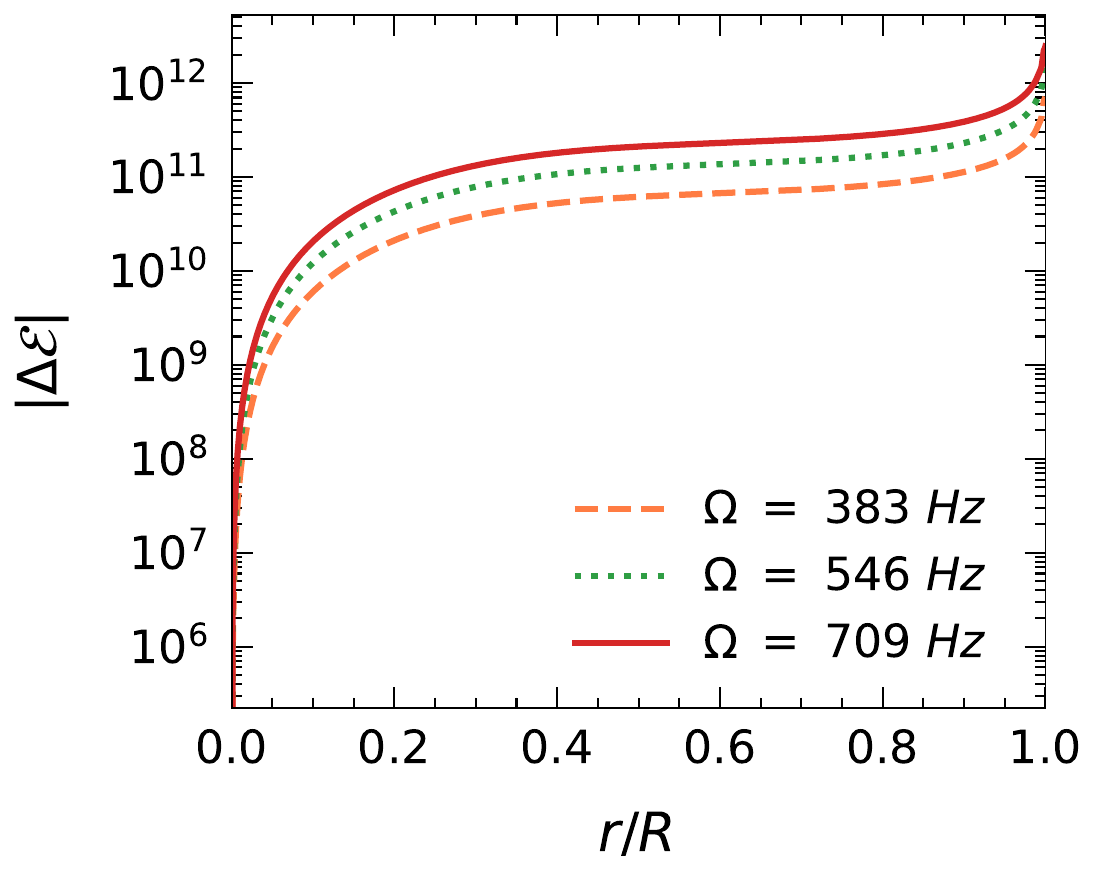}}
	\vspace{-0.3cm}
	\caption{Radial variation of the rotation corrected energy density 
part of slow-rotating stars with the angular frequencies of $383$ Hz,
$546$ Hz and $709$ Hz for the Model I (top left), Model II (top right),
Model III (bottom left) and Model IV (bottom right).}
	\label{fig:comparison_ene_den_with_non_rot_part}
\end{figure}
\begin{figure}[!h]
	\centerline{
		\includegraphics[scale = 0.3]{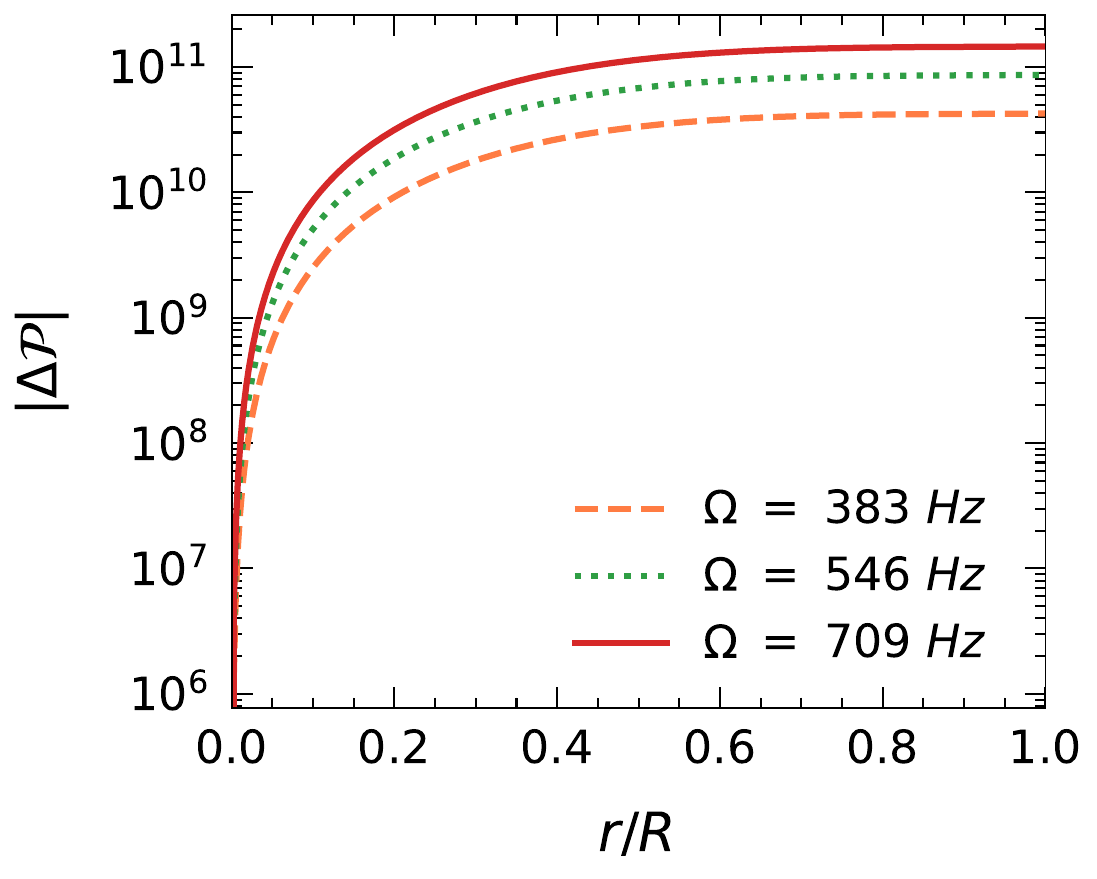}\hspace{0.5cm}
		\includegraphics[scale = 0.3]{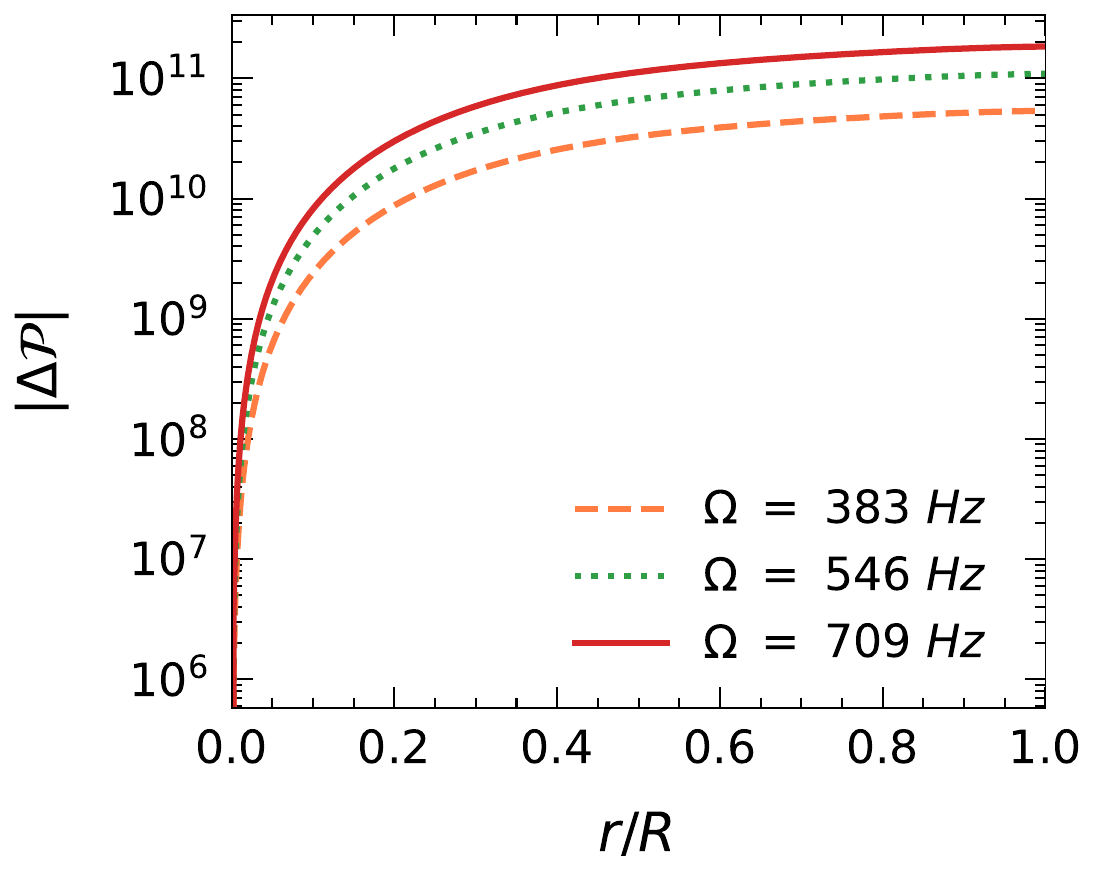}}
	\vspace{0.3cm}
	\centerline{
		\includegraphics[scale = 0.3]{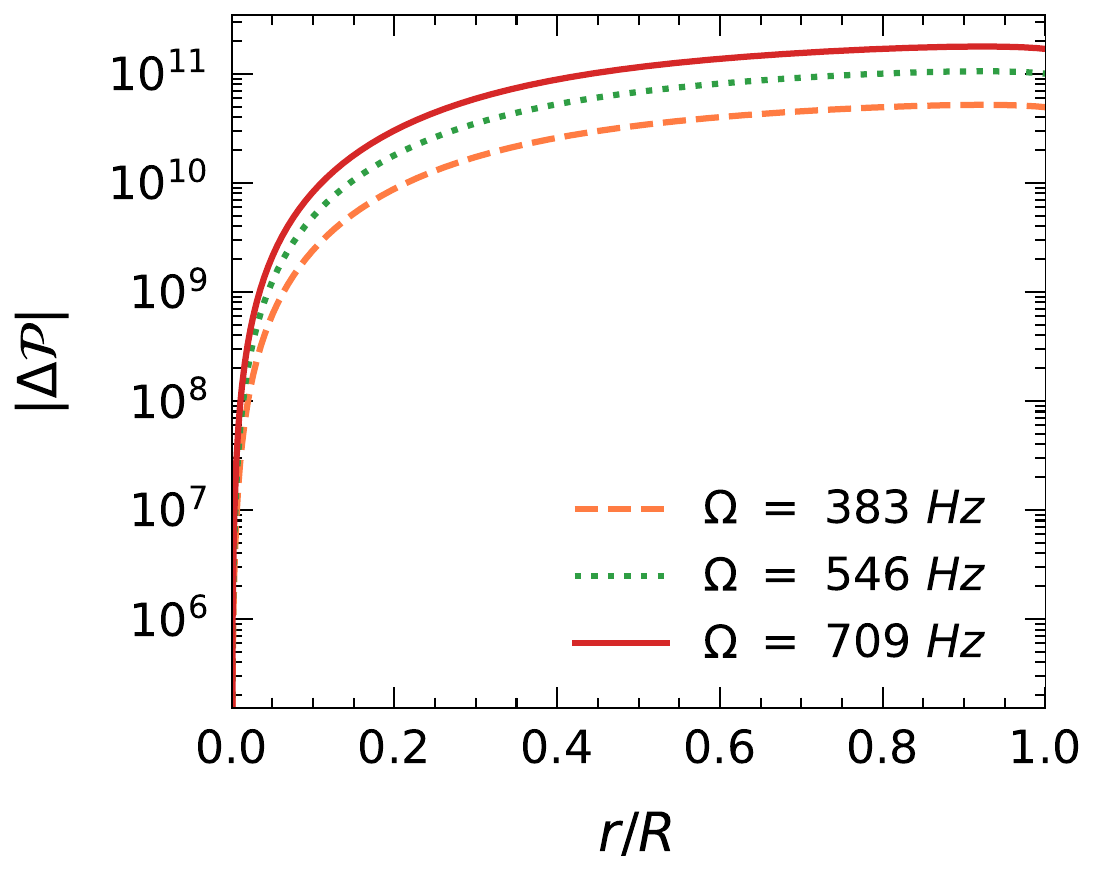}\hspace{0.5cm}
		\includegraphics[scale = 0.3]{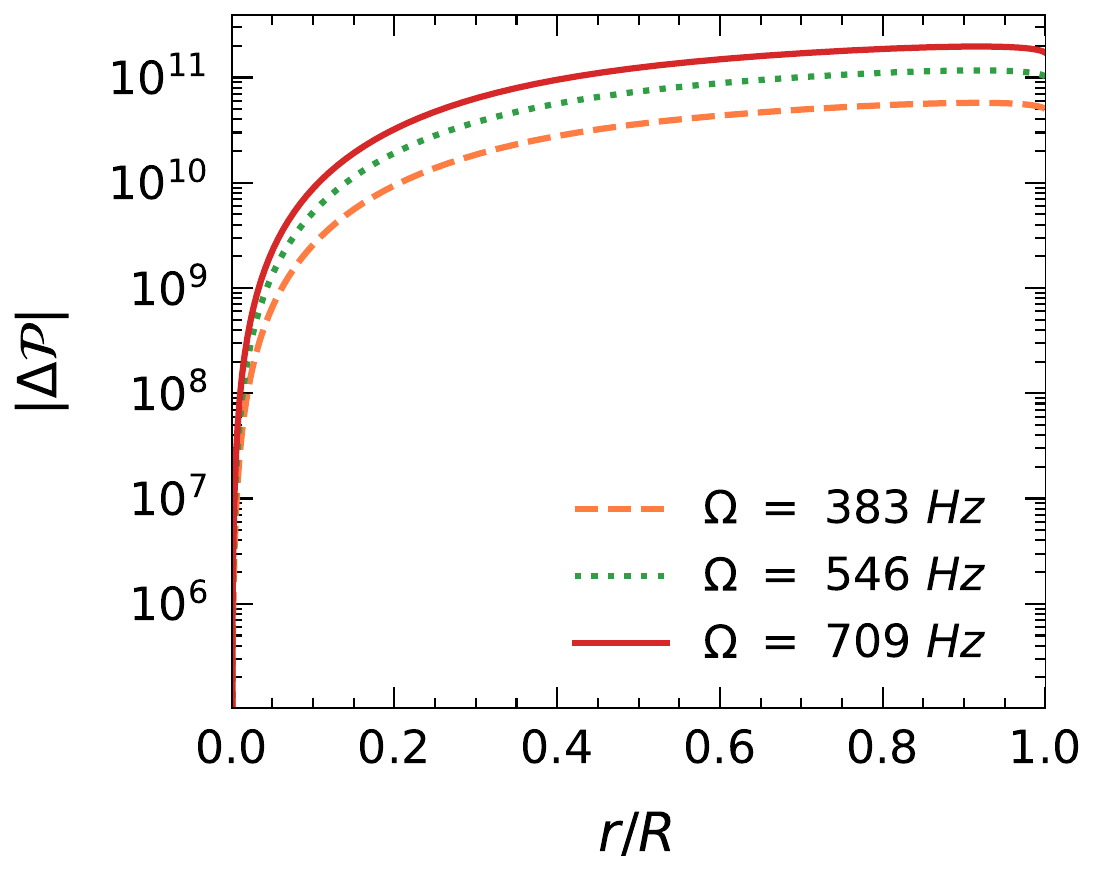}}
	\vspace{-0.3cm}
	\caption{Radial variation of the rotation corrected pressure part of
		slow-rotating stars with the angular frequencies of $383$ Hz, 
$546$ Hz and $709$ Hz for the Model I (top left), Model II (top right), Model 
III (bottom left) and Model IV (bottom right).}
	\label{fig:comparison_pressure_with_non_rot_part}
\end{figure}
Due to rotation, there is also variation in the number 
density, energy density and pressure distribution across the star's radius. 
From Figs.~\ref{fig:comparison_num_den_with_non_rot_part},
\ref{fig:comparison_ene_den_with_non_rot_part} and 
\ref{fig:comparison_pressure_with_non_rot_part} respectively, it is observed 
that the changes in number density, energy density and pressure from 
non-rotating stars are significantly large towards the surface of the stars. 
At the surface of the stars, the number density varies by the order 
of $10^{-4} - 10^{-3}$ with the increase of angular frequencies of the 
stars for Models I, II and III. For Model IV, the number density varies by the 
order of $10^{-3} -10^{-2}$ $n_{0}$ with the increase of angular frequencies of 
the stars. Furthermore, at the surface of the stars, the variation of energy 
density is of the order of $\sim 10^{11}$ $\text{gcm}^{-3}$ for Model I, II and III, 
and $10^{11} - 10^{12}$ $\text{gcm}^{-3}$ for Model IV, while the magnitude of the variation 
increases with increasing angular frequencies. We know that for a 
self-bound star, the pressure at the surface vanishes \cite{alcock_1986}. Near 
surface of such self-bound stars, the order of pressure value 
varies between $10^{10} - 10^{11}$ $\text{gcm}^{-3}$ with an increase of angular frequencies 
for all models. Almost all models exhibit a similar variation pattern. 
Moreover, it is found that the maximum absolute 
relative variation of pressure is the lowest for Model II, which are $0.2039$, 
$0.4144$ and $0.6986$ for $383$ Hz, $546$ Hz and $709$ Hz, respectively. 
Whereas the Model IV shows the highest maximum absolute relative variation. 
The respective 
values are $4.5574$, $9.2630$ and $15.6202$. Similarly, the maximum value of 
absolute relative variation of energy density is the lowest for Model II and 
the highest for Model IV. The maximum value of relative variation changes 
within the order of $10^{-4} - 10^{-5}$ for Model II, and within the order of 
$10^{-4} - 10^{-3}$ for Model IV. Also, the maximum value of the absolute 
relative variation of the baryon number density is the lowest for Model II and 
the highest for Model IV. The order of variation for Model II is within the 
range of $10^{-5} - 10^{-4}$, and for Model IV the order is $\sim 10^{-3}$. 
For all the 
models, the values of maximum absolute relative variation increase with the 
increase in angular frequencies. The distributions of variations 
of number density, energy density and pressure across the radius of the stars 
are almost uniform, excluding the case of no variation at the center or very
near the center of the stars.
\subsection{Quasiradial Oscillations}

In this study, we only consider the fluid perturbation in radial direction for 
the analysis of QROs using the HTC formalism \cite{hartle_1972} as discussed in Section~\ref{secIV_b}.
We have already calculated the eigenfrequencies of radial oscillations for all 
the models, as given in Table~\ref{table:eigen-frequencies}. For rotating 
stars, those radial oscillations are affected by the centrifugal force of 
rotation, due to which there arises an imaginary part of eigenfrequencies 
which tells us the stability of the star and the half-life of damping 
vibration energy as discussed in Section~\ref{secIV_b}. We calculate the QRO 
frequencies of the stars for different angular frequencies and list them in 
Tables~\ref{table:rotCorrEigFreq_I}, \ref{table:rotCorrEigFreq_II}, 
\ref{table:rotCorrEigFreq_III} and \ref{table:rotCorrEigFreq_IV} for Model I, 
II, III and IV, respectively. The number of modes for each model remains the 
same while there are real and imaginary parts for each mode. Furthermore, in 
our study we consider the stars with maximum possible mass for a stable 
configuration while ignoring all other stable and unstable masses from the 
mass and radius relation as shown in Fig.~\ref{fig:M_R_plot_comparison}. From 
Table~\ref{table:rotCorrEigFreq_I}, \ref{table:rotCorrEigFreq_II}, 
\ref{table:rotCorrEigFreq_III} and \ref{table:rotCorrEigFreq_IV}, one can 
observe that the imaginary part of each mode for every model is positive, 
which confirms that the oscillation modes of the rotating stars that have been 
considered here are all 
stable. Also, it can be observed that the oscillation frequencies have shifted 
from the oscillation frequencies of non-rotating stars. Additionally, only 
the imaginary part of the oscillation frequencies varies with angular 
frequencies.
\begin{table}[!h]
	\begin{center}
		\caption{Modified eigenfrequencies of radial oscillations due 
to rotation of the stars for the EoS Model I. All frequencies are in kHz.}
		\vspace{3pt}
		\setlength{\tabcolsep}{15pt}
		\begin{tabular}{cccc}
			\hline\hline\\[-8pt]
			\rule{0pt}{15pt} \shortstack{Modes\\(Order $n$)}
			& 383 Hz & 546 Hz & 709 Hz \\[2pt]
			\hline\hline\\[-12pt]
			$f (0)$		   & 0.49 + 1.20j    & 0.49 + 1.23j    & 0.49 + 1.22j   \\
			$p_{1} (1) $   & 22.27 + 2.66j   & 22.27 + 2.73j   & 22.27 + 2.71j  \\
			$p_{2} (2) $   & 36.58 + 4.03j   & 36.58 + 4.14j   & 36.58 + 4.11j  \\
			$p_{3} (3) $   & 50.18 + 5.39j   & 50.18 + 5.54j   & 50.18 + 5.50j  \\
			$p_{4} (4) $   & 63.54 + 6.75j   & 63.54 + 6.94j   & 63.54 + 6.88j  \\
			$p_{5} (5) $   & 76.77 + 8.11j   & 76.77 + 8.33j   & 76.77 + 8.26j  \\
			$p_{6} (6) $   & 89.93 + 9.46j   & 89.93 + 9.73j   & 89.93 + 9.64j  \\
			$p_{7} (7) $   & 103.05 + 10.82j & 103.05 + 11.12j & 103.05 + 11.02j \\
			$p_{8} (8) $   & 116.13 + 12.17j & 116.13 + 12.51j & 116.13 + 12.40j \\
			$p_{9} (9) $   & 129.20 + 13.52j & 129.20 + 13.90j & 129.20 + 13.78j \\
			$p_{10} (10)$  & 142.26 + 14.87j & 142.26 + 15.29j & 142.26 + 15.15j \\
			$p_{11} (11)$  & 155.30 + 16.23j & 155.30 + 16.68j & 155.30 + 16.53j \\
			$p_{12} (12)$  & 168.34 + 17.58j & 168.34 + 18.07j & 168.34 + 17.91j \\
			$p_{13} (13)$  & 181.37 + 18.93j & 181.37 + 19.46j & 181.37 + 19.28j \\
			$p_{14} (14)$  & 194.39 + 20.28j & 194.39 + 20.84j & 194.39 + 20.66j \\[3pt]
			\hline
		\end{tabular}
		\label{table:rotCorrEigFreq_I}
	\end{center}
\end{table}

\begin{table}[h]
	\begin{center}
		\caption{Modified eigenfrequencies of radial oscillations due 
to rotation of the stars for the EoS Model II. All frequencies are in kHz.}
		\vspace{3pt}
		\setlength{\tabcolsep}{15pt}
		\begin{tabular}{cccc}
			\hline\hline\\[-8pt]
			\rule{0pt}{15pt} \shortstack{Modes\\(Order $n$)}
			& 383 Hz & 546 Hz & 709 Hz \\[2pt]
			\hline\hline\\[-12pt]
			$f (0)$		   & 0.16 + 0.08j   & 0.16 + 0.10j  & 0.16 + 0.12j   \\
			$p_{1} (1) $   & 19.84 + 0.20j  & 19.84 + 0.26j  & 19.84 + 0.31j  \\
			$p_{2} (2) $   & 32.24 + 0.30j  & 32.24 + 0.40j  & 32.24 + 0.48j  \\
			$p_{3} (3) $   & 44.05 + 0.41j  & 44.05 + 0.54j  & 44.05 + 0.64j  \\
			$p_{4} (4) $   & 55.67 + 0.51j  & 55.67 + 0.67j  & 55.67 + 0.80j  \\
			$p_{5} (5) $   & 67.18 + 0.62j  & 67.18 + 0.81j  & 67.18 + 0.96j  \\
			$p_{6} (6) $   & 78.65 + 0.72j  & 78.65 + 0.94j  & 78.65 + 1.12j  \\
			$p_{7} (7) $   & 90.08 + 0.82j  & 90.08 + 1.08j  & 90.08 + 1.27j \\
			$p_{8} (8) $   & 101.49 + 0.92j & 101.49 + 1.21j & 101.49 + 1.43j \\
			$p_{9} (9) $   & 112.88 + 1.02j & 112.88 + 1.34j & 112.88 + 1.59j \\
			$p_{10} (10)$  & 124.27 + 1.12j & 124.27 + 1.48j & 124.27 + 1.74j \\
			$p_{11} (11)$  & 135.65 + 1.22j & 135.65 + 1.61j & 135.65 + 1.90j \\
			$p_{12} (12)$  & 147.02 + 1.33j & 147.02 + 1.74j & 147.02 + 2.05j \\
			$p_{13} (13)$  & 158.38 + 1.43j & 158.38 + 1.87j & 158.38 + 2.21j \\
			$p_{14} (14)$  & 169.75 + 1.53j & 169.75 + 2.00j & 169.75 + 2.36j \\
			$p_{15} (15)$  & 181.11 + 1.63j & 181.11 + 2.12j & 181.11 + 2.51j \\
			$p_{16} (16)$  & 192.47 + 1.73j & 192.47 + 2.26j & 192.47 + 2.66j \\[3pt]
			\hline
		\end{tabular}
		\label{table:rotCorrEigFreq_II}
	\end{center}
\end{table}

\begin{table}[h]
	\begin{center}
		\caption{Modified eigenfrequencies of radial oscillations due 
to rotation of the stars for the EoS Model III. All frequencies are in kHz.}
		\vspace{3pt}
		\setlength{\tabcolsep}{15pt}
		\begin{tabular}{cccc}
			\hline\hline\\[-8pt]
			\rule{0pt}{15pt} \shortstack{Modes\\(Order $n$)}
			& 383 Hz & 546 Hz & 709 Hz \\[2pt]
			\hline\hline\\[-12pt]
			$f (0)$		   & 0.43 + 0.18j   & 0.43 + 0.19j  & 0.43 + 0.18j   \\
			$p_{1} (1) $   & 19.21 + 0.40j  & 19.21 + 0.44j  & 19.21 + 0.44j   \\
			$p_{2} (2) $   & 30.97 + 0.64j  & 30.97 + 0.70j  & 30.97 + 0.70j   \\
			$p_{3} (3) $   & 42.17 + 0.88j  & 42.17 + 0.95j  & 42.17 + 0.96j   \\
			$p_{4} (4) $   & 53.17 + 1.12j  & 53.17 + 1.21j  & 53.17 + 1.21j   \\
			$p_{5} (5) $   & 64.08 + 1.36j  & 64.08 + 1.47j  & 64.08 + 1.47j   \\
			$p_{6} (6) $   & 74.95 + 1.60j  & 74.95 + 1.73j  & 74.95 + 1.72j   \\
			$p_{7} (7) $   & 85.80 + 1.83j  & 85.80 + 1.98j  & 85.80 + 1.98j  \\
			$p_{8} (8) $   & 96.62 + 2.07j  & 96.62 + 2.24j  & 96.62 + 2.23j   \\
			$p_{9} (9) $   & 107.44 + 2.31j & 107.44 + 2.49j & 107.44 + 2.48j  \\
			$p_{10} (10)$  & 118.24 + 2.54j & 118.24 + 2.75j & 118.24 + 2.73j  \\
			$p_{11} (11)$  & 129.05 + 2.78j & 129.05 + 3.00j & 129.05 + 2.99j  \\
			$p_{12} (12)$  & 139.84 + 3.02j & 139.84 + 3.26j & 139.84 + 3.24j  \\
			$p_{13} (13)$  & 150.64 + 3.25j & 150.64 + 3.51j & 150.64 + 3.49j  \\
			$p_{14} (14)$  & 161.43 + 3.49j & 161.43 + 3.76j & 161.43 + 3.74j  \\
			$p_{15} (15)$  & 172.21 + 3.72j & 172.21 + 4.02j & 172.21 + 3.99j  \\
			$p_{16} (16)$  & 183.00 + 3.96j & 183.00 + 4.27j & 183.00 + 4.24j  \\
			$p_{17} (17)$  & 193.79 + 4.19j & 193.79 + 4.52j & 193.79 + 4.49j  \\[3pt]
			\hline
		\end{tabular}
		\label{table:rotCorrEigFreq_III}
	\end{center}
\end{table}

\begin{table}[h]
	\begin{center}
		\caption{Modified eigenfrequencies of radial oscillations due 
to rotation of the stars for the EoS Model IV. All frequencies are in kHz.}
		\vspace{3pt}
		\setlength{\tabcolsep}{15pt}
		\begin{tabular}{cccc}
			\hline\hline\\[-8pt]
			\rule{0pt}{15pt} \shortstack{Modes\\(Order $n$)}
			& 383 Hz & 546 Hz & 709 Hz \\[2pt]
			\hline\hline\\[-12pt]
			$f (0)$		   & 1.03 + 0.09j   & 1.03 + 0.09j  &  1.03 + 0.06j   \\
			$p_{1} (1) $   & 15.55 + 0.23j  & 15.55 + 0.24j  & 15.55 + 0.23j   \\
			$p_{2} (2) $   & 24.94 + 0.38j  & 24.94 + 0.39j  & 24.94 + 0.37j   \\
			$p_{3} (3) $   & 33.85 + 0.53j  & 33.85 + 0.55j  & 33.85 + 0.51j   \\
			$p_{4} (4) $   & 42.59 + 0.69j  & 42.59 + 0.71j  & 42.59 + 0.65j   \\
			$p_{5} (5) $   & 51.25 + 0.86j  & 51.25 + 0.87j  & 51.25 + 0.79j   \\
			$p_{6} (6) $   & 59.87 + 1.02j  & 59.87 + 1.04j  & 59.87 + 0.93j   \\
			$p_{7} (7) $   & 68.46 + 1.19j  & 68.46 + 1.20j  & 68.46 + 1.07j  \\
			$p_{8} (8) $   & 77.04 + 1.37j  & 77.04 + 1.37j  & 77.04 + 1.21j   \\
			$p_{9} (9) $   & 85.61 + 1.54j  & 85.61 + 1.54j  & 85.61 + 1.35j  \\
			$p_{10} (10)$  & 94.17 + 1.72j  & 94.17 + 1.72j  & 94.17 + 1.50j  \\
			$p_{11} (11)$  & 102.72 + 1.89j & 102.72 + 1.89j & 102.72 + 1.64j  \\
			$p_{12} (12)$  & 111.27 + 2.07j & 111.27 + 2.06j & 111.27 + 1.78j  \\
			$p_{13} (13)$  & 119.82 + 2.25j & 119.82 + 2.24j & 119.82 + 1.93j  \\
			$p_{14} (14)$  & 128.37 + 2.43j & 128.37 + 2.41j & 128.37 + 2.07j  \\
			$p_{15} (15)$  & 136.91 + 2.60j & 136.91 + 2.59j & 136.91 + 2.22j  \\
			$p_{16} (16)$  & 145.46 + 2.78j & 145.46 + 2.76j & 145.46 + 2.36j  \\
			$p_{17} (17)$  & 154.00 + 2.97j & 154.00 + 2.94j & 154.00 + 2.51j  \\
			$p_{18} (18)$  & 162.54 + 3.15j & 162.54 + 3.11j & 162.54 + 2.65j  \\
			$p_{19} (19)$  & 171.08 + 3.33j & 171.08 + 3.29j & 171.08 + 2.79j  \\
			$p_{20} (20)$  & 179.62 + 3.51j & 179.62 + 3.47j & 179.62 + 2.94j  \\
			$p_{21} (21)$  & 188.16 + 3.69j & 188.16 + 3.64j & 188.16 + 3.08j  \\
			$p_{22} (22)$  & 196.70 + 3.87j & 196.70 + 3.82j & 196.70 + 3.23j  \\[3pt]
			\hline
		\end{tabular}
		\label{table:rotCorrEigFreq_IV}
	\end{center}
\end{table}

\begin{figure}[!h]
	\centerline{
	    \includegraphics[scale = 0.3]{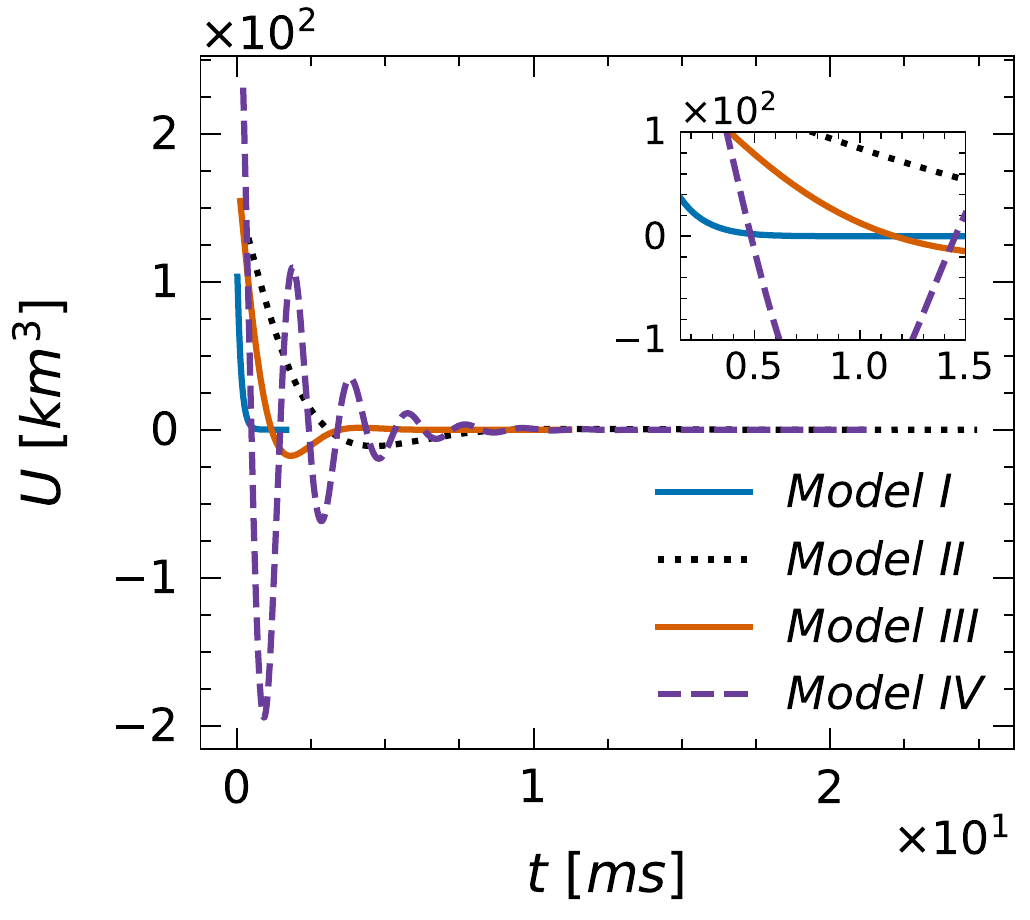}\hspace{0.5cm}
	    \includegraphics[scale = 0.3]{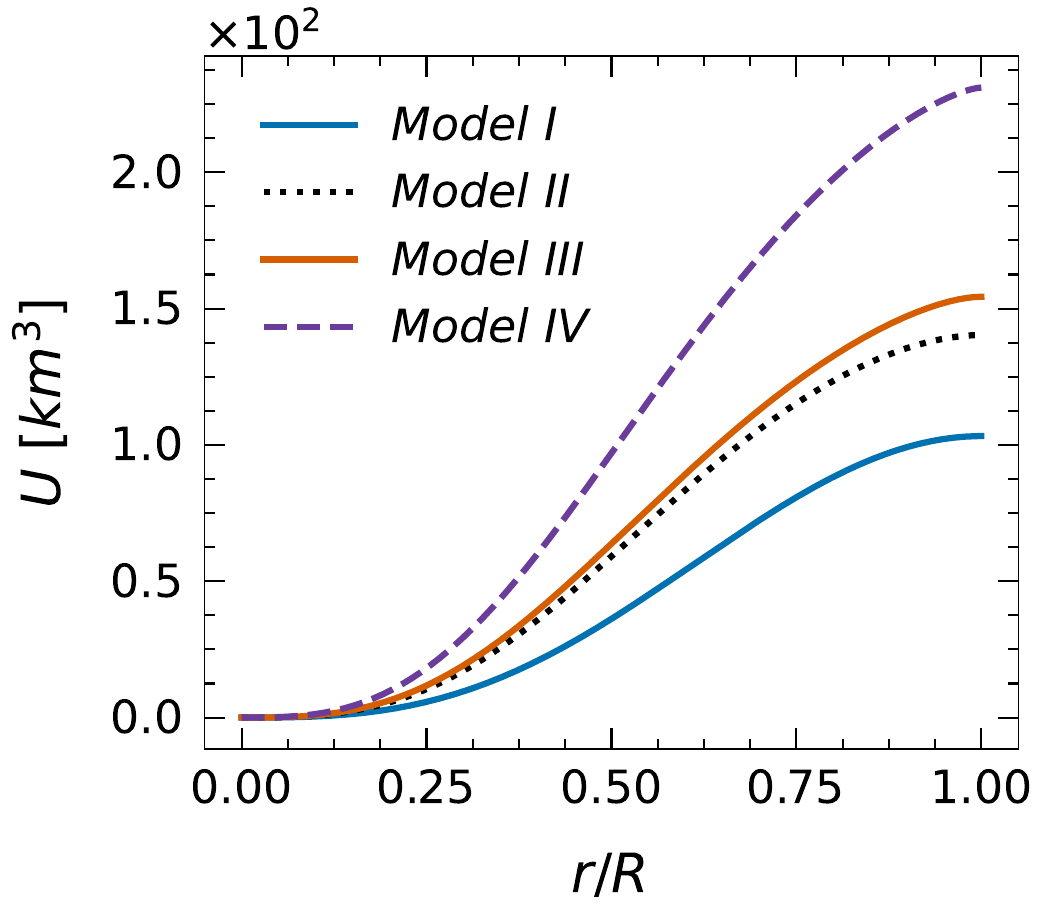}}
            \vspace{0.3cm}
       \centerline{
	    \includegraphics[scale = 0.3]{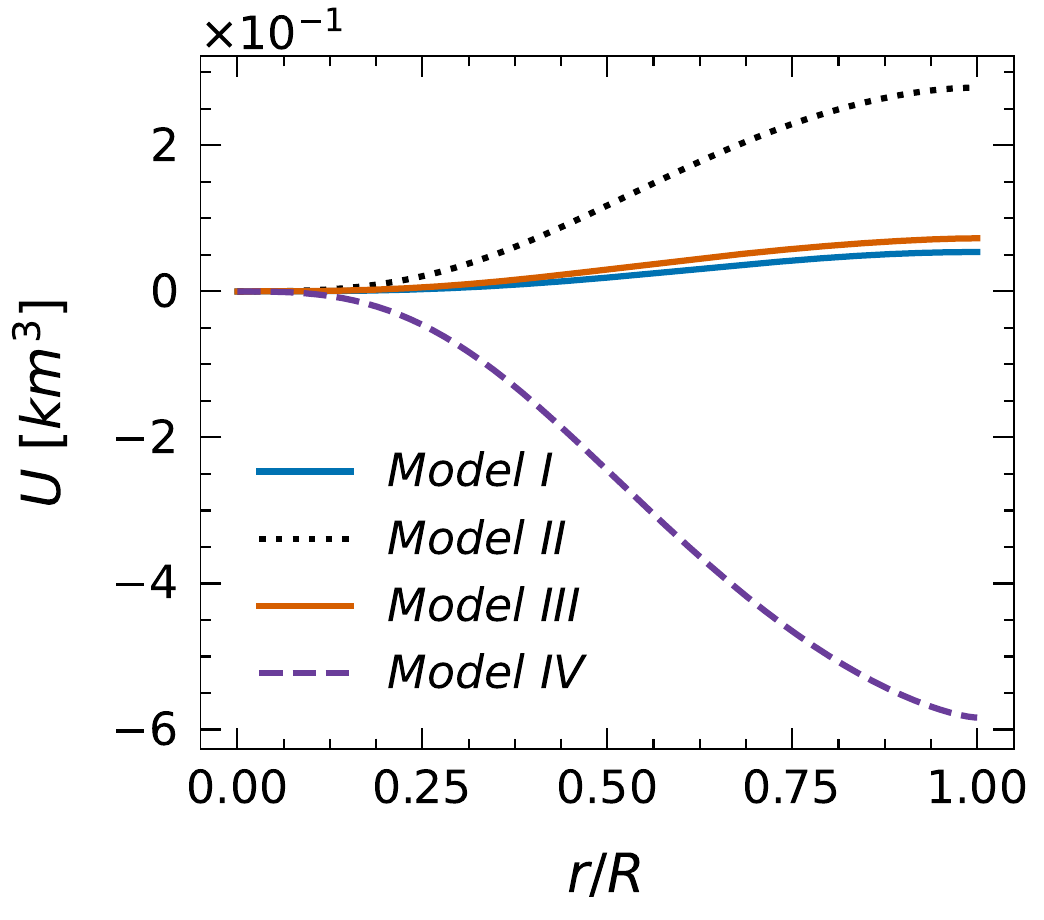}\hspace{0.5cm}
	    \includegraphics[scale = 0.3]{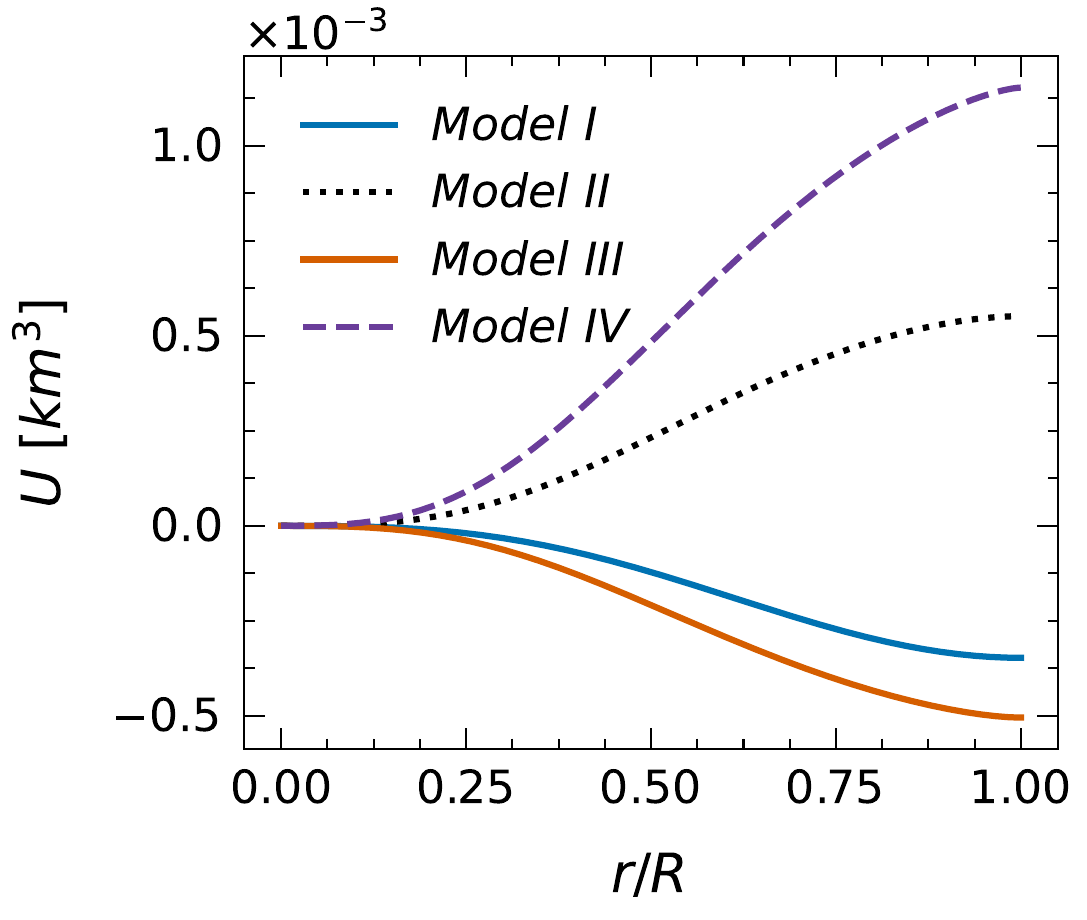}}
            \vspace{-0.3cm}
	\caption{Temporal evolution of radial displacement parameter $U(r)$ 
at the surface of slow-rotating stars for all stellar models with 
rotation frequency of $383$ Hz, up to half-life of damping vibration energy 
(top-left plot). Also, the top-right, bottom-left, and bottom-right plots show 
the radial variations of $U(r)$ at $\sim 0 \tau$, $0.5 \tau$, and $1.0 \tau$, 
respectively. Here, $\tau$ represents the half-life of the oscillations.}
\label{fig:radial_and_temporal_variation_of_lagrangian_disp_383Hz}
\end{figure}

\begin{figure}[!h]
	\centerline{
	    \includegraphics[scale = 0.3]{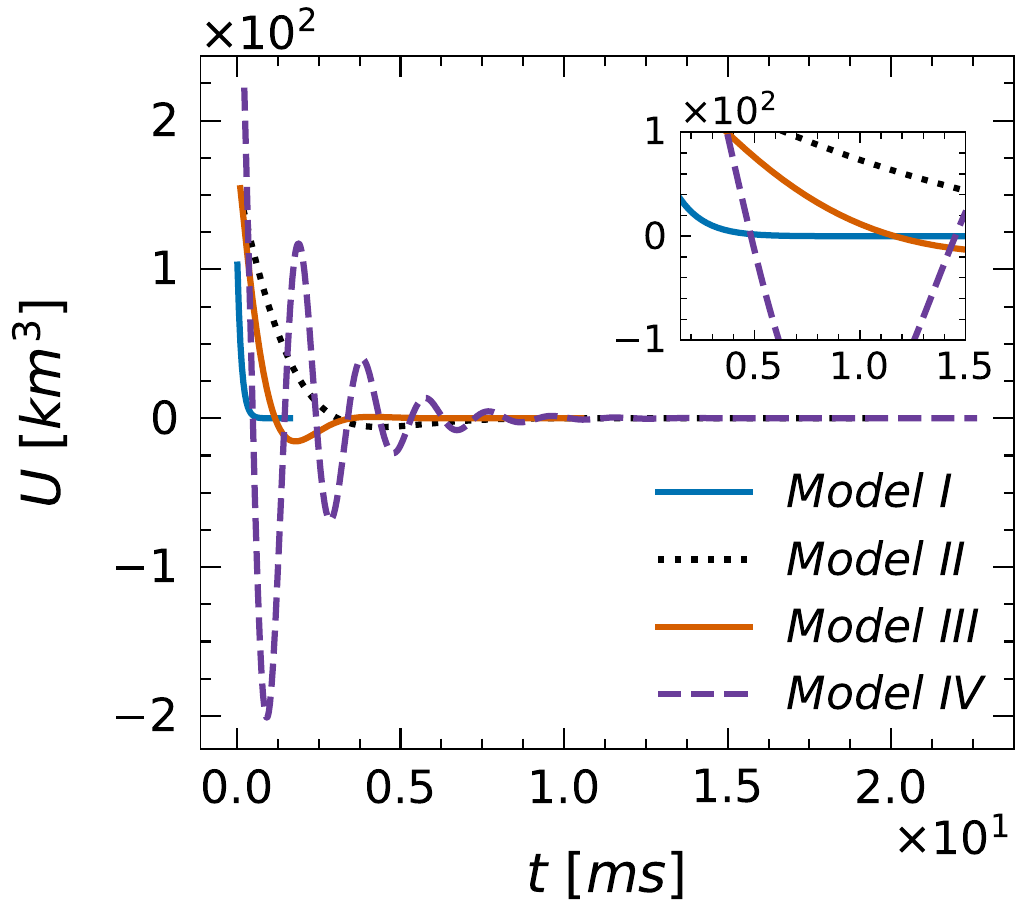}\hspace{0.5cm}
	    \includegraphics[scale = 0.3]{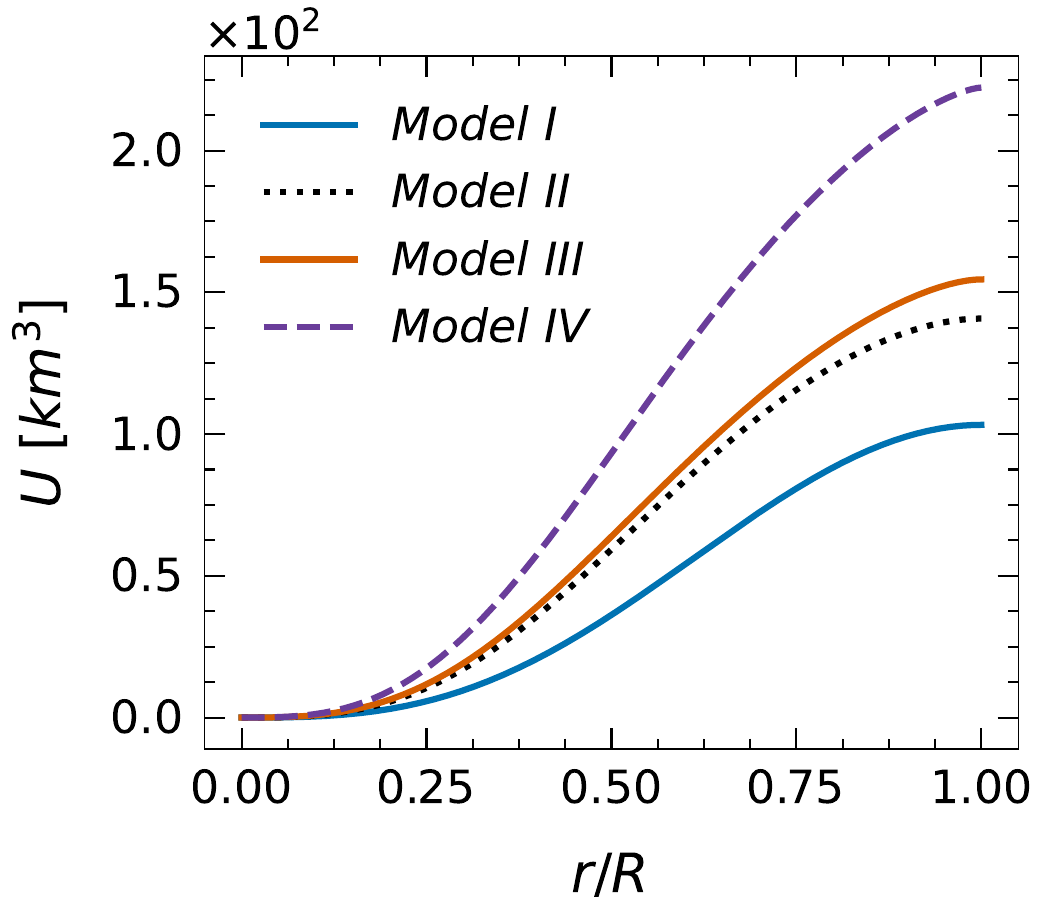}}
            \vspace{0.3cm}
         \centerline{
	     \includegraphics[scale = 0.3]{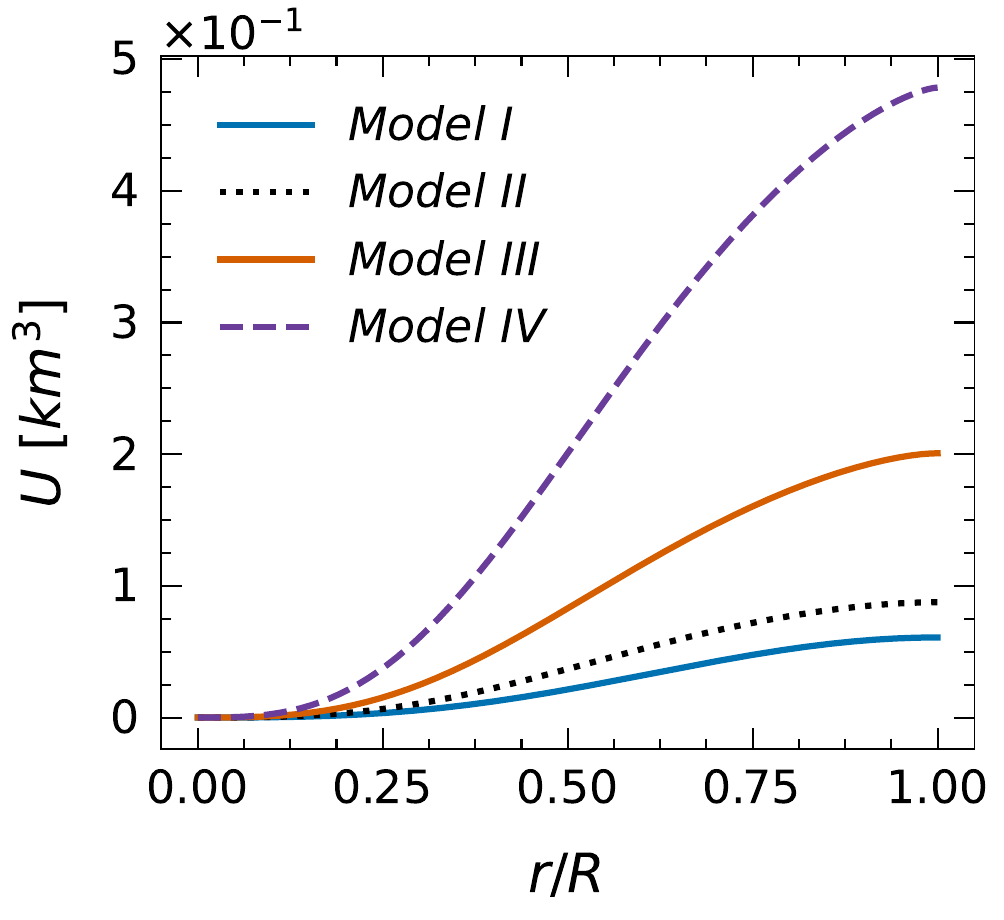}\hspace{0.5cm}
	     \includegraphics[scale = 0.3]{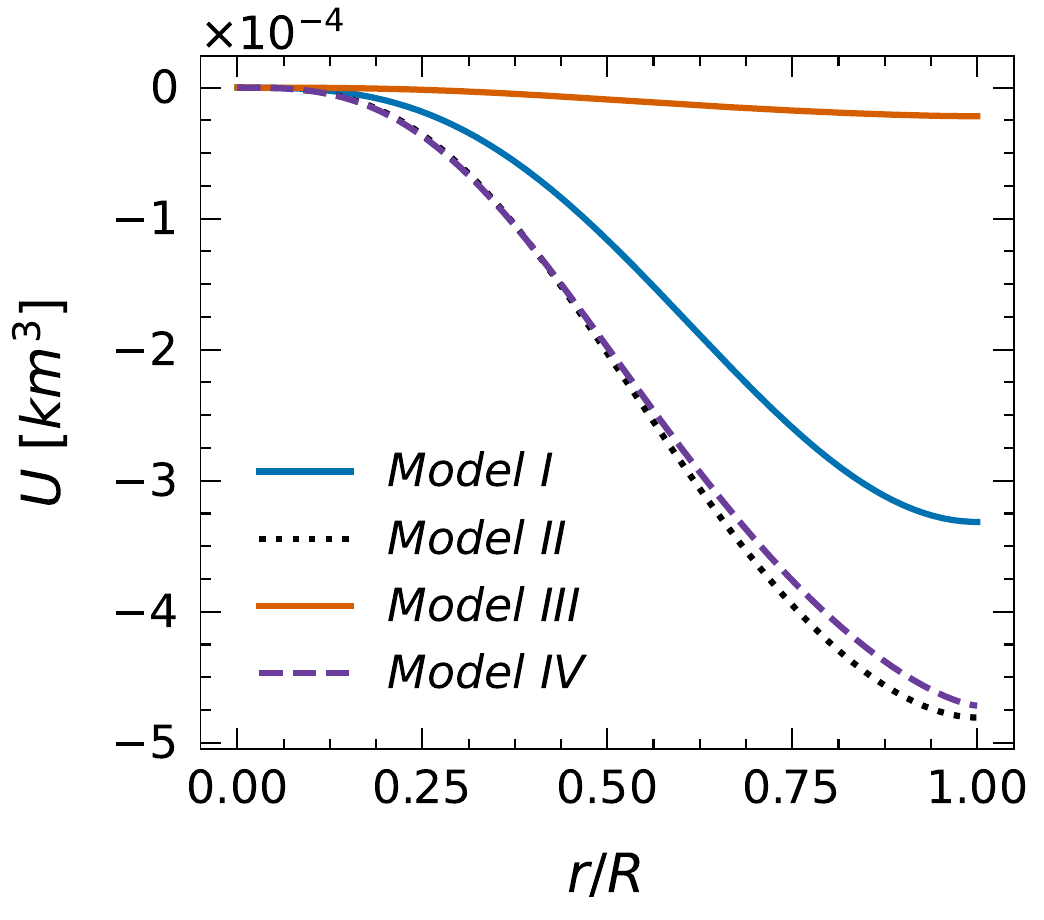}}
             \vspace{-0.3cm}
	\caption{Temporal evolution of radial displacement parameter $U(r)$ at 
the surface of slow-rotating stars for all stellar models with rotation 
frequency of $546$ Hz, up to half-life of damping vibration energy (top-left 
plot). Also, the top-right, bottom-left, and bottom-right plots show the 
radial variations of $U(r)$ at $\sim 0 \tau$, $0.5 \tau$, and $1.0 \tau$, 
respectively. Here, $\tau$ represents the half-life of the oscillations.}
	\label{fig:radial_and_temporal_variation_of_lagrangian_disp_546Hz}
\end{figure}

\begin{figure}[!h]
	\centerline{
	     \includegraphics[scale = 0.3]{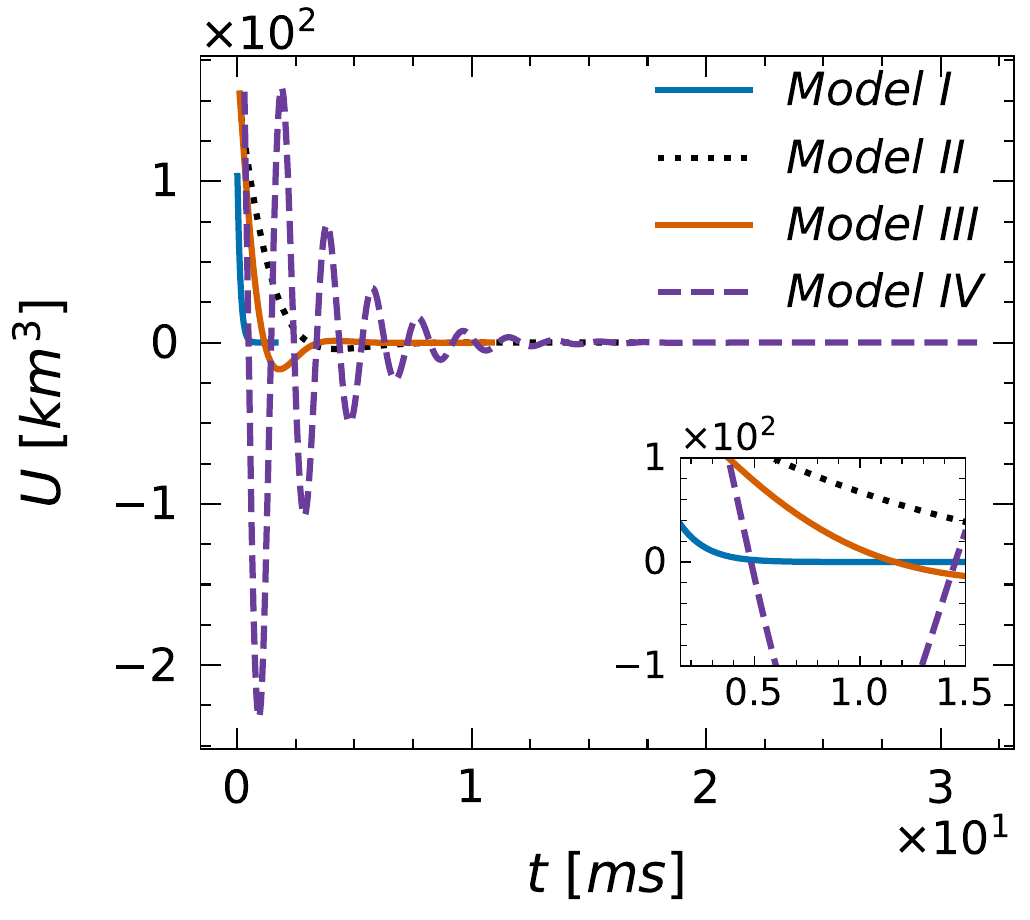}\hspace{0.5cm}
	     \includegraphics[scale = 0.3]{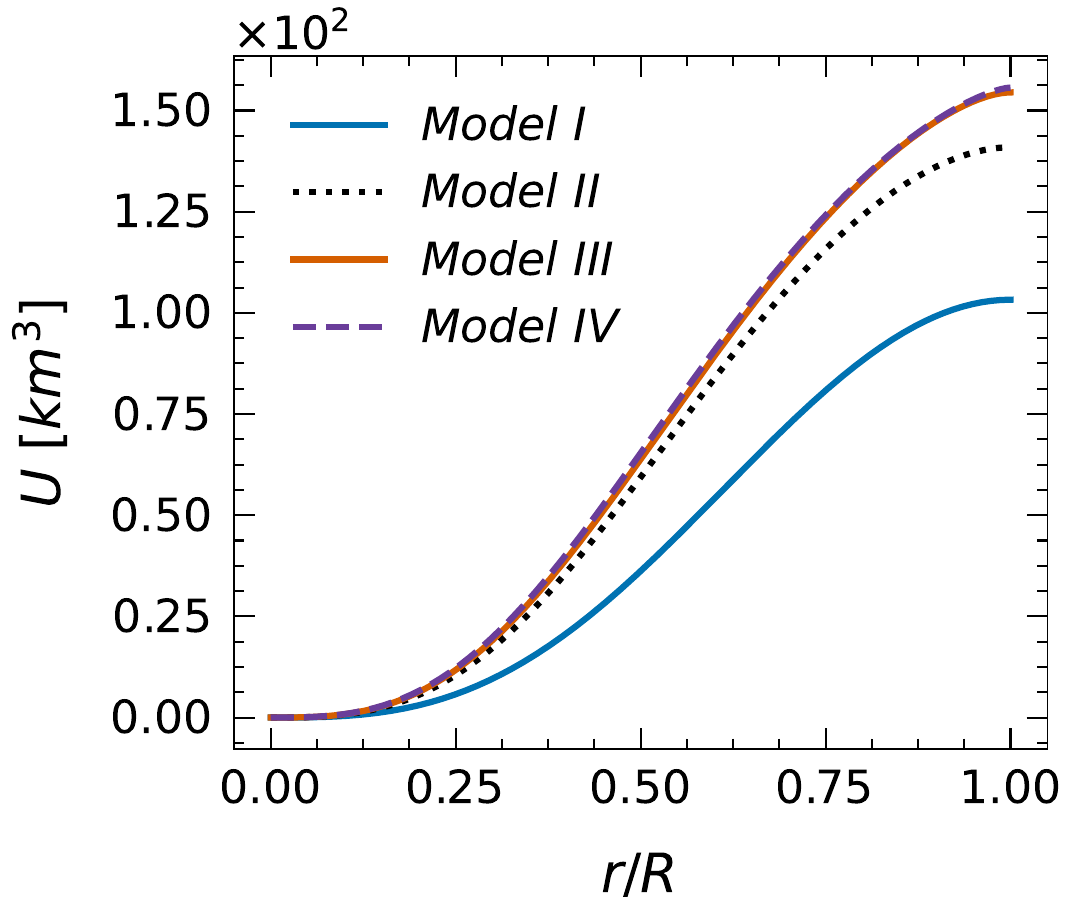}}
             \vspace{0.3cm}
        \centerline{
	     \includegraphics[scale = 0.3]{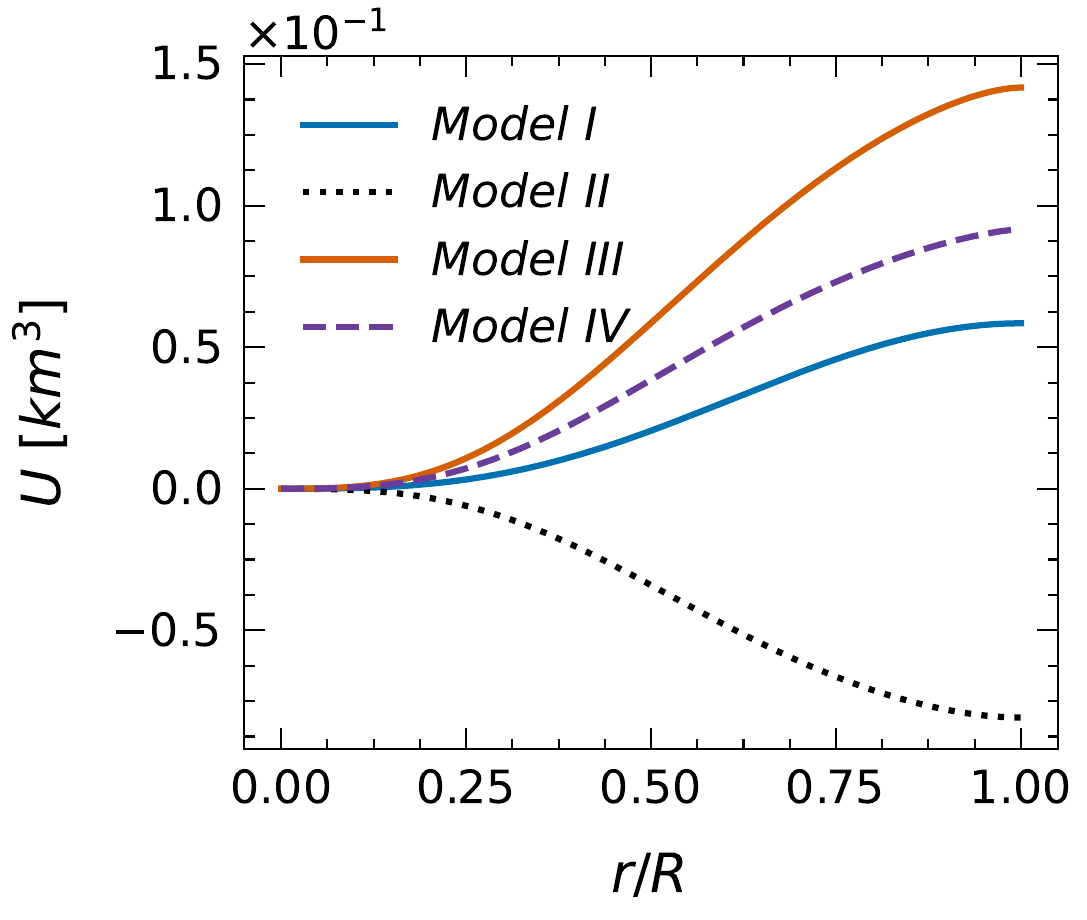}\hspace{0.5cm}
	     \includegraphics[scale = 0.3]{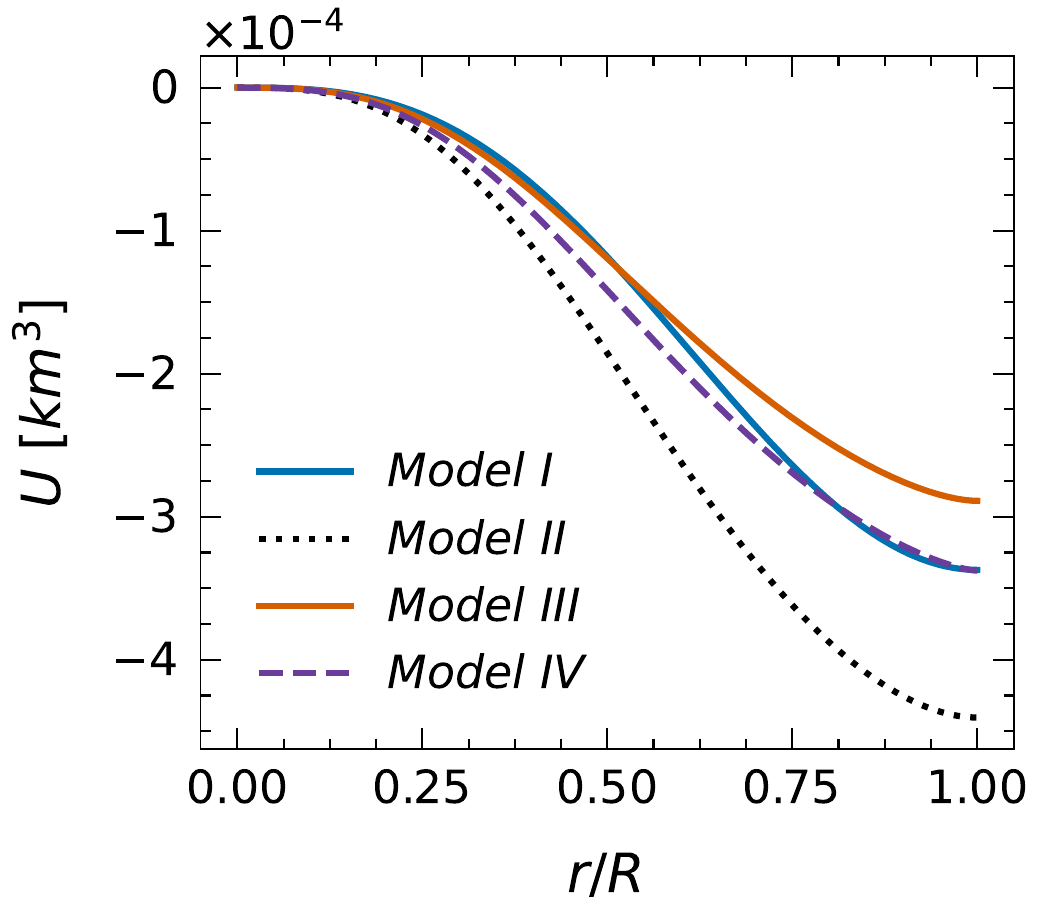}}
             \vspace{-0.3cm}
	\caption{Temporal evolution of radial displacement parameter $U(r)$ at
the surface of slow-rotating stars for all stellar models with rotation
frequency of $709$ Hz, up to half-life of damping vibration energy (top-left 
plot). Also, the top-right, bottom-left, and bottom-right plots show the 
radial variations of $U(r)$ at $\sim 0 \tau$, $0.5 \tau$, and $1.0 \tau$, 
respectively. Here, $\tau$ represents the half-life of the oscillations.}
	\label{fig:radial_and_temporal_variation_of_lagrangian_disp_709Hz}
\end{figure}

The imaginary part of the QRO frequencies also gives the half-life of 
the damping vibration energy for a stable configuration. We calculate the 
half-life for each mode for all the models and list them in 
Tables~\ref{table:oscHalfLife_I} and \ref{table:oscHalfLife_II}. From these 
tables, it can be observed that the half-life of damping vibration energy 
decreases with increasing oscillation frequencies, and the $f$-mode for every 
angular frequency for all models has the highest half-life. Also, half-life 
varies with the variation of the angular frequencies of the stars. We plot 
the temporal and radial variations of the $f$-mode in 
Figs.~\ref{fig:radial_and_temporal_variation_of_lagrangian_disp_383Hz}, 
\ref{fig:radial_and_temporal_variation_of_lagrangian_disp_546Hz} and 
\ref{fig:radial_and_temporal_variation_of_lagrangian_disp_709Hz}, as $f$-mode 
has the longest half-life than that of any other mode. All other modes still 
have their significance as they can help us to study the interior of the stars. 
But from the observational perspective, here $f$-mode survives for the longest 
period while having frequency value favorable with different detectors. 
\begin{table}[!h]
	\begin{center}
		\caption{Half-life times of damping vibrational energies 
obtained by using the complex part of the oscillation frequencies for all 
three angular frequencies of stars for EoS models I and II. All values are in 
milliseconds (ms).}
		\vspace{3pt}
		\setlength{\tabcolsep}{15pt}
		\begin{tabular}{cccc|ccc}
			\hline\hline\\[-8pt]
			\rule{0pt}{15pt} \shortstack{Modes\\(Order $n$)}
			& \multicolumn{3}{c}{Model I}
			& \multicolumn{3}{c}{Model II}\\[5pt]
			
			\cline{2-4}\cline{5-7}
			\rule{0pt}{10pt}
			& 383 Hz & 546 Hz & 709 Hz
			& 383 Hz & 546 Hz & 709 Hz\\[2pt]
			\hline\hline\\[-12pt]
			$f (0)      $  & 1.6668 & 1.6224 & 1.6380 & 24.9528 & 19.5155 & 17.1628 \\
			$p_{1} (1)  $  & 0.7526 & 0.7321 & 0.7385 & 10.1403 & 7.7018  & 6.4932 \\
			$p_{2} (2)  $  & 0.4961 & 0.4826 & 0.4868 & 6.5742  & 4.9915  & 4.2063 \\
			$p_{3} (3)  $  & 0.3708 & 0.3607 & 0.3639 & 4.8893  & 3.7143  & 3.1323 \\
			$p_{4} (4)  $  & 0.2962 & 0.2882 & 0.2907 & 3.8997  & 2.9643  & 2.5018 \\
			$p_{5} (5)  $  & 0.2467 & 0.2400 & 0.2421 & 3.2474  & 2.4698  & 2.0860 \\
			$p_{6} (6)  $  & 0.2114 & 0.2056 & 0.2075 & 2.7843  & 2.1186  & 1.7905 \\
			$p_{7} (7)  $  & 0.1849 & 0.1799 & 0.1815 & 2.4383  & 1.8560  & 1.5695 \\
			$p_{8} (8)  $  & 0.1643 & 0.1599 & 0.1613 & 2.1694  & 1.6520  & 1.3977 \\
			$p_{9} (9)  $  & 0.1479 & 0.1439 & 0.1452 & 1.9548  & 1.4891  & 1.2604 \\
			$p_{10} (10)$  & 0.1345 & 0.1308 & 0.1320 & 1.7792  & 1.3558  & 1.1481 \\
			$p_{11} (11)$  & 0.1233 & 0.1199 & 0.1210 & 1.6329  & 1.2446  & 1.0544 \\
			$p_{12} (12)$  & 0.1138 & 0.1107 & 0.1117 & 1.5091  & 1.1506  & 0.9751 \\
			$p_{13} (13)$  & 0.1057 & 0.1028 & 0.1037 & 1.4029  & 1.0699  & 0.9070 \\
			$p_{14} (14)$  & 0.0986 & 0.0960 & 0.0968 & 1.3109  & 0.9999  & 0.8479 \\
			$p_{15} (15)$  &  $\_$  &  $\_$  &  $\_$  & 1.2303  & 0.9387  & 0.7962 \\
			$p_{16} (16)$  &  $\_$  &  $\_$  &  $\_$  & 1.1592  & 0.8846  & 0.7506 \\[3pt]
			\hline
		\end{tabular}
		\label{table:oscHalfLife_I}
	\end{center}
\end{table}

\begin{table}[!h]
	\begin{center}
		\caption{Half-life times of damping vibrational energies 
obtained by using the complex part of the oscillation frequencies for all 
three angular frequencies of stars for EoS models III and IV. All values are 
in milliseconds (ms).}
		\vspace{3pt}
		\setlength{\tabcolsep}{15pt}
		\begin{tabular}{cccc|ccc}
			\hline\hline\\[-8pt]
			\rule{0pt}{15pt} \shortstack{Modes\\(Order $n$)}
			& \multicolumn{3}{c}{Model III}
			& \multicolumn{3}{c}{Model IV} \\[5pt]
			
			\cline{2-4}\cline{5-7}
			\rule{0pt}{10pt}
			& 383 Hz & 546 Hz & 709 Hz
			& 383 Hz & 546 Hz & 709 Hz \\[2pt]
			\hline\hline\\[-12pt]
			$f (0)      $  & 11.2066 & 10.5233 & 10.8585 & 21.2295 & 22.6265 & 31.5558  \\
			$p_{1} (1)  $  & 4.9511  & 4.5462  & 4.5158  & 8.5722  & 8.1991  & 8.5663   \\
			$p_{2} (2)  $  & 3.1266  & 2.8755  & 2.8636  & 5.2721  & 5.0852  & 5.3970   \\
			$p_{3} (3)  $  & 2.2766  & 2.0971  & 2.0938  & 3.7492  & 3.6484  & 3.9397   \\
			$p_{4} (4)  $  & 1.7893  & 1.6501  & 1.6506  & 2.8851  & 2.8269  & 3.0960   \\
			$p_{5} (5)  $  & 1.4742  & 1.3607  & 1.3629  & 2.3331  & 2.2982  & 2.5456   \\
			$p_{6} (6)  $  & 1.2539  & 1.1581  & 1.1612  & 1.9522  & 1.9311  & 2.1589   \\
			$p_{7} (7)  $  & 1.0912  & 1.0083  & 1.0119  & 1.6745  & 1.6622  & 1.8726   \\
			$p_{8} (8)  $  & 0.9661  & 0.8931  & 0.8969  & 1.4638  & 1.4572  & 1.6524   \\
			$p_{9} (9)  $  & 0.8669  & 0.8017  & 0.8056  & 1.2987  & 1.2959  & 1.4779   \\
			$p_{10} (10)$  & 0.7863  & 0.7274  & 0.7313  & 1.1661  & 1.1661  & 1.3363   \\
			$p_{11} (11)$  & 0.7195  & 0.6658  & 0.6697  & 1.0574  & 1.0593  & 1.2193   \\
			$p_{12} (12)$  & 0.6633  & 0.6139  & 0.6177  & 0.9668  & 0.9701  & 1.1209   \\
			$p_{13} (13)$  & 0.6152  & 0.5696  & 0.5733  & 0.8902  & 0.8945  & 1.0372   \\
			$p_{14} (14)$  & 0.5737  & 0.5312  & 0.5349  & 0.8245  & 0.8296  & 0.9650   \\
			$p_{15} (15)$  & 0.5375  & 0.4978  & 0.5014  & 0.7678  & 0.7734  & 0.9023   \\
			$p_{16} (16)$  & 0.5056  & 0.4683  & 0.4718  & 0.7182  & 0.7242  & 0.8471   \\
			$p_{17} (17)$  & 0.4773  & 0.4422  & 0.4456  & 0.6745  & 0.6808  & 0.7983   \\
			$p_{18} (18)$  &  $\_$   &  $\_$   &  $\_$   & 0.6357  & 0.6423  & 0.7549   \\
			$p_{19} (19)$  &  $\_$   &  $\_$   &  $\_$   & 0.6011  & 0.6078  & 0.7160   \\
			$p_{20} (20)$  &  $\_$   &  $\_$   &  $\_$   & 0.5700  & 0.5769  & 0.6808   \\
			$p_{21} (21)$  &  $\_$   &  $\_$   &  $\_$   & 0.5420  & 0.5489  & 0.6490   \\
			$p_{22} (22)$  &  $\_$   &  $\_$   &  $\_$   & 0.5165  & 0.5235  & 0.6201   \\[3pt]
			\hline
		\end{tabular}
		\label{table:oscHalfLife_II}
	\end{center}
\end{table}

In Figs.~\ref{fig:radial_and_temporal_variation_of_lagrangian_disp_383Hz}, \ref{fig:radial_and_temporal_variation_of_lagrangian_disp_546Hz} and 
\ref{fig:radial_and_temporal_variation_of_lagrangian_disp_709Hz}, we plot the 
temporal evolutions of the $f$-mode oscillations on the surface of the stars 
for different models at angular frequencies $383$ Hz, $546$ Hz 
and $709$ Hz, respectively, within the range of half-life of each $f$-mode 
oscillation (top-left). The time axis starts from near zero, where the 
absolute amplitude of $U(r)$ is the highest. From 
Fig.~\ref{fig:radial_and_temporal_variation_of_lagrangian_disp_383Hz},  it is
seen that for Model I, the oscillation starts at an absolute amplitude of 
$1.0325\times10^{2}$ km$^{3}$ and ends at $1.6668$ ms (as shown in Table~\ref{table:oscHalfLife_I}) with an amplitude of $3.4710\times10^{-4}$ km$^{3}$. For 
Model IV, the $f$-mode has the highest half-life of $21.2295$ ms as shown in 
Table~\ref{table:oscHalfLife_II} for the stars with angular frequency $383$ 
Hz. The amplitude of $U(r)$ started at $2.3107\times10^{2}$ km$^{3}$ and damped 
to a value $1.1524\times10^{-3}$ km$^{3}$ at the half-life of 
oscillation. Similarly, for all other models at different angular 
frequencies, on the surface of the stars, the $f$-mode started at the amplitude 
of the order of $10^{2}$ km$^{3}$ and damped to a half-life with an amplitude 
of the order of $10^{-3} - 10^{-5}$ km$^{-3}$. We list the amplitudes of $U(r)$ 
at near $0$ s ($0.0 \tau$), at half of half-life ($0.5 \tau$) and at half-life 
($1.0 \tau$) in the Table~\ref{table:absAmpQROs}. At the surface, the 
absolute value of $U(r)$ remains almost the same order of magnitude across all 
the considered models and angular frequencies at each of the selected time 
intervals $0.0\tau$, $0.5\tau$, and $1.0\tau$. From the plots, 
it is observed that for Models II, III and IV, the $f$-mode decays and becomes
saturated after producing a certain number of oscillations. However, for Model 
I, it shows no oscillation and is saturated exponentially. Additionally, the 
stars with the highest mass-radius relation, that is the Model IV, show very 
high oscillations between $0 \tau$ and $10$ ms. The absolute value of $U(r)$ 
decays to the order of $10^{-1}$ km$^{3}$ at $10$ ms for all three angular 
frequencies. The radial variation of $U(r)$ for different models at different 
angular frequencies varies for these three time values, which can be seen from 
the top-right, bottom-left and bottom-right plots of Figs.~\ref{fig:radial_and_temporal_variation_of_lagrangian_disp_383Hz}, \ref{fig:radial_and_temporal_variation_of_lagrangian_disp_546Hz}, and \ref{fig:radial_and_temporal_variation_of_lagrangian_disp_709Hz}.

\begin{table}[!h]
	\begin{center}
		\caption{Absolute values of $U(r)$ at the surface of the stars, 
for different models at different angular frequencies at near $0$ s 
($0.0 \tau$), at half of the half-life ($0.5 \tau$) and at the half-life 
($1.0 \tau$), where $\tau$ is the half-life of the oscillations. Here all the 
values of $U(r)$ are in km$^{3}$.}\vspace{3pt}
		\setlength{\tabcolsep}{15pt}
		\begin{tabular}{ccccc}
			\hline \hline\\[-12pt]
			\rule[24pt]{0pt}{0pt}
			\shortstack{Models \\[8pt]}
			& \shortstack{$\Omega$ \\[2pt] (Hz)}
			& $0\tau$
			& $0.5\tau$
			& $1.0\tau$ \\[8pt]
			\hline\hline
			\rule[10pt]{0pt}{0pt}
			  & 383 & 1.0325e2 & 5.3997e-2 & 3.4710e-4 \\[2pt]
			I & 546 & 1.0325e2 & 6.0870e-2 & 3.3148e-4 \\[2pt]
			  & 709 & 1.0325e2 & 5.8459e-2 & 3.3716e-3 \\[2pt]
			
			\hline
			\rule[10pt]{0pt}{0pt}
			
			   & 383 & 1.4031e2 & 2.7918e-1 & 5.5127e-4 \\[2pt]
			II & 546 & 1.4076e2 & 8.7590e-2 & 4.8076e-4 \\[2pt]
			   & 709 & 1.4092e2 & 8.0829e-2 & 4.4041e-4 \\[2pt]
			
			\hline
			\rule[10pt]{0pt}{0pt}
			
			    & 383 & 1.5435e2 & 7.2754e-2 & 5.0423e-4 \\[2pt]
			III & 546 & 1.5456e2 & 2.0069e-1 & 2.1928e-5 \\[2pt]
			    & 709 & 1.5446e2 & 1.4171e-1 & 2.8889e-4 \\[2pt]
			
			\hline
			\rule[10pt]{0pt}{0pt}
			
			   & 383 & 2.3107e2 & 5.8366e-1 & 1.1524e-3  \\[2pt]
			IV & 546 & 2.2221e2 & 4.7840e-1 & 4.7162e-4 \\[2pt]
			   & 709 & 1.5566e2 & 9.1664e-2 & 3.3769e-4 \\[2pt]
			
			\hline\hline
		\end{tabular}
		\label{table:absAmpQROs}
	\end{center}
\end{table}


\section{Summary and Conclusion}\label{secVI}
In this work, our primary focus is to investigate different QS models based
on three different QM EoSs and study their radial and quasiradial 
oscillations. Under ordinary conditions, quarks 
remain confined inside the hadrons, such as neutrons and protons. However, at 
sufficiently high densities, deconfinement and chiral symmetry restoration 
may take place, allowing the formation of stable deconfined QM that 
may exist in the form of pure QSs. In the first part of our work, we calculate 
the mass, radius, and compactness of different stellar models by solving the 
TOV equations. Subsequently, we compute the radial oscillation frequencies up 
to $100$ kHz and analyze their radial perturbation through the dimensionless 
quantity $\xi$, and the dimensionless pressure perturbation parameter $\eta$. 
We also analyze the radial oscillations in the radial direction using the 
quantity $U(r)$. The radial oscillations do not have much significance for the
non-rotating configuration. While in rotating configuration, those radial 
oscillations are affected by centrifugal force and can emit GWs through the 
decay of vibrational energy. We use HTC formalism for the slowly rotating relativistic star to calculate the QROs and 
the physical 
parameters of stellar models modified due to rotation. In this formalism, 
the perturbative approach has been used, which gives different rotational 
terms, such as spherically symmetric mass, radius and pressure corrections' 
terms, and terms for the frame-dragging, angular frequencies, angular momentum, 
eccentricity, and mass quadrupole moment. Based on this formalism, the QRO 
frequencies are determined by the parameters of the non-rotating, 
pulsating–non-rotating, and rotating–non-pulsating configurations, whereas the 
fluid motion, characterized by $U(r)$, depends solely on the 
pulsating–nonrotating configuration. We study the $f$-mode of QROs, since it 
has the highest oscillation decay 
half-time as well as frequency range within observational significance. 
Further, we calculate the dimensionless quadrupole moment and dimensionless 
moment of inertia, which are almost independent of the stellar angular 
frequencies. These dimensionless quantities are useful as they give a 
universal relationship between the moment of inertia, the Love number and the 
quadrupole moment (I-Love-Q). These quantities can be measured based on the 
surface properties of a stellar body by an external observer without 
considering the angular frequency of the object. This can help us in 
constraining the different EoSs just by measuring at least two of these three 
parameters.

It is observed that the mass and radius of the stars increase when we 
consider massive quark EoS in comparison to massless quark flavor EoS for the 
nearly equal central baryon number density. Additionally, it is observed
 that the mass and radius of stars increase with increasing temperature of the 
stars, while the central baryon number density decreases.

The radial variation of $\xi$ and $\eta$ for Model I is observed to be 
significant than other models only for the first three modes of radial 
oscillations. For the last 
three modes, there is no significant difference in the radial variation of 
$\xi$ and $\eta$ for the models. Whereas there is no significant difference 
in the amplitude of $U(r)$ for all the models for both the first and last 
three modes of radial oscillations.

Since the $f$-mode is considered to be the observationally most significant 
mode for isolated stars in the absence of any violent astrophysical events, 
we further investigate the QROs of slowly rotating relativistic stars at the 
equatorial plane. It is observed that for the rotating stars, a higher compact 
star produces less frame-dragging as the relation 
$\bar{\omega} = \Omega - \omega$ has a higher value for the stars with 
smaller values of mass, radius and compactness. For a star rotating with a 
constant angular velocity $\Omega$, $\bar{\omega}$ increases as shown in 
Fig.~\ref{fig:frameDraggingForAllEOSForAllThreeFrequency} if $\omega$, the 
angular velocity of the local inertial frame decreases. Which means the local 
inertial frame gets twisted more for the stars with higher mass, radius and 
compactness throughout the radial length of the stars. But there is no change 
in the amount of twisting of the inertial frame for different angular 
velocities of the stars. Hence, the twisting of the local inertial frame is 
independent of the angular velocity of the star, while it depends on how 
big and massive a star is. However, the matter distribution inside the star 
is affected by the rotation of the star. The mass distribution towards the 
surface increases due to rotation. The stars with higher mass and radius 
have higher mass realignment at the surface due to the rotation for all 
angular velocities, such that the mass realignment at the surface increases 
with increasing angular velocity of the stars. Due to rotation, the stars' 
microscopic properties such as baryon number density distribution throughout 
the stellar interior changes. For all stars, the change in distribution of 
number density increases towards the surface of the stars. Also, the number 
density distribution changes more for the stars with higher angular velocity. 
Similarly, the change in pressure distribution and energy density distribution 
of the stars increases towards the surface and also the change in these 
distributions increases with increasing angular velocity of the stars. 
Furthermore, the stars with higher mass and radius values are affected more 
by rotation. Their number density distribution, pressure distribution and 
energy density distribution have changed more from their non-rotating 
configurations.

The normalized I-Love-Q quantities \cite{yagi_2013} were calculated 
using the rotation-corrected masses and radii rather than the non-rotating 
ones. Although rotational deformation increases with angular frequency, the 
dimensionless moment of inertia $\bar{I}$ and quadrupole moment $\bar{Q}$ 
remain nearly independent of spin within the slow-rotation regime, supporting 
the approximate universality of normalized rotational observables. Model-IV 
exhibits the highest eccentricity, indicating comparatively stronger surface 
deformation. However, its values of $\bar{I}$ and $\bar{Q}$ suggest a more 
centrally concentrated configuration with stronger self-gravity, which 
suppresses rotational deformation. Consequently, despite exhibiting larger 
surface oblateness, Model-IV stars are comparatively more resistant to global 
deformation due to stronger relativistic gravitational binding. $\bar{I}$ and
$\bar{Q}$ remove the dominant spin contribution and are governed primarily by 
the stellar compactness, making them useful quantities for testing theoretical 
EoS models when combined with future observational constraints on compact 
stars' masses, radii and rotational properties.

The eigenfrequencies of radial oscillations shift from their non-rotating 
values due to the rotation of the stars. Because of rotation, we get the 
eigenfrequencies as complex numbers. The real part gives the shifted 
eigenfrequencies from a non-rotating configuration, while the imaginary part 
tells the stability and 
decay time of the vibrational oscillations. The imaginary part of the 
eigenfrequency varies with the variation of the angular velocity of the stars. 
Furthermore, the study of QROs for all four models shows that the $f$-mode 
has the longest half-time of vibrational decay among all other overtones. Also, 
the comparison of models shows that the EoS where quark flavors are 
considered massless has the lowest half-life for $f$-mode than the models with 
massive quark flavors. For bigger and massive stars, more noticeable surface 
oscillations for a longer period of time are observed before the amplitude of 
oscillation become much smaller to be able to detect. Moreover, for the stars 
with the massive quark flavor models, with increasing temperature of the star, 
the half-life of vibrational decay of $f$-mode increases. 

Despite the evidence of the existence of QM and possible QS candidates, there 
is no direct detection of QSs yet. In our work, we studied the $f$-mode of 
radial oscillations affected by the centrifugal force of slowly rotating 
relativistic stars. Their vibrational frequency is below $2$ kHz, which is 
actually within the observational range of current GW observations. But our 
calculation shows that these oscillations are short-lived, of the order of 
$\sim 1$ ms -- $32$ ms, which shows that these oscillations are not 
continuous. Also, it will be helpful if we can calculate the strain produced 
at the background to differentiate them from other GWs from different sources, 
which is not included in this work. Additionally, the models of QSs used in 
this study can be assumed as the quark core of HSs and study the mixed phase 
within it to get a more realistic view of NSs with a quark core inside them.
\newpage

\section*{Acknowledgments}
UDG is thankful to the Inter-University Centre for Astronomy and Astrophysics 
(IUCAA), Pune, India for the Visiting Associateship of the institute.

\appendix
\section{Driving Function and Weight Function}\label{appen1}
For the calculation of $\sigma^{(2)}$ we need the driving function 
$\mathfrak{D}$ and weight function $\mathfrak{W}$. The different terms of 
$\mathfrak{D}$ are given as
\begin{align}
	A_{1} &= m_{0} r^{-4} \Gamma (\epsilon + p),\\[5pt]
	A_{2} &= \frac{1}{2} p_{0}^{*} \Gamma (\epsilon + p) r^{-3} e^{\lambda + \nu/2} \left[\frac{\epsilon + p}{\gamma p} - \frac{\epsilon}{p}\right]\left(1 - e^{-\lambda}\right),\\[5pt]
	A_{3} &= -\frac{2}{3} \bar{\omega} \bar{\omega}^{\prime} e^{-\nu/2} \left[\Gamma(\epsilon+p) + 2 \frac{(\Gamma p)^{2}}{\epsilon + p}\right],\\[5pt]
	A_{4} &= -\frac{1}{12} \bar{\omega}^{\prime\,2} r e^{-\nu/2} \Gamma (\epsilon+p),\\[5pt]
	A_{5} &= \frac{2}{3} \bar{\omega}^3 r^{-1} e^{\lambda - \nu/2} \Gamma p \Bigg[-\frac{1}{2} \left(3 e^{-\lambda} - 1\right) \frac{\epsilon + p}{p} + \frac{1}{2} \frac{\Gamma p}{\epsilon + p} \left(1 - 5 e^{-\lambda}\right) + \frac{1}{2} \Gamma \left(1 - e^{-\lambda}\right) \left(1 - \frac{1}{\gamma}\right)\nonumber \\[5pt]
	& + 4 \pi r^{2} \Gamma p \left(1 - \frac{1}{\gamma} + \frac{p}{\epsilon + p}\right) - r e^{-\lambda} \frac{\Gamma^{\prime} p}{\epsilon + p} \Bigg],\\
	B_{1} &= m_{0} r^{-5} e^{3\lambda +\nu/2} \Bigg[ \Gamma (\epsilon + p) \Biggl\{-\frac{1}{2} \left(1 - e^{-\lambda}\right) + 4 \pi r^{2} p \left(1 + 2 e^{-\lambda}\right) + 64 \pi^{2} r^{4} p^{2}\Biggr\} + (\epsilon + p + \Gamma p) \times \nonumber\\[5pt]
	& \Biggl\{-1 - 3 e^{-\lambda} - 16 \pi r^{2} p \left(1 + \frac{1}{2} e^{-\lambda}\right) - 64 \pi^{2} r^{4} p^{2}\Biggr\} + r e^{-\lambda} \Gamma^{\prime} p \left(1 + 8 \pi r^{2} p\right) - 2 (\epsilon + p) (\sigma^{(0)})^{2} r^{2} e^{-\lambda - \nu}\Bigg], \\[5pt]
	B_{2} &= 2 h_{0} r^{-2} (\epsilon + p) e^{\lambda - \nu/2} (\sigma^{(0)})^{2},\\[5pt]
	B_{3} &= p_{0}^{*} (\epsilon + p) r^{-4} e^{2\lambda + \nu/2} \Bigg[ \left(\frac{\epsilon + p}{\gamma p} - \frac{\epsilon}{p}\right) \Biggl\{-(\sigma^{(0)})^{2} r^{2} e^{-\lambda - \nu} - \frac{1}{4} \left(1 - e^{-\lambda}\right)\left(1 + 7 e^{-\lambda}\right)\Biggr\} \nonumber\\[5pt]
	& + 4 \pi \Gamma^{\prime} p r^{3} e^{-\lambda} - 2 \pi p r^{2} \Biggl\{\left(1 + e^{-\lambda}\right)\left(2 + \Gamma\right) + 8 \pi r^{2} p \left(1 + \Gamma\right)\Biggr\}\nonumber \\[5pt]
	& - 2\pi (\epsilon + p) r^{2} \Biggl\{\left(1 - e^{\lambda}\right)\Gamma + \left(1 + e^{-\lambda}\right) \left(2 - \Gamma\right)\frac{1}{\Gamma} + 8 \pi p r^{2} \left(1 - \Gamma\right) \left(1 + \frac{1}{\gamma}\right)\Biggr\}\Bigg], \\[5pt]
	B_{4} &= 4 \bar{\omega} \bar{\omega}^{\prime} e^{-\nu/2} \left(\epsilon + p + \frac{1}{3} \Gamma p\right), \\[5pt]
	B_{5} &= \frac{2}{3} \bar{\omega}^{\prime\,2} e^{\lambda - \nu/2} \Bigg[ \pi r^{2} \left(1 - \frac{\Gamma}{2}\right) p \left(\epsilon + p\right) + \pi r^{2} \Gamma p^{2}\nonumber \\[5pt]
	& + \frac{1}{16} \Gamma (\epsilon + p) \left(1 - e^{-\lambda}\right) + \frac{1}{8} \left(\epsilon + p + \Gamma p\right) \left(1 + 7 e^{-\lambda}\right) - \frac{1}{8} r \Gamma^{\prime} p e^{-\lambda}\Bigg], \\[5pt]
	B_{6} &= \frac{2}{3} \bar{\omega}^{2} r^{-2} e^{\lambda-\nu/2} \Bigg[ - (\epsilon + p - \Gamma p) (\sigma^{(0)})^{2} r^{2}  e^{-\nu} + (\epsilon  + p) \Biggl\{\frac{31}{4} e^{-\lambda} - \frac{5}{2} - \frac{1}{4} e^{\lambda} + \frac{1}{2} \Gamma \left(e^{-\lambda} - 1\right)\Biggr\}\nonumber \\[5pt] 
	& + \Gamma p \Biggl\{-\frac{11}{4} e^{-\lambda} + \frac{3}{2} + \frac{e^{\lambda}}{4}\Biggr\} + 4 \pi r^{2} (\epsilon + p) p \left(3 + e^{\lambda}\right) \left(\frac{1}{2}\Gamma + 1\right)\nonumber \\[5pt] 
	& + 4 \pi r^{2} \Gamma p^{2} \left(1 + e^{\lambda}\right) + 16 \pi r^{4} p^{2} e^{\lambda} \left((\Gamma -1) (\epsilon + p) + \Gamma p\right) + r \Gamma^{\prime} p e^{-\lambda} \Bigg],
\end{align}
where $\Gamma$ is adiabatic index for pulsations of the rotating configuration and is given by
\begin{equation}
	\Gamma = \frac{\mathcal{N}}{\mathcal{P}}\,\frac{d\mathcal{P}}{d\mathcal{N}}.
\end{equation}
Thus, the driving function is defined as
\begin{equation}
	\mathfrak{D} = U(r)^{\prime}\left(A_{1} + A_{2} + A_{3} + A_{4} + A_{5}\right) + U(r) \left(B_{1} + B_{2} + B_{3} + B_{4} + B_{5} + B_{6}\right). 
\end{equation}
Also, the weight function is defined as
\begin{equation}
	\mathfrak{W} = e^{(\nu + 3\lambda)/2} r^{-2} (\epsilon + p).
\end{equation}

\end{document}